\documentclass[twocolumn]{aastex61}
\usepackage{natbib}
\usepackage{color}
\usepackage{graphicx}
\bibpunct{(}{)}{;}{a}{}{,}
\newcommand\taug{{\tau_g}}
\newcommand\tauin{{\tau_{in}}}
\newcommand\eagle{EAGLE}

\received{July 1, 2016}
\revised{September 27, 2016}
\accepted{\today}
\submitjournal{ApJ}

\shorttitle{Origin of the gas metallicity vs. size relation}
\shortauthors{S\'anchez Almeida \& Dalla Vecchia}

\begin{document}

\title{The origin of the relation between metallicity and size in star-forming galaxies}

\correspondingauthor{J. S\'anchez Almeida}
\email{jos@iac.es}

\author[0000-0003-1123-6003]{J. S\'anchez Almeida}
\affil{Instituto de Astrof\'\i sica de Canarias, La Laguna, Tenerife, E-38200, Spain}
\affil{Departamento de Astrof\'\i sica, Universidad de La Laguna}

\author[0000-0002-2620-7056]{C. Dalla Vecchia}
\affil{Instituto de Astrof\'\i sica de Canarias, La Laguna, Tenerife, E-38200, Spain}
\affil{Departamento de Astrof\'\i sica, Universidad de La Laguna}




\begin{abstract}
For the same stellar mass,  physically smaller star-forming galaxies   are also metal richer \citep{2008ApJ...672L.107E}. What causes the relation remains unclear. The central star-forming galaxies in the EAGLE cosmological numerical simulation reproduce the observed trend. We use them to explore the origin of the relation assuming that the physical mechanism responsible for the anti-correlation between size and gas-phase metallicity is the same in the simulated and the observed galaxies. We consider the three most likely causes: (1) metal-poor gas inflows feeding the star-formation process, (2) metal-rich gas outflows particularly efficient in shallow gravitational potentials, and (3) enhanced efficiency of the star-formation process in compact galaxies.  Outflows (2) and enhanced star-formation efficiency (3) can be discarded. Metal-poor gas inflows (1) cause the correlation in the simulated galaxies. Galaxies grow in size with time, so those that receive gas later are both metal poorer and larger, giving rise to the observed anti-correlation. As expected within this explanation, larger galaxies have younger stellar populations. We explore the variation with redshift of the relation, which is maintained up to,  at least, redshift 8. 
\end{abstract}




\section{Introduction} \label{sec:intro}

Based on some 44,000 star-forming galaxies from the Sloan Digital Sky Survey (SDSS), \citet{2008ApJ...672L.107E} found a relation connecting stellar mass ($M_\star$),  galaxy size (as parameterized by the half-light radius), and gas-phase metallicity. They discovered that at fixed $M_\star$, physically smaller galaxies are also metal richer. The metallicity changes by 0.1 dex when the galaxy size changes by a factor of 2. The authors discard observational biases due to the finite size of the central region used to estimate metallicities, and to the Hubble type dependence of the radius. A similar relationship between size and gas-phase metallicity was also found by \citet{2012ApJ...750..142B} and \citet{2015ApJ...800...91H}, and it remains in place at redshift  $\simeq 1.4$ as measured by \citet{2012PASJ...64...60Y,2014MNRAS.437.3647Y}. \cite{2004ApJ...613..898T} and \citet{2015ApJ...810..151W} observed that, given $M_\star$,  galaxies with higher stellar surface density are also metal richer, which implies a relation between metallicity and size in qualitative agreement with all these other works.

The physical cause of the observed relation remains unclear. \citet{2008ApJ...672L.107E} considered and discarded both metal poor gas inflows and metal rich gas outflows. Their argument was based on comparing with the simple chemical evolution models by \citet{2008MNRAS.385.2181F}, and the fact that they do not satisfactory explain the observed correlation.  \citeauthor{2008ApJ...672L.107E} favor differences in star-formation efficiencies. Small galaxies are denser and exhaust their gas faster, and thus become metal enriched sooner. On the other hand,  \citet{2014A&ARv..22...71S} pointed out that the relation is a natural  outcome of the gas accretion driven star-formation (SF) process. In the stationary state the gas-phase metallicity is set by the efficiency of the outflows, which changes systematically with halo mass, that is to say, with the depth of the gravitational potential the baryons have to escape from. The gravitational binding energy depends on the distance to the center of the gravitational well, therefore, at a fixed mass, winds escape easier from larger galaxies. Finally, \citet{2012PASJ...64...60Y} invoke metal poor gas accretion driven by mergers with no further elaboration.   

Whatever the explanation may be, it is telling us about  
the basic physics underlying the star-formation process, since the observed anti-correlation between size and metallicity is likely a  fundamental property of galaxies. Although less studied than the others, the relation found by \citet{2008ApJ...672L.107E} belong to the realm of the well-known empirical relations linking global properties of star-forming galaxies, including the {\em main sequence} \citep[scaling between star-formation rate, SFR, and $M_\star$; e.g.,][]{2007ApJ...660L..43N,2007ApJ...670..156D}, the mass-metallicity relation \citep[e.g.,][]{1981ARA&A..19...77P,2004ApJ...613..898T}, the {\em fundamental metallicity relation}  \citep[connecting $M_\star$, gas-phase metallicity, and SFR;][]{2010MNRAS.408.2115M,2010A&A...521L..53L}, or even the lopsidedness-metallicity relation \citep{2009ApJ...691.1005R,2011ApJ...743...77M}. All together provide the main observational constraints  to understand the subtleties  of the mechanism by which galaxies form and grow. It is generally accepted that galaxies grow in a self-regulated process controlled by gas accretion and feedback from SF and black holes \citep[e.g.,][]{2010ApJ...718.1001B,2011MNRAS.416.1354D,2012MNRAS.421...98D,2012RAA....12..917S,2013ApJ...772..119L,2017ASSL..430...67S}. However, the way in which the global properties of galaxies emerge from the underlying physical processes is not properly understood yet.
    
Here we revisit the problem of explaining the anti-correlation between galaxy size and gas metallicity. We use the EAGLE cosmological numerical simulations  \citep{2015MNRAS.446..521S,2015MNRAS.450.1937C,2016A&C....15...72M}.
The model includes a pressure-based law for star formation \citep{2008MNRAS.383.1210S}, line cooling in photoionisation equilibrium \citep{2009MNRAS.393...99W}, stellar evolution \citep{2009MNRAS.399..574W}, thermal supernova feedback \citep{2012MNRAS.426..140D}, and black holes growth and feedback \citep{2015MNRAS.454.1038R}. An extensive description of the model, its calibration and the hydrodynamics solver are given in \cite{2015MNRAS.446..521S}, \cite{2015MNRAS.450.1937C} and \cite{2015MNRAS.454.2277S}, respectively.
They represent  state-of-the-art  cosmological numerical simulations that self-consistently include baryon physics. 
(Other simulations of this kind are described in, e.g.,  \citet[][]{2014MNRAS.445..581H,2017arXiv170206148H,2014MNRAS.442.1545C,2014Natur.509..177V}, or \citet{2018MNRAS.475..676S}.)
As we will show, the EAGLE galaxies reproduce the trend observed by \citet{2008ApJ...672L.107E} and, thus, assuming that they are grasping the essentials of the physical process giving rise to the anti-correlation, we study them to identify what can be causing the observed trend. Obviously, the appropriateness of the explanation depends on whether this working hypothesis is correct, which is a caveat affecting the whole paper.

The EAGLE model galaxies have already proven their potential to reproduce some of the well known scale relations, which encouraged us to use them in the present context. Explicitly,  \citet{2015MNRAS.446..521S} demonstrate that the simulations reproduce a correlation between stellar mass and gas-phase metallicity in agreement with observed data at redshift zero. \citet{2016MNRAS.459.2632D} show the existence of a relation between gas fraction, stellar mass, and current SFR, with the galaxies distributed on a plane in the 3D spaced defined by these three parameters.  \citet{2017MNRAS.472.3354D} analyze the fundamental metallicity relation, finding a very good matching with observations up to redshift 5.  They also find that the physical parameter that best correlates with metallicity is gas fraction. 
EAGLE galaxies also provide a relation between mean size and stellar mass in agreement with the relation observed in the local universe \citep{2015MNRAS.446..521S,2015MNRAS.450.1937C,2017MNRAS.471L..11D}.  This result is reassuring for our analysis, which relies on galaxy sizes.

The paper is organized as follows. Section~\ref{sec:eagle} shows how the EAGLE numerical simulations produce galaxies slightly more metallic when their stellar sizes are smaller. Such difference quantitatively agrees with the difference observed by  \citet{2008ApJ...672L.107E} -- Sect.~\ref{sec:eagle}. Thus, the physical mechanism responsible for the difference of gas metallicity between small and large galaxies in the simulations may  also responsible for the difference in observed galaxies. The question arises as to what is this physical mechanism. Section~\ref{sec:framework}  explores the three obvious possibilities, namely, metal-poor gas inflows feeding the SF process, metal-rich gas outflows particularly efficient in shallow gravitational potentials, and enhanced efficiency of the SF in compact galaxies. This first exploration is based on simple analytical {\em bathtub} models (Sect.~\ref{sec:framework} and App.~\ref{sec:appb}). Then we study these three possibilities in the EAGLE galaxies. It is clear that outflows (Sect.~\ref{sec:well}) and SF efficiency controlled by density (Sect.~\ref{sec:mdensity}) can be discarded as the underlying cause. We are left with metal-poor gas inflows, which  cause the correlation in the model galaxies. Galaxies grow in size with time, so those that receive gas later are both metal poorer and larger, giving rise to the observed correlation (Sect.~\ref{sec:thisisit}).  We also explore the variation with redshift of the relation between size and gas-phase metallicity, which is shown to be maintained and even strengthened at higher redshifts (up to redshift 8; Sect.~\ref{sec:redshift}). The increase in galaxy size with time is a central ingredient of our explanation. The physical cause of this growth is extensively discussed in the literature, and we examine the various possibilities in Sect.~\ref{sec:conclusions}. The general results and conclusions are also summarized in Sect.~\ref{sec:conclusions}.      

\section{The size metallicity relation in the EAGLE simulation}\label{sec:eagle}

The suite of \eagle\ simulations is described in detail by \citet{2015MNRAS.446..521S} and \citet{2015MNRAS.450.1937C}.
We use the run covering the largest volume, namely  L100N1504, which corresponds to a 100 comoving-Mpc cube and goes all the way from redshift 127 to redshift 0. It has initial gas mass particles of $\sim 2\times 10^6\,M_\odot$ and  initial dark matter particles of $\sim 10^7\,M_\odot$.  All the relevant physical parameters employed in the present study were retrieved by querying the \eagle\ SQL\footnote{Structured Query Language} web interface\footnote{\tt http://galaxy-catalogue.dur.ac.uk:8080/Eagle/} \citep[][]{2016A&C....15...72M}.

\begin{figure*}
\begin{center}
\includegraphics[width=0.42\linewidth]{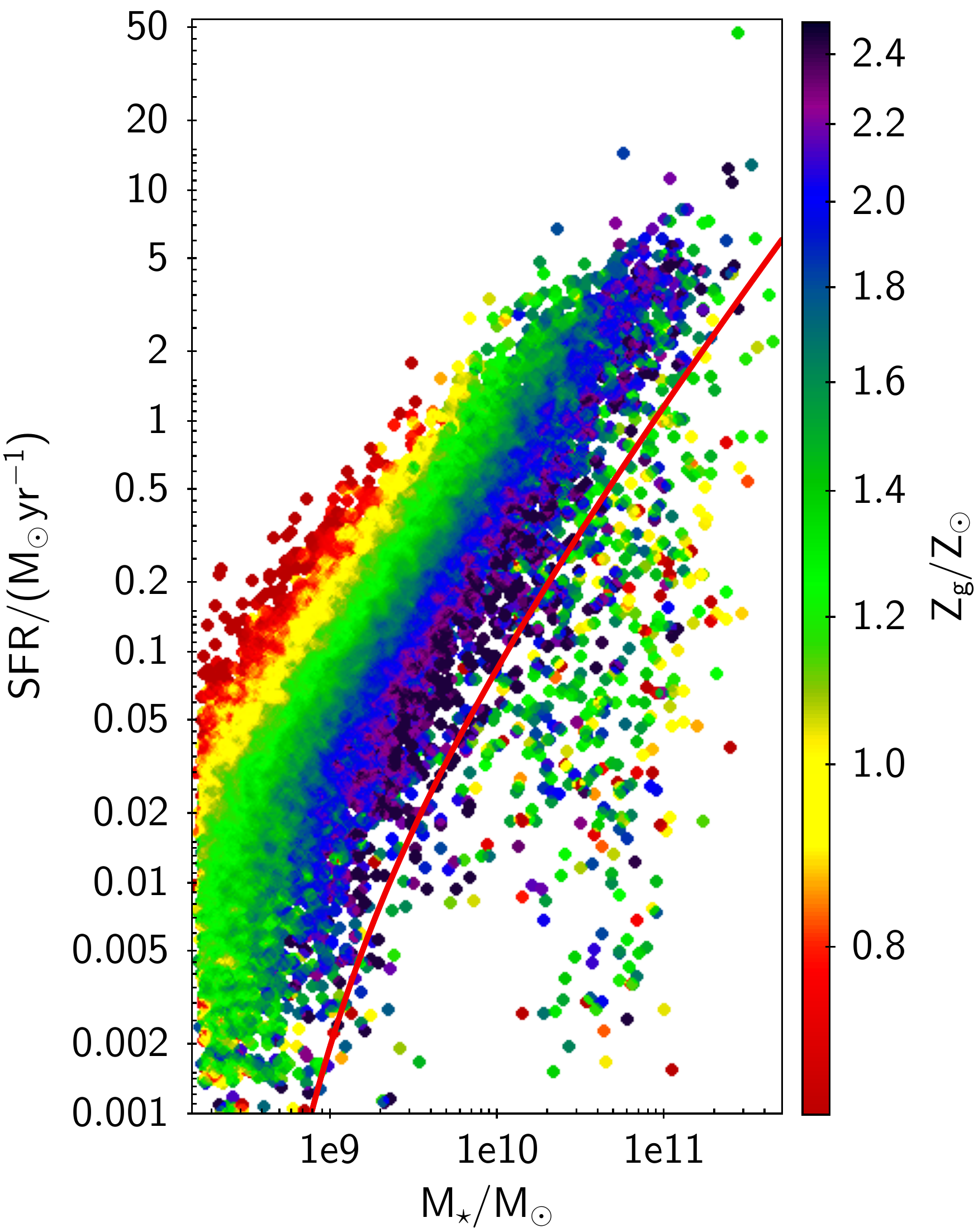} 
\includegraphics[width=0.42\linewidth]{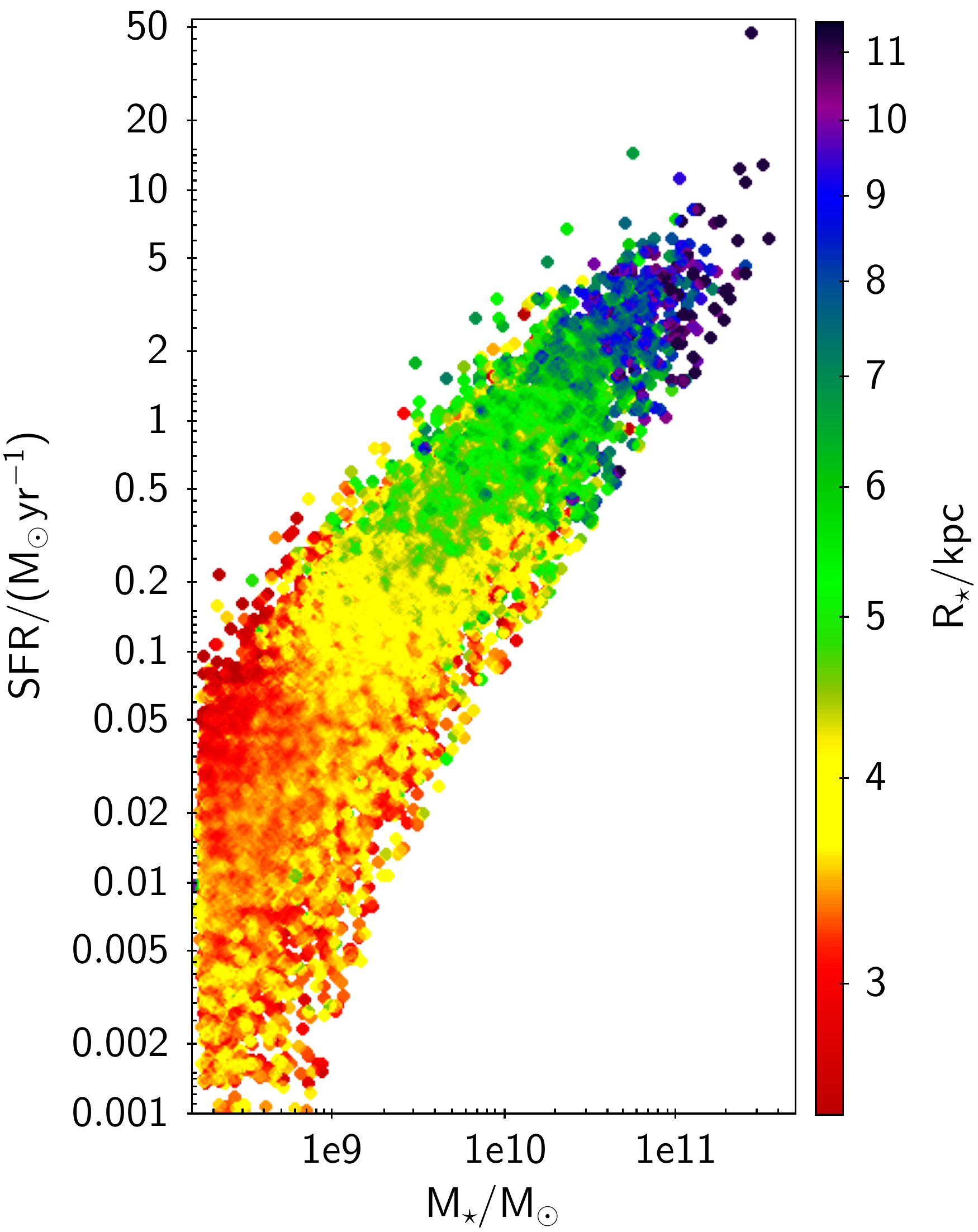}
\end{center}
\caption{Left: main sequence, i.e. SFR versus $M_\star$, color-coded with gas metallicity. It includes the full population of central galaxies in \eagle\  at redshift zero. Our study only considers star-forming galaxies, selected as those above the solid red line. Given $M_\star$, the larger the SFR the lower metallicity. Right: {\em cleaned} main sequence color-coded with half stellar-mass radius. Masses are referred to the solar mass, $M_\odot$, and metallicities to the solar metallicity, $Z_\odot$, assumed to be 0.02, here and throughout the paper.
}
\label{eq:ellison2_save_ssfr}
\end{figure*}

The selection of all central galaxies at redshift zero renders 16671 objects. This set was further filtered out to remove objects clearly out the main sequence, i.e., the well defined relation between $M_\star$ and SFR followed by star-forming galaxies \citep[e.g.,][]{2007ApJ...660L..43N,2007ApJ...670..156D}. Figure~\ref{eq:ellison2_save_ssfr}, left panel, shows the full set and the divide we use, so that only the 13410 galaxies above the line are retained for further analysis. The resulting trimmed main sequence is shown in Fig.~\ref{eq:ellison2_save_ssfr}, right panel.
Our analysis is based on the star-forming gas metallicity ($Z_g$, which traces dense gas), and the half-mass radius ($R_\star$) computed for the mass within a spherical 100 kpc aperture. We checked that the conclusions in the paper do not qualitative change when using a spherical 30 kpc aperture to measure $R_\star$. The \eagle\ database does not provide half-light radii, and we use $R_\star$ as a proxy for them. Here and throughout the paper, $Z_g$ has been referred to the solar metallicity, which we take as $Z_\odot =0.02$ to comply with the nucleosynthetic yields used in EAGLE \citep[e.g.,][]{2001A&A...370..194M,1998A&A...334..505P}.

\begin{figure}
\includegraphics[width=0.42\textwidth]{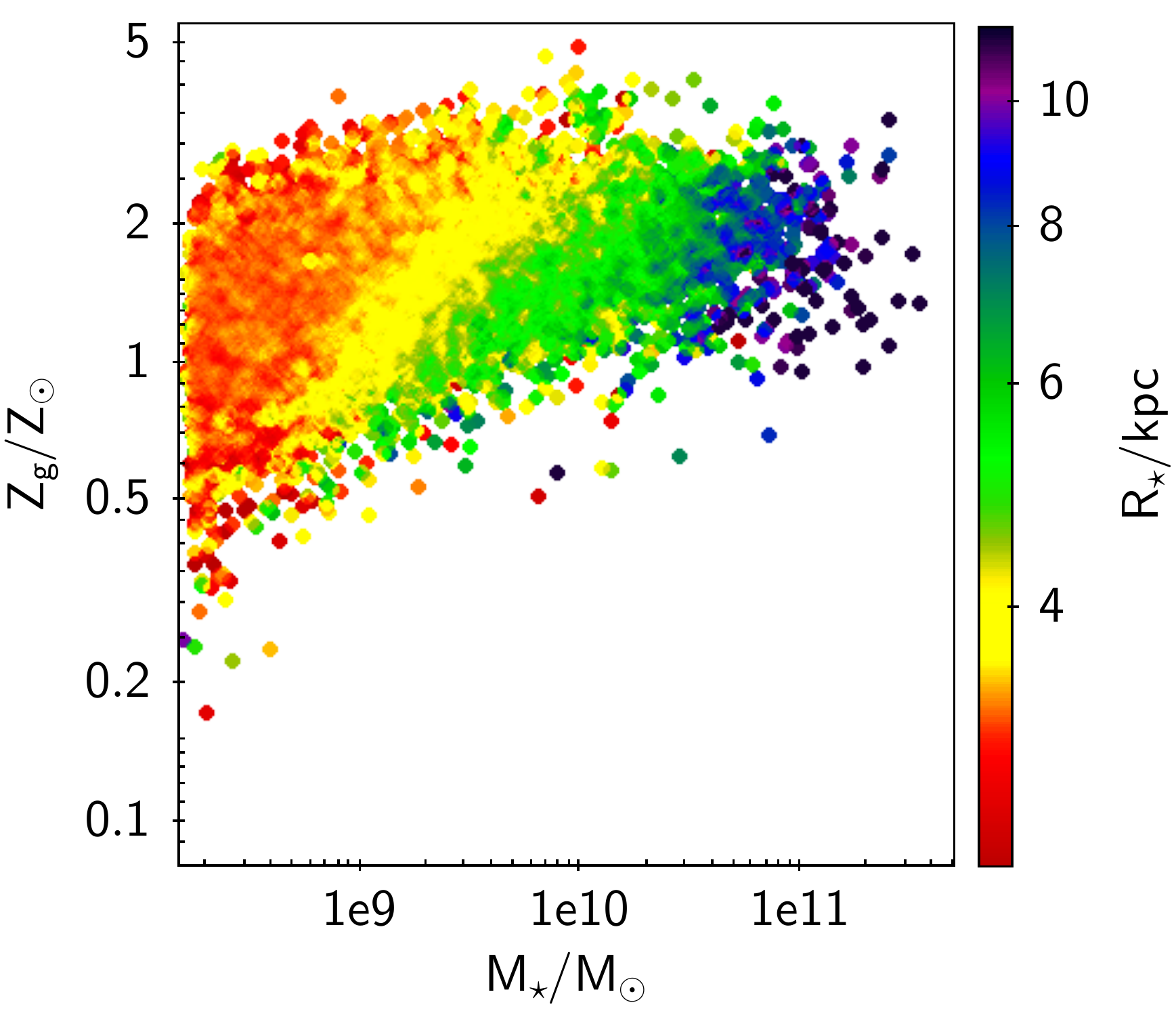}
\caption{Gas-phase metallicity versus stellar mass relation color-coded according to the mean value of the galaxy radius at each location of the plane. Metallicities are referred to the solar metallicity, masses to the solar mass, and sizes to one kpc. For a given $M_\star$, smaller galaxies tend to be more metallic, in qualitative agreement with observations.} 
\label{fig:ellison2o}
\end{figure}

Figure~\ref{fig:ellison2o} shows the gas metallicity versus $M_\star$ relation for the galaxies in EAGLE. The points are color-coded according to the mean  half-mass radius of all the galaxies having the same $Z_g$ and $M_\star$. The simulation clearly reveals an anti-correlation between size and metallicity; fixed $M_\star$, the smaller the galaxy the larger its metallicity. 

\begin{figure}
\plotone{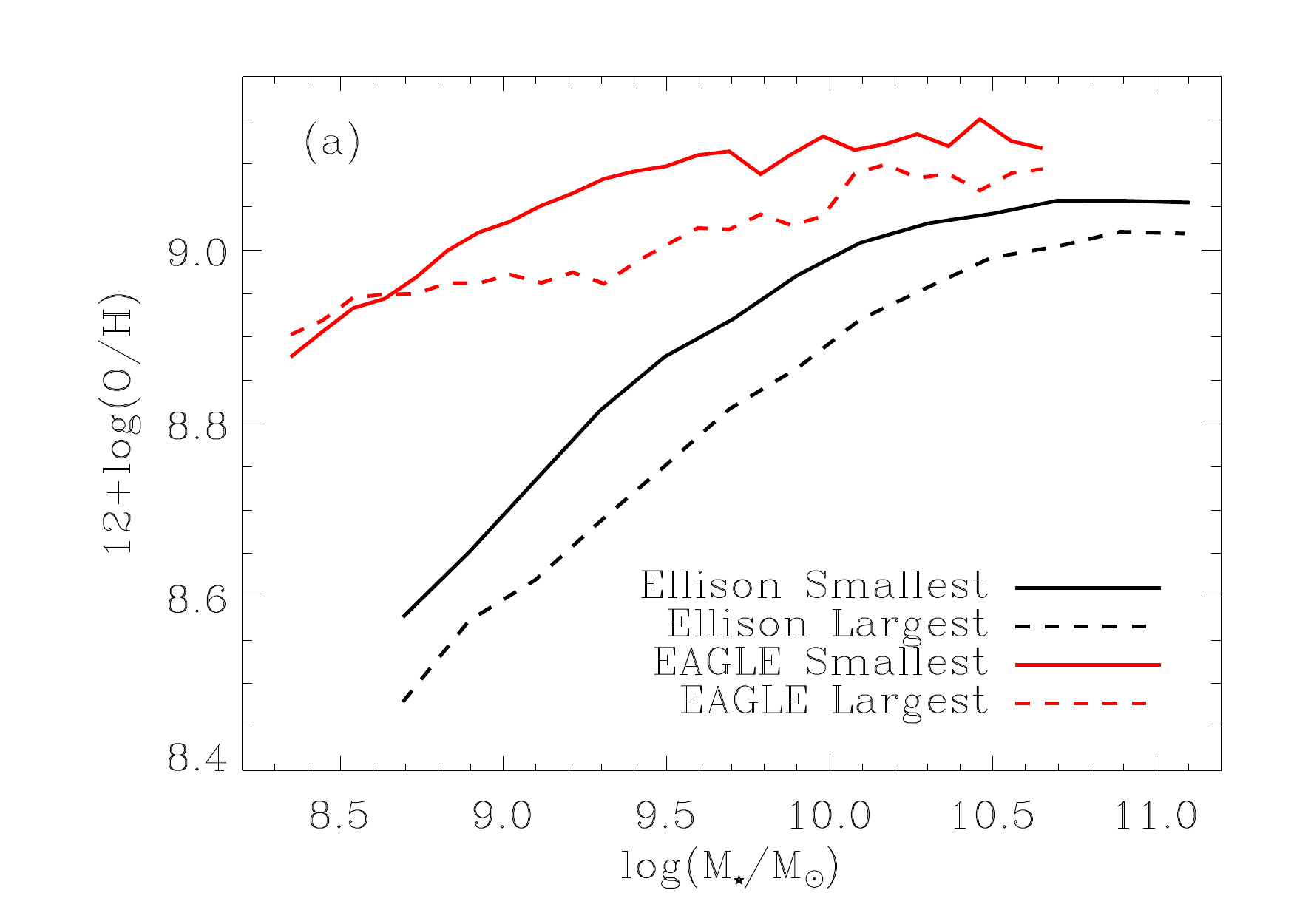}
\plotone{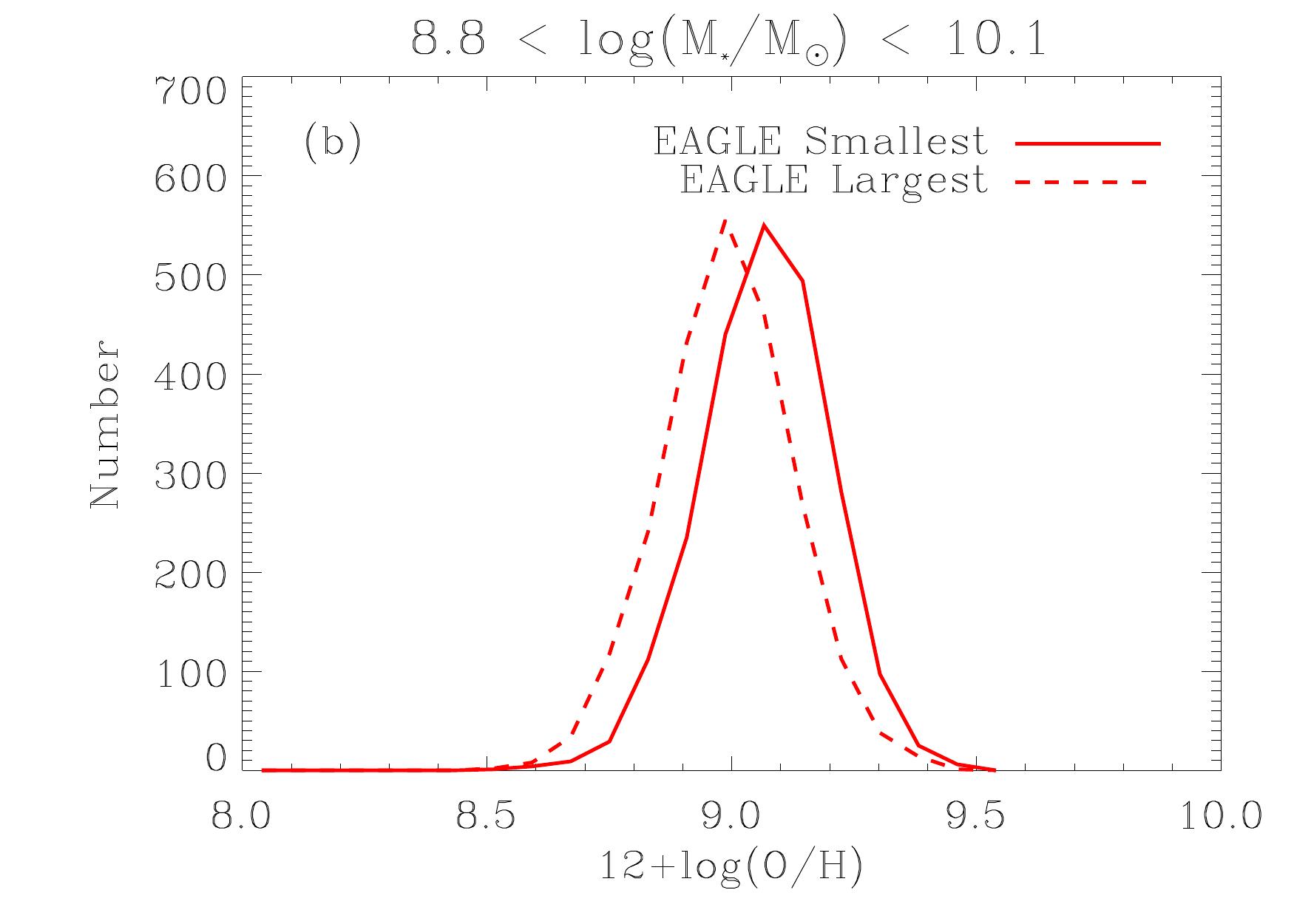}
\plotone{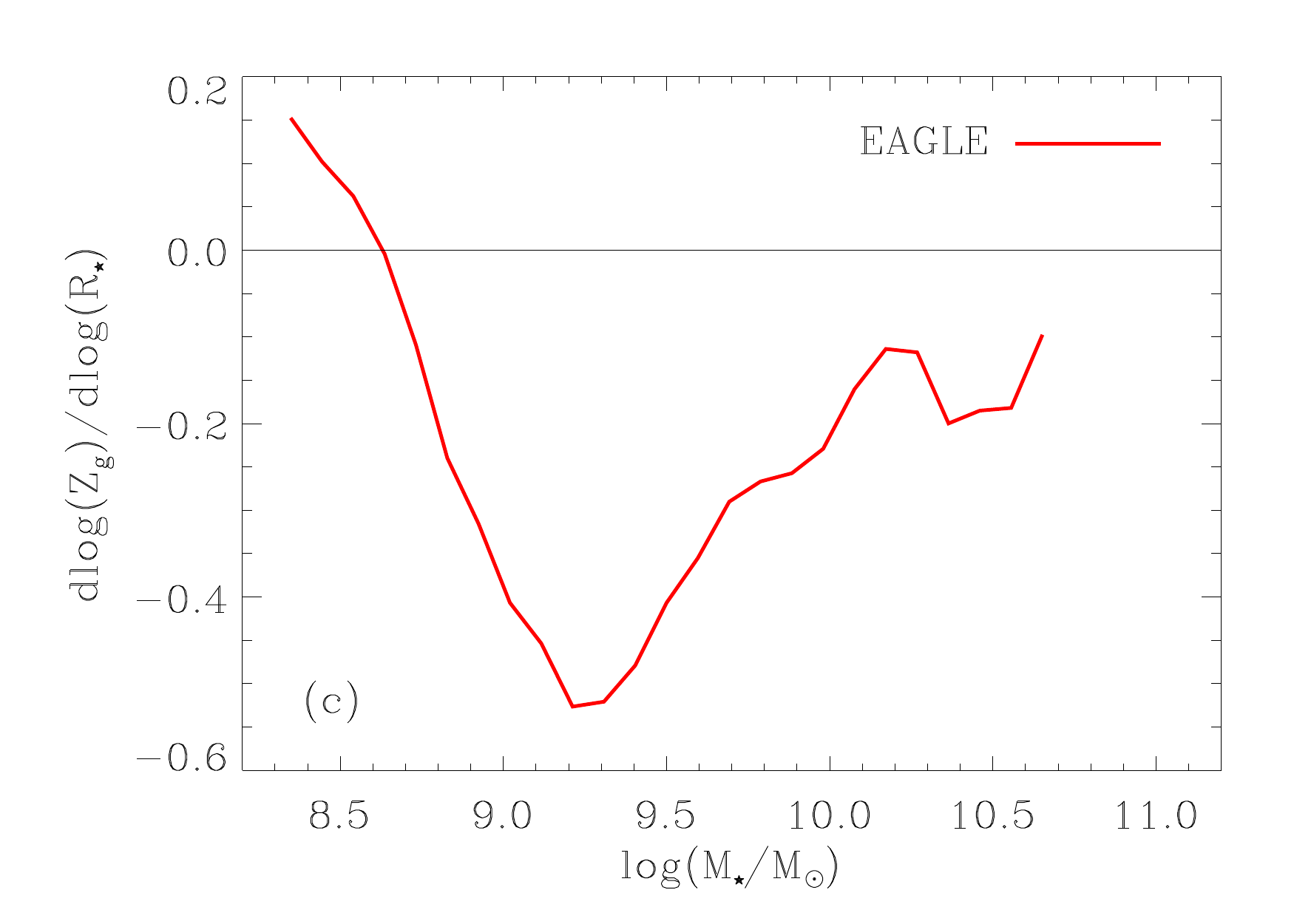}
\caption{(a) Gas-phase metallicity versus stellar mass for the smallest and largest galaxies in a given mass bin, including the galaxies observed by \citet[][the black lines]{2008ApJ...672L.107E} and the \eagle\ simulation (the red lines).  The simulated metallicities have been shifted vertically by an arbitrary amount so that they do not overlap with the observed metallicities.
(b) Distribution of metallicities for the model galaxies with stellar masses within $8.8 < \log(M_\star/M_\odot) < 10.1$. 
(c) Logarithmic derivative of metallicity with respect to the effective stellar radius at a fixed stellar mass for the galaxies in the \eagle\ simulation.}
\label{fig:ellison4}
\end{figure}

The correlation in \eagle\ closely resembles the correlation observed by \citet{2008ApJ...672L.107E}. They  split $\sim$44,000 SDSS-DR4 star-forming galaxies in bins of equal mass. The objects in each bin were divided in terciles according to their half-light radius -- the largest galaxies, the intermediate-size galaxies, and the smallest galaxies.  The mean metallicity was computed for each tercile, finding the relation metallicity--stellar mass to differ for the three sets, in the sense that the largest galaxies tend to have the smallest metallicities. The curves found by \citeauthor{2008ApJ...672L.107E} for the 1st and 3rd  terciles are reproduced in Fig.~\ref{fig:ellison4}a, the black lines. We repeat the same exercise with the EAGLE  data represented in Fig.~\ref{fig:ellison2o}. The curves predicted by the simulations are also included in  Fig.~\ref{fig:ellison4}a (the red lines). Leaving aside a global factor, the fact that the EAGLE simulation shows a metallicity--stellar mass relation too flat compared with the observed relation \citep[see][Fig.~13]{2015MNRAS.446..521S}, and the use of half-mass radii rather than half-light radii, the difference of metallicity between small and large model galaxies, of the order of 0.1 dex, is in good agreement with observations (Fig.~\ref{fig:ellison4}a).  

The uncertain scaling factor stems from biases in both observations and simulations. On the one hand, the measurements of O/H were carried out using a strong line ratio method, and this procedure introduces non-negligible systematic errors \citep[e.g.,][]{2008ApJ...681.1183K}. On the other hand, the metallicity in the simulations depends on the adopted nucleosynthetic yields, which also have significant uncertainties. In addition,
\citet{2008ApJ...672L.107E} measure the metallicity in terms of the oxygen abundance whereas we employ the mass fraction in metals. In order to show them in Fig.~\ref{fig:ellison4}a, the EAGLE abundances have been transformed to O/H  assuming a constant value for the ratio $({\rm O/H})/Z_g$. We opted for the solar composition given by \citet{2009ARA&A..47..481A}, however, this scaling is also arbitrary. In order to discard biases arising from the use of $Z_g$ rather than O/H, Fig. ~\ref{fig:ellison4}a was repeated with the actual O/H from the EAGLE simulation. The result is almost identical to the curves on display. In fact, O/H is expected to be a good tracer of $Z_g$ since O is the major contributor to the mass in metals.  

The curves  $12+\log({\rm O/H})$ versus $M_\star$ in EAGLE are flatter than observed (Fig.~\ref{fig:ellison4}a).
Such difference  seems to be caused by the limited resolution of the numerical simulation. As shown by \citet{2015MNRAS.446..521S}, other EAGLE runs at higher spatial resolution produce significantly lower metallicity and a steeper relation, in closer agreement with observations.  The merging of the curves for small and large galaxies at $\log(M_\star/M_\odot)\sim 8.5$ also reinforces this view --  the effect is highest for the lowest mass bins that are represented by fewer particles in the simulation.

The distribution of metallicities among the galaxies of the 1st and 3rd terciles is represented in Fig.~\ref{fig:ellison4}b.
Figure~\ref{fig:ellison4}c shows the derivative of the (log) metallicity with respect to the (log) stellar effective radius. It has been estimated from the differences in metallicity and radius of the galaxies in the upper and lower terciles. $d\log (Z_g)/d\log (R_\star)\geq -0.6$, with the minimum at $\log(M_\star/M_\odot)\simeq -9.3$.
The increase of the slope toward low masses is produced by the aforementioned limited resolution.


\section{Interpretative framework}\label{sec:framework}

The mean metallicity of a galaxy is primarily regulated by (1) the efficiency of the star-formation process that produces the metals, (2) the presence of outflows carrying away these metals, and (3) the existence of inflows of metal-poor gas fueling the star-formation \citep[e.g.,][and Appendix~\ref{sec:appb}]{1972NPhS..236....7L,1990MNRAS.246..678E,2007ApJ...658..941D}.  We want to determine if one (or several) of these processes is responsible for the existence of an anti-correlation between gas metallicity and stellar size in the \eagle\ model galaxies and, in doing so, to identify a plausible physical scenario that explains the relation observed by \citeauthor{2008ApJ...672L.107E} Therefore, we need to examine how these three key processes depend on the galaxy size. 

\begin{enumerate}
\item The efficiency of the star-formation is related to the galaxy size through the gas density, since denser galaxies are more efficient transforming gas into  stars as reflected by the Kennicutt-Schmidt relation \citep[e.g.,][]{1998ApJ...498..541K}. Given a stellar mass, those galaxies that are more efficient forming stars consume gas sooner, and what is leftover becomes more metallic by mixing with supernova (SN) ejecta. This effect gives rise to an anti-correlation between gas metallicity and stellar size in qualitative agreement with the observations of \citeauthor{2008ApJ...672L.107E} 
\item Since outflows take gas and metals out of the galaxies, they modulate the gas-phase metallicity. The effectiveness in carrying away gas is related to the power of the winds and the  depth of the gravitational potential to be overcome.  When the winds are powered by stars, the supply of energy and momentum is set only by the SFR, independently of the galaxy size.  However, galaxy size enters into the equation through the depth of the gravitational potential. For a given mass, the depth increases with decreasing size and, therefore, smaller galaxies are expected to have less effective winds and so to retain more metals. Thus, for a fixed mass, smaller galaxies become more metallic.     
\item Galaxies grow in size with time \citep[e.g.,][]{2016ApJ...828...27N,2016MNRAS.461.2728M,2017MNRAS.465..722F}. Thus, galaxies with late star-formation (SF) are systematically bigger than those formed earlier. If the SF is driven by metal-poor gas accretion, differences in the recent gas accretion rate produce the type of observed relation. Given a stellar mass, younger galaxies are bigger and, since they still preserve recently accreted metal-poor gas, they are metal poorer as well.
\end{enumerate}

In order to  understand the physical bases for the correlation in the simulation (Figs.~\ref{fig:ellison2o} and \ref{fig:ellison4}), we will rely on a simple self-regulated galaxy model\footnote{Often known as {\em bathtub model}.}, where galaxies are characterized by a  stellar mass, a  gas mass, a  metallicity, and so on \citep[e.g.,][]{2008MNRAS.385.2181F,2010ApJ...718.1001B,2012MNRAS.421...98D,2013ApJ...772..119L,2014MNRAS.443.3643P,2014A&ARv..22...71S}. This kind of toy-model is broadly used in the literature because, despite its apparent simplicity, it includes all the key physical ingredients and their interrelations, and often reveals the underlaying physical processes in a way hard to disclose in the full numerical solutions. We use it to work out the expected variation of metallicity with size if the anti-correlation is created by the depth of the gravitational potential (Sect.~\ref{sec:depthwell}), the density of the galaxy (Sect.~\ref{sec:denstar}), and the recent accretion of gas on an already grown-up galaxy (Sect.~\ref{sect:lateacc}).
The predictions derived from these sections will be used later on to analyze, favor or discard each of the plausible mechanisms.  

\subsection{Depth of the gravitational well}\label{sec:depthwell}
Under the hypothesis  of stationary gas infall, the toy-model predicts that the gas metallicity reaches a constant value $Z_{g0}$ set only by the stellar yield $y$ (the mass of new metals eventually ejected per unit mass locked into stars), the mass return fraction $R$ (the fraction of mass in stars that returns
to the interstellar medium), and the so-called mass loading factor $w$,
\begin{equation}
\Delta Z_{g0}=Z_{g0}-Z_{in}=\frac{y\,(1-R)}{1-R+w},
\label{eq:z0}
\end{equation}
with $Z_{in}$ the metallicity of the accreted gas ($\ll Z_{g0}$ so that  $\Delta Z_{g0} \simeq Z_{g0}$). $w$ is defined as the constant of proportionality between the gas outflow rate produced by the starburst, $\dot{M}_{out}$, and its SFR,
\begin{equation}
\dot{M}_{out} = w\, {\rm SFR}.
\label{eq:massload}
\end{equation}
Equation (\ref{eq:z0}) is derived in Appendix~\ref{sec:appb} and corresponds to Eq.~(\ref{eq:z0app}). According to Eq.~(\ref{eq:z0}),  differences in $w$  lead to differences in $Z_{g0}$. $w$ depends on the depth of the gravitational potential, which we parameterize in terms of the escape velocity, $v_{esc}$, defined as the velocity whose kinetic energy balances the (negative) gravitational energy \citep[e.g.,][Eq.~{[2.31]}]{2008gady.book.....B}. 
Galaxies with the same mass but smaller radius reside in a  deeper gravitational potential. Then the winds produced by stellar feedback will be less effective to escape, lowering $w$. This is the explanation proposed in \citet[][]{2014A&ARv..22...71S}.  Using Eq.~(\ref{eq:z0}) with $Z_{in} \ll Z_{g0}$, one can write down the expected variation of $Z_{g0}$ with $v_{esc}$ as  
\begin{equation}
\frac{d\log Z_{g0}}{d\log v_{esc}}=\frac{\beta\,w}{1-R+w},
\label{eq:tonta}
\end{equation}
with $\beta$ parameterizing the relation between $w$ and $v_{esc}$,
\begin{equation}
w\propto v_{esc}^{-\beta}.
\label{eq:vscpdef}
\end{equation}
The actual value of $\beta$ is unknown but it is expected to go from 2 for {\em energy driven} winds to 1 for {\em momentum driven} winds \citep[e.g.,][]{2005ApJ...618..569M}. Explaining the mass-metallicity relation in terms of varying $w$ with $v_{esc}$ favors low values of  $\beta$; 0.5\,--\,0.9 \citep[e.g.,][with $v_{esc}^2\propto M_\star$]{2013MNRAS.430.2891D,2013ApJ...765..140A}. Therefore, even in the most favorable case for $w$ to be important ($w \gg 1$), Eq.~(\ref{eq:tonta}) leads to,
\begin{equation}
\frac{d\log Z_{g0}}{d\log v_{esc}}\leq 1.
\label{eq:limit}
\end{equation}
  
There is an additional constraint to be satisfied if changes in $w$ are responsible for the observed correlation between gas-phase metallicity and size. In this case, the ratio between SFR and metallicity has to be independent of the mass loading factor and, thus, independent of the depth of the gravitational potential. According to the simple model in Appendix~\ref{sec:appb}, this ratio is set only by the current gas accretion rate, $\dot{M}_{in0}$, $y$, and $R$ (Eq.~[\ref{eq:ultimate}]), namely,
\begin{equation}
\frac{\rm SFR_{0}}{\Delta Z_{g0}}=\frac{\dot{M}_{in0}}{y\,(1-R)}.
\label{eq:ultimate2}
\end{equation}

\subsection{Density of the galaxy}\label{sec:denstar}

\begin{figure}
\includegraphics[width=0.45\textwidth]{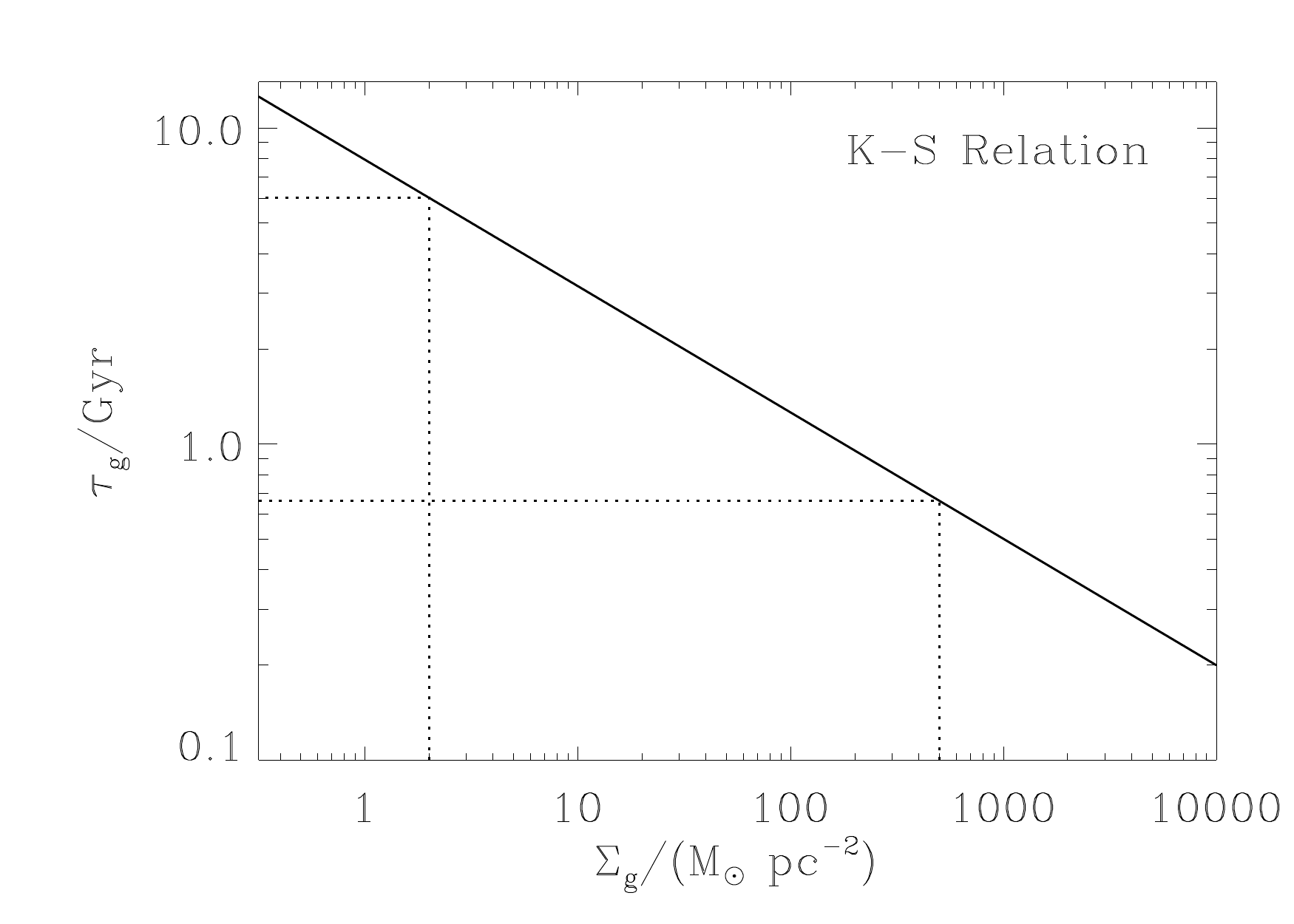}
\caption{Gas consumption timescale, $\tau_g$, as a function of the gas surface density, $\Sigma_g$, according to the parametrization of the Kennicutt-Schmidt relation in \citet{2012ARA&A..50..531K}. The vertical dotted lines point out the range of stellar surface densities of the model galaxies in EAGLE. If this is also the range of the gas surface densities ($2\, M_\odot\,{\rm pc}^{-2}$ to $5\times 10^2 M_\odot\,{\rm pc}^{-2}$), it yields the range of timescales between 6\,Gyr and  0.7\, Gyr indicated in the figure by the horizontal dotted lines.}
\label{fig:kstimescale}
\end{figure}

The {\em gas consumption timescale}, $\tau_g$, is defined as the ratio between the gas mass, $M_g$, and the SFR,
\begin{equation}
\tau_g=M_g\big/{\rm SFR}.
\label{eq:taug}
\end{equation} 
This timescale depends on the surface gas density, so that denser systems have shorter time-scales \citep[e.g.,][]{1998ApJ...498..541K}. Using the Kennicutt-Schmidt  relation as parameterized by \citet[][]{2012ARA&A..50..531K}, $\tau_g$ increases from 0.5\,Gyr to 5\,Gyr when the gas surface density decreases from $5\times 10^2\, M_\odot\,{\rm pc}^{-2}$ to $2\,M_\odot\,{\rm pc}^{-2}$ (see Fig.~\ref{fig:kstimescale}).  For a given mass, smaller systems are denser and so they should consume the gas faster. Thus, considering an ensemble of galaxies with similar stellar mass and accreting gas, those smaller are expected to be denser, to consume the gas faster and, consequently, to become metal richer sooner. This mechanism requires the galaxies to be outside  equilibrium because, as we pointed out in the previous section, the equilibrium gas-phase metallicity is independent of the gas consumption timescale, and is set only by stellar physics and the mass loading factor (Eq.~[\ref{eq:z0}]).

\begin{figure}
\includegraphics[width=0.45\textwidth]{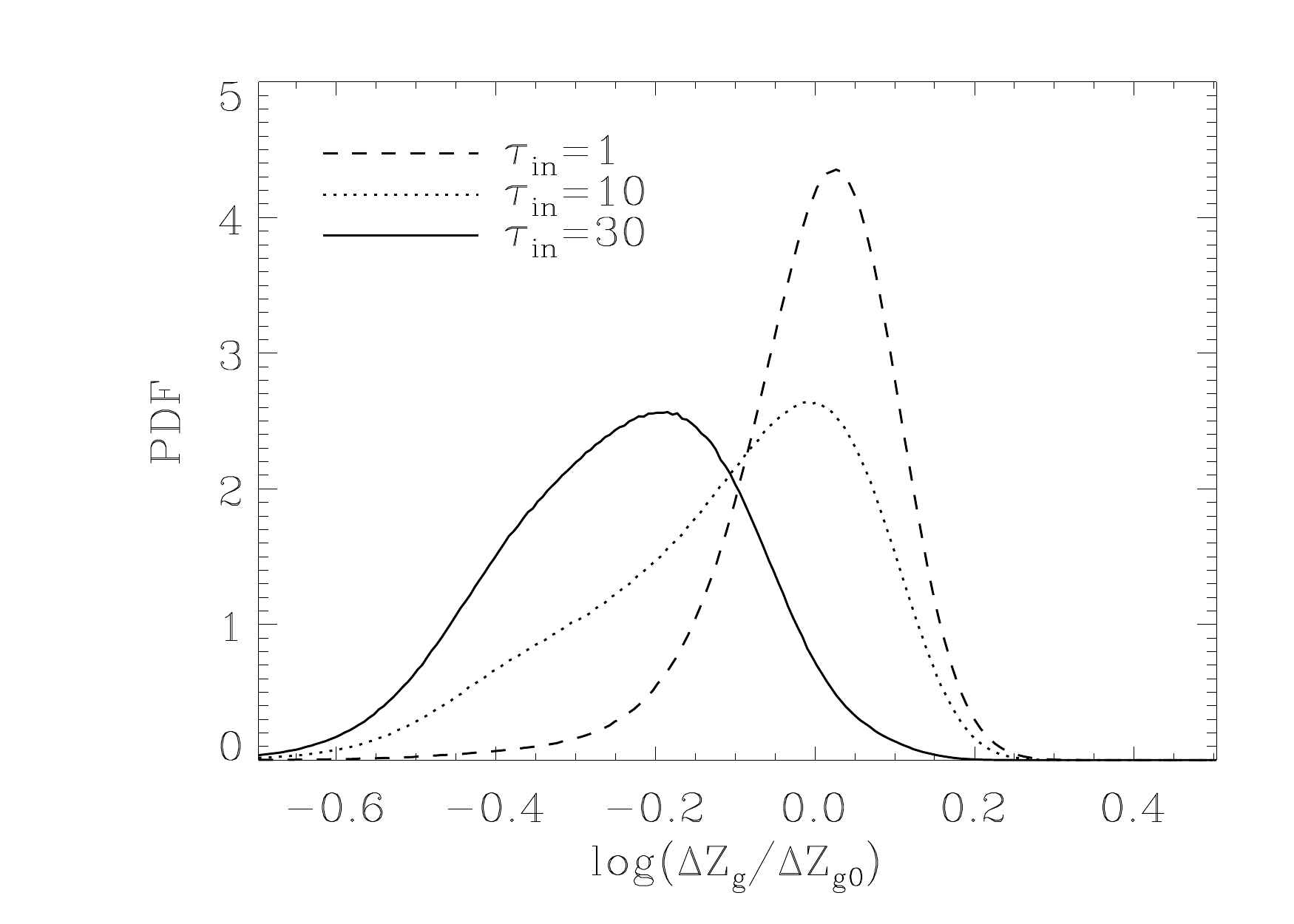}
\caption{Distribution of gas-phase metallicities to be expected if a set of galaxies receive a fixed amount of gas and they are observed at random times after the accretion episode. The dotted line shows the PDF when the gas depletion timescale is ten times larger than the dashed line (see the inset). The timescale is thirty times larger in the case of the solid line, and the galaxies have not had time to reach the equilibrium  metallicity. The equilibrium metallicity is assumed to differ for the different galaxies, with a random Gaussian distribution of standard deviation 0.2 times the mean value. $\Delta Z_{g0}$ stands for the equilibrium metallicity.
}
\label{fig:make_histogram}
\end{figure}

In Appendix~\ref{sec:appb}, we derive the variation of the metallicity with time,  $\Delta Z_g(t)$, when the model galaxy receives an amount of gas. This metallicity depends on $\tau_g$ through an {\em effective gas consumption} timescale $\tauin$ (Eq.~[\ref{2ndeq}]),
\begin{equation}
\tauin=\frac{\taug}{1-R+w}.
\label{eq:caveat}
\end{equation}
The dependence is given in Eq.~(\ref{my_metal2}) and it is shown in Fig.~\ref{fig:single_burst}. The probability density function (PDF) of the metallicity that the galaxy presents during its time evolution, $P(\Delta Z_g)$, is proportional to the timespan spent by the galaxy at each metallicity, i.e.,
\begin{equation}
P(\Delta Z_g)\propto \big|\frac{dt}{d\Delta Z_g}\big|.
\end{equation}  
Since $t(\Delta Z_g)$ is bivalued (it is the inverse to the function shown in Fig.~\ref{fig:single_burst}),  $P(\Delta Z_g)$ has an involved analytical expression. We evaluate it numerically using a Monte-Carlo simulation as follows. We consider a population of galaxies going through gas accretion events, which are detected at random times from the time of accretion. We assume that all of them received the same amount of pristine gas, and have the same gas depletion timescale. Using Eq.~(\ref{my_metal2}), we compute the gas-phase metallicity of each object at the time of observation, and the corresponding PDF is inferred from them.  The results for three populations that differ in their gas depletion timescale is represented in Fig.~\ref{fig:make_histogram}.  Note the extended tails of the distributions with large gas consumption timescales.  The solid line corresponds to a time-scale thirty times longer than the case of the dashed line when the galaxies have not had time to reach the equilibrium metallicity.  The equilibrium metallicity in this Monte-Carlo simulation is assumed to differ for the different galaxies, following a random Gaussian distribution with its standard deviation 0.2 times its mean value.

The expected increase in metallicity associated with an increase in stellar mass density, $\rho_\star$, can be estimated splitting the derivative as
\begin{equation}
\frac{d\log Z_g}{d\log\rho_\star}=\frac{d\log Z_g}{d\log\tau_g}\,\frac{d\log\tau_g}{d\log\Sigma_g}\,\frac{d\log\Sigma_g}{d\log\rho_\star}.
\end{equation}  
The first term in the right-hand-side of the equation can be evaluated using the simulations shown in Fig.~\ref{fig:make_histogram}, and it amounts to some -0.2 when comparing the dashed and the solid lines.  The second term follows from the Kennicutt-Schmidt relation; using the one in \citet{2012ARA&A..50..531K} it results -0.4. The third term turns out to be 2/3 assuming the surface density of the gas $\Sigma_g$ to be proportional to the surface density of stars $\Sigma_\star$, and then working out the scaling between surface density and volume density\footnote{ $d\log\Sigma_\star/d\log\rho_\star=2/3$ for objects of the same mass and varying size.}. All in all, the logarithmic derivative becomes,
\begin{equation}
\frac{d\log Z_g}{d\log\rho_\star}\simeq 5\times 10^{-2}.
\label{eq:for_my_guts}
\end{equation}
The terms involved in the evaluation of this equation are uncertain. When the surface density is low, larger exponents in the Kennicutt-Schmidt relation are favored \citep[e.g.,][and references therein]{2008AJ....136.2846B,2010MNRAS.407.2091G}, and the estimated derivative could easily be twice larger than the value in Eq.~(\ref{eq:for_my_guts}).

Since the metallicity depends on the gas consumption timescale as expressed in Eq.~(\ref{eq:caveat}), all that is said above also applies to the changes in $\tauin$ induced by variations in the mass loading factor $w$. The reader should keep in mind, however, that the trend is opposite to the one described in Sect.~\ref{sec:depthwell}. According to Eq.~(\ref{eq:caveat}), an increase in $w$ decreases $\tauin$ and thus increases the metallicity. The chain rule yields the change in metallicity when $v_{esc}$ varies through $\tau_{in}$, i.e.,
\begin{equation}
\frac{d\log Z_g}{d\log v_{esc}}=\frac{d\log Z_g}{d\log\tauin}\,\frac{d\log\tauin}{d\log w}\,\frac{d\log w}{d\log v_{esc}}=-0.2.
\label{eq:odd}
\end{equation}
The numerical value has been worked out using Eqs.~(\ref{eq:vscpdef}) and (\ref{eq:caveat}), and plugging in the parameters used to evaluate the derivative in Eq.~(\ref{eq:limit}). 
This dependence of $Z_g$ on $v_{esc}$ through $\tau_{\rm in}$ may qualitatively explain the correlation shown in the lower panels of Fig.~\ref{fig:z_vs_vscp}, which will be discussed later on.

\subsection{Differences in the recent gas accretion rate}\label{sect:lateacc}

\begin{figure}
\includegraphics[width=0.45\textwidth]{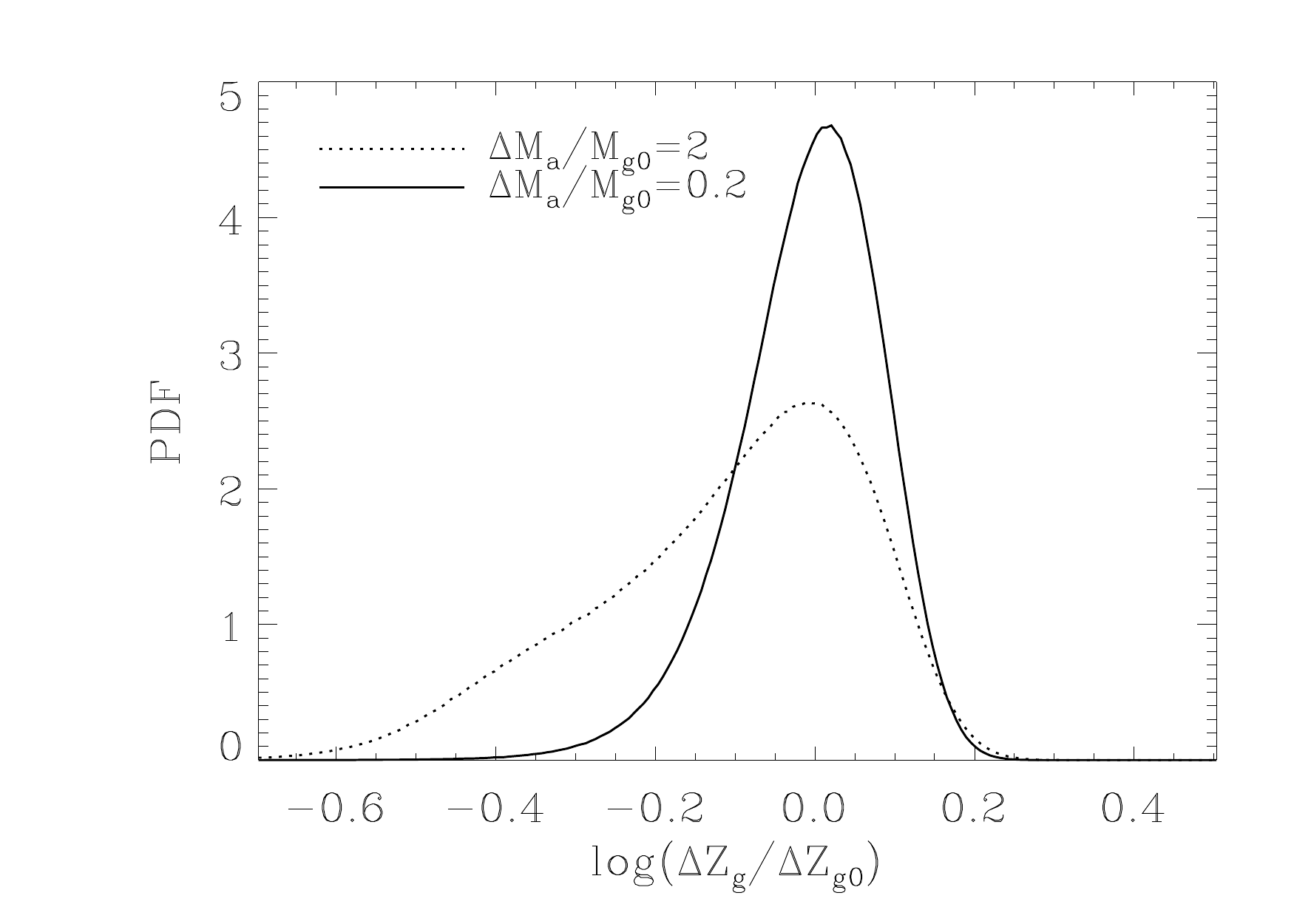}
\caption{Distribution of metallicities to be expected if a set of galaxies receive gas and they are observed at random times after the accretion episode. The depletion timescale is the same in both cases. They differ in the accreted mas; the mass is ten times smaller  in the solid line compared with the dashed line (see the inset, with the symbols defined in the text). The dotted line here is identical to the dotted line in Fig.~\ref{fig:make_histogram}.
}
\label{fig:make_histogram2}
\end{figure}

Galaxies with late star-formation tend to be larger. At the same time, galaxies outside equilibrium have their gas-phase metallicity in proportion to the gas recently accreted, i.e., in proportion to their recent $\dot{M}_{\rm in}$. The metal-poor gas acquired through accretion is still in place, so, the gas mass is metal-poorer in objects with larger current $\dot{M}_{\rm in}$. This can be shown using the simple model described in the previous section.  By increasing the gas mass per clump, one increases $\dot{M}_{\rm in}$. Thus, more massive clumps produce metallicity distributions biased and skewed toward lower metallicities, as illustrated by Fig.~\ref{fig:make_histogram2}.  

The expected change of gas-phase metallicity due to this process can also be estimated using the toy model worked out in Appendix~\ref{sec:appb}. 
On the one hand, the drop of metallicity after accreting  a gas mass $\Delta M_a$ scales with the accreted gas mass  as, 
\begin{equation}
Z_g/Z_{g0}\simeq \frac{1}{1+C\, \Delta M_a/M_{g0}},
\end{equation} 
with $M_{g0}$ the mass of gas already present in the object, and $C$ a time dependent parameter which is of the order of one after the accretion event  (see Eq.~[\ref{my_metal2}], with $Z_{in} << Z_g < Z_{g0}$). On the other hand, $\dot{M}_{\rm in}\sim \Delta M_{a}/\tauin$, therefore, 
\begin{equation}
\frac{d\log Z_g}{d\log \dot{M}_{\rm in}}\simeq \frac{d\log Z_g}{d\log \Delta M_a}\simeq \frac{-C\, \Delta M_a/M_{g0}}{1+C\, \Delta M_a/M_{g0}}\sim  -0.5.
\label{eq:logder}
\end{equation} 
The logarithmic derivative has been evaluated assuming $ \Delta M_a\sim M_{g0}$, and $C\sim 1$. Its actual value can go all the way from -1 (when $\Delta M_a\gg M_{g0}$) to 0 (when $\Delta M_a\ll M_{g0}$).


\section{Is the correlation between size and gas-phase metallicity due to differences in the depth of the gravitational potential of the galaxies?}\label{sec:well}

The short answer to the above question is {\em no}. The long answer is elaborated in this section showing that the escape velocity and the metallicity are not correlated in the EAGLE galaxies. 

In order to parameterize the depth of the gravitational potential, we use the escape velocity at the half stellar-mass radius,
\begin{equation}
v_{esc}(R_\star)=\sqrt{2\,|\Phi(R_\star)|},
\end{equation}
with $\Phi(R_\star)$ the gravitational potential at $R_\star$. We model $\Phi$ as a combination of the potential due to dark-matter (DM), $\Phi_{DM}$, and the potential due to stars, $\Phi_\star$,
\begin{equation}
\Phi(R_\star)=\Phi_{DM}(R_\star) + \Phi_\star(R_\star).
\end{equation} 
The EAGLE database does not directly provide the depth of the gravitational potential. We infer the DM component from the DM mass and the half-mass radius of the DM assuming the density to drop with radius following a NFW profile  (Navarro, Frenk and White \citeyear[][]{1996ApJ...462..563N}). The procedure is sketched in Appendix~\ref{sec:appa}.  The stellar contribution is also inferred from the stellar mass and the half-mass stellar radius, this time assuming that the stellar density drops exponentially with radius. We follow the work by \citet{2015MNRAS.448.2934S}. We assume  the galaxy disks to be thin (scale-height of 0.2 kpc) with the escape velocity computed in the plane of the disk. The results are rather insensitive to these assumptions since the potential is dominated by the DM component. This is also the reason why the gas mass in not included in the computation of $v_{esc}$, because it represents only a minute fraction of the total mass.

\begin{figure}
\includegraphics[width=0.42\textwidth]{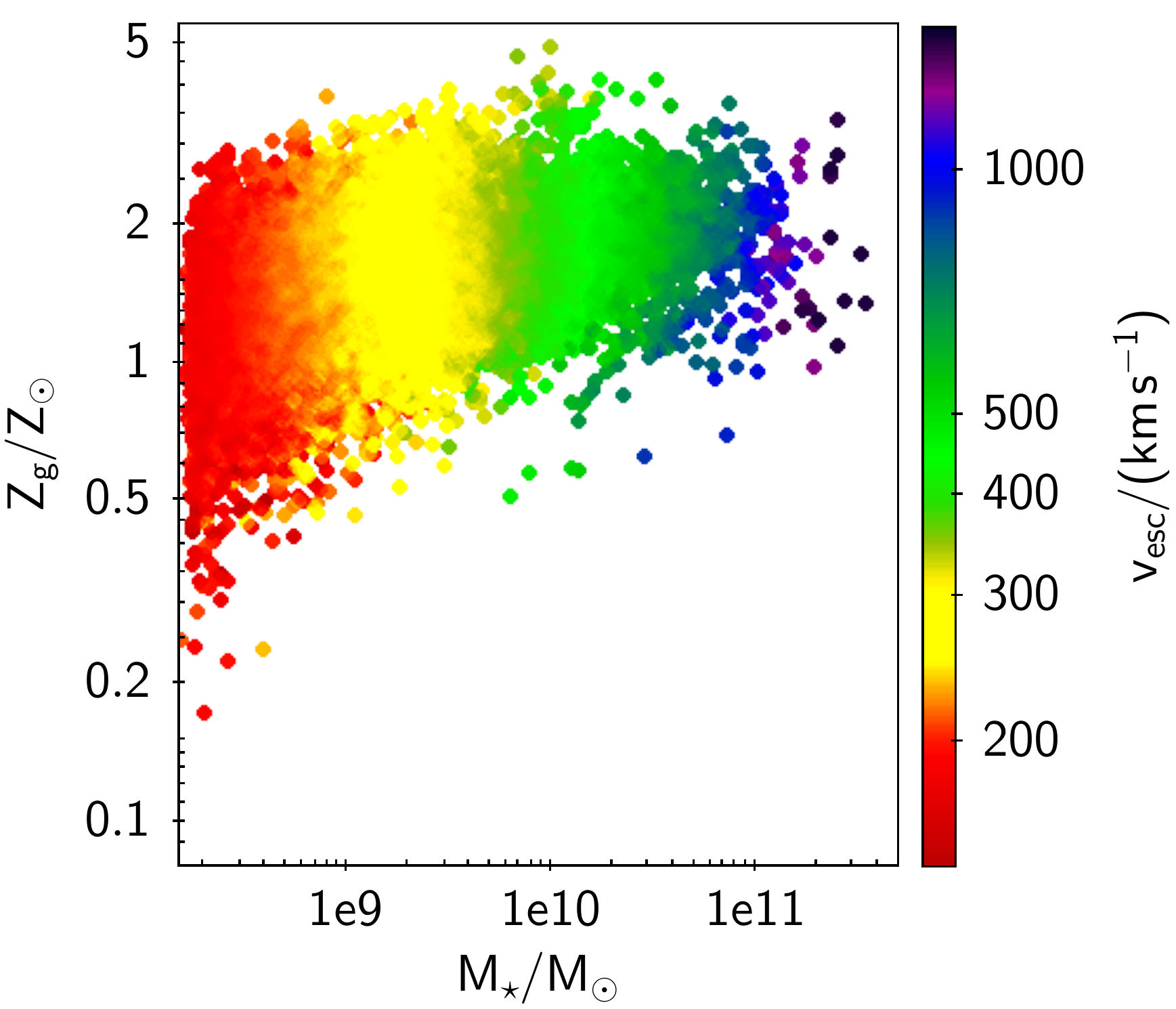}
\caption{Gas-phase metallicity versus stellar mass  color-coded according to the escape velocity from the half stellar mass radius $v_{esc}$. There is no obvious correlation between $Z_g$ and $v_{esc}$. The gravitational potential used to compute $v_{esc}$ includes both DM and stars, although $v_{esc}$ is mostly set by the DM mass. Velocities are given in km\,s$^{-1}$.}
\label{fig:ellison2_save}
\end{figure}
\begin{figure*}
\includegraphics[width=0.33\textwidth]{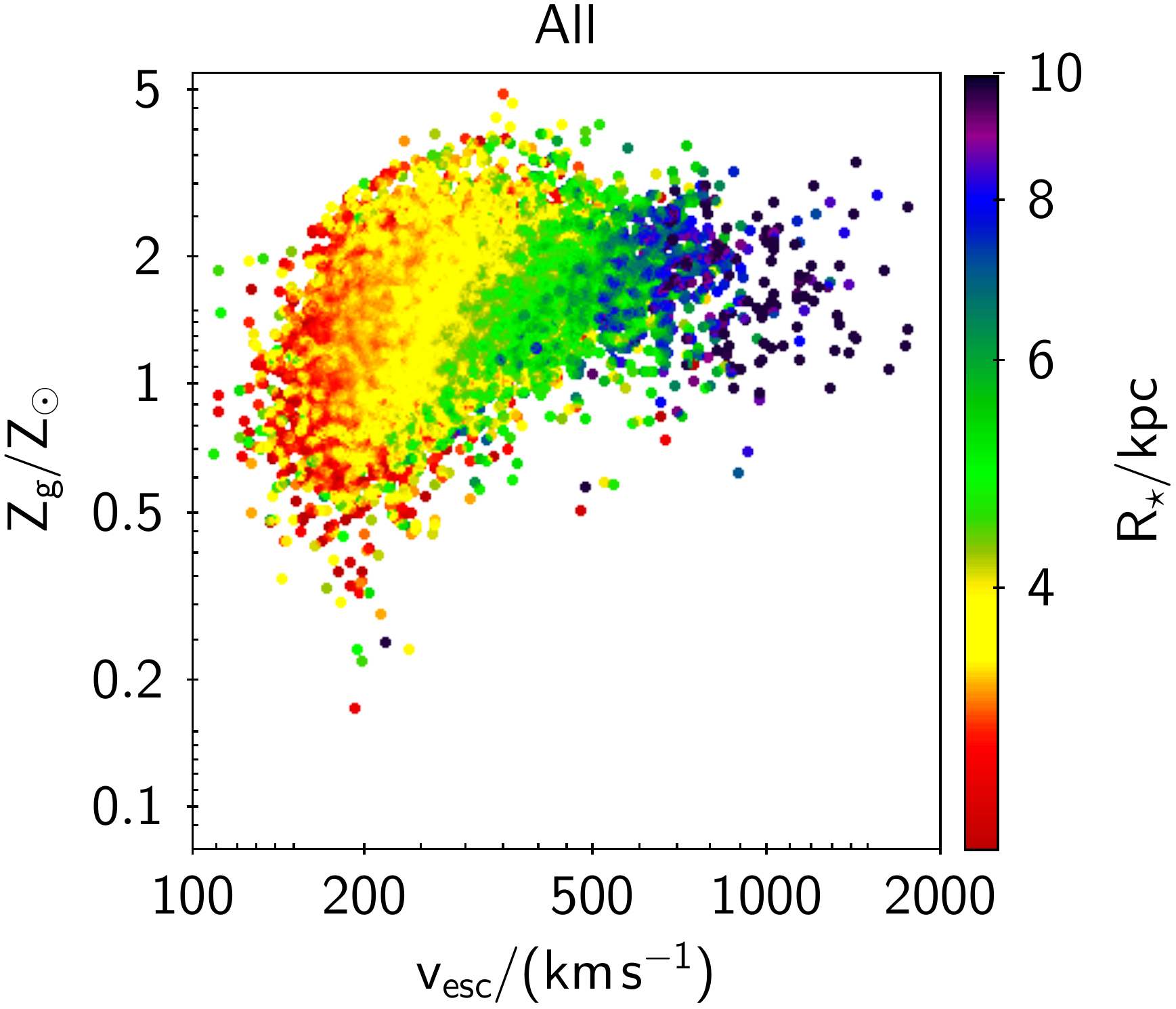}\hskip 0.2cm%
\includegraphics[width=0.33\textwidth]{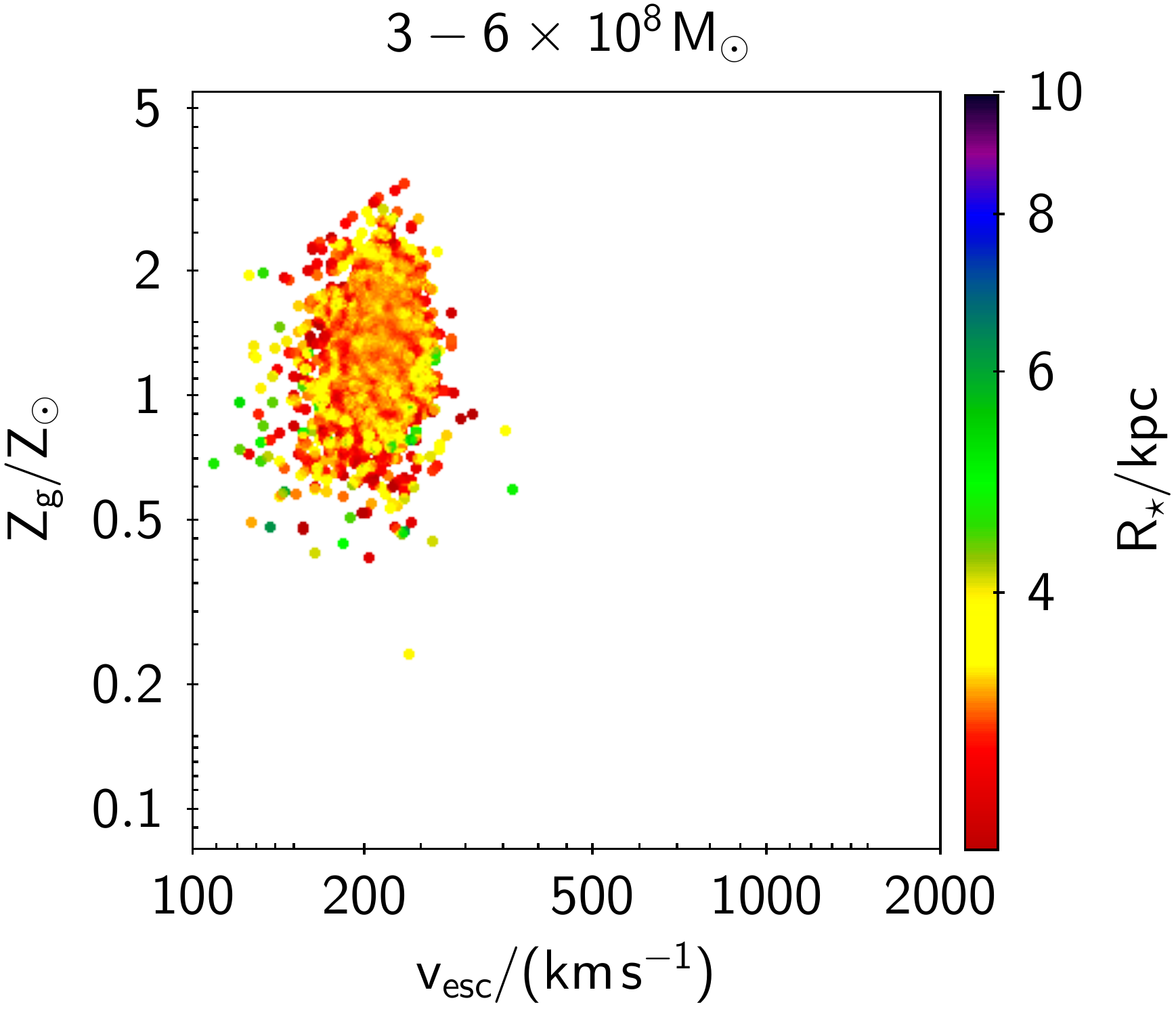}\hskip 0.2cm%
\includegraphics[width=0.33\textwidth]{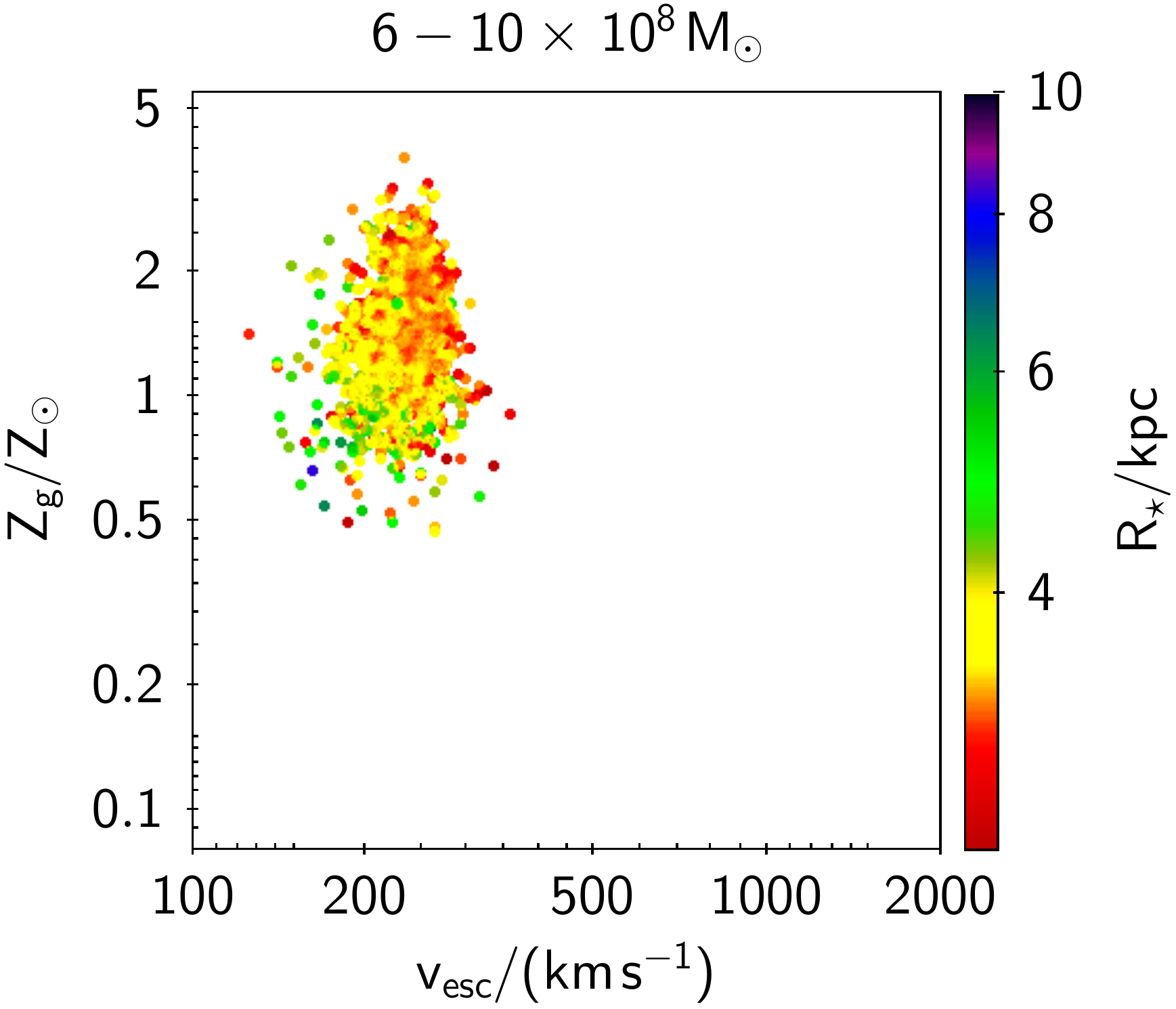}\vskip 0.2cm%
\includegraphics[width=0.33\textwidth]{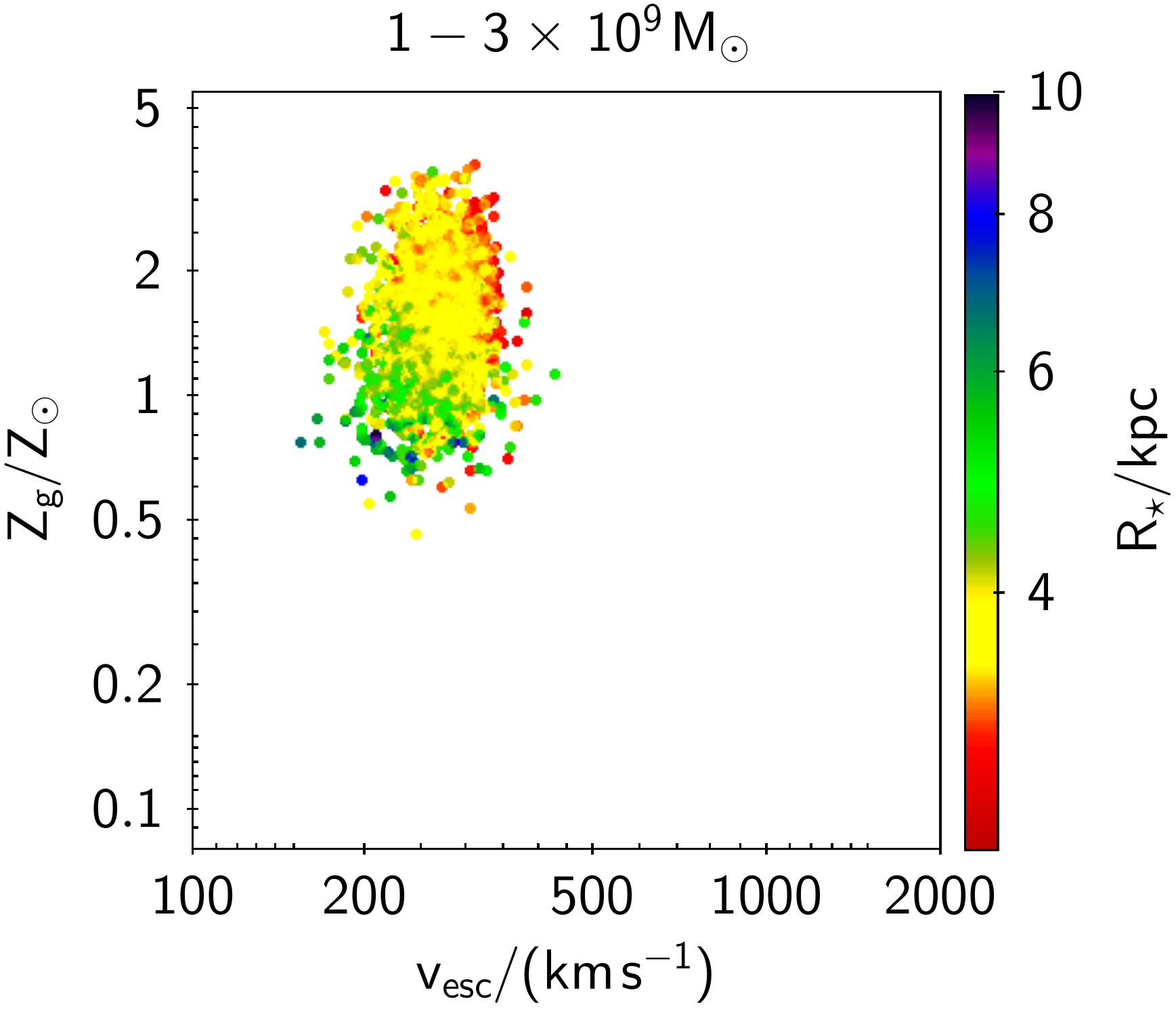}\hskip 0.2cm%
\includegraphics[width=0.33\textwidth]{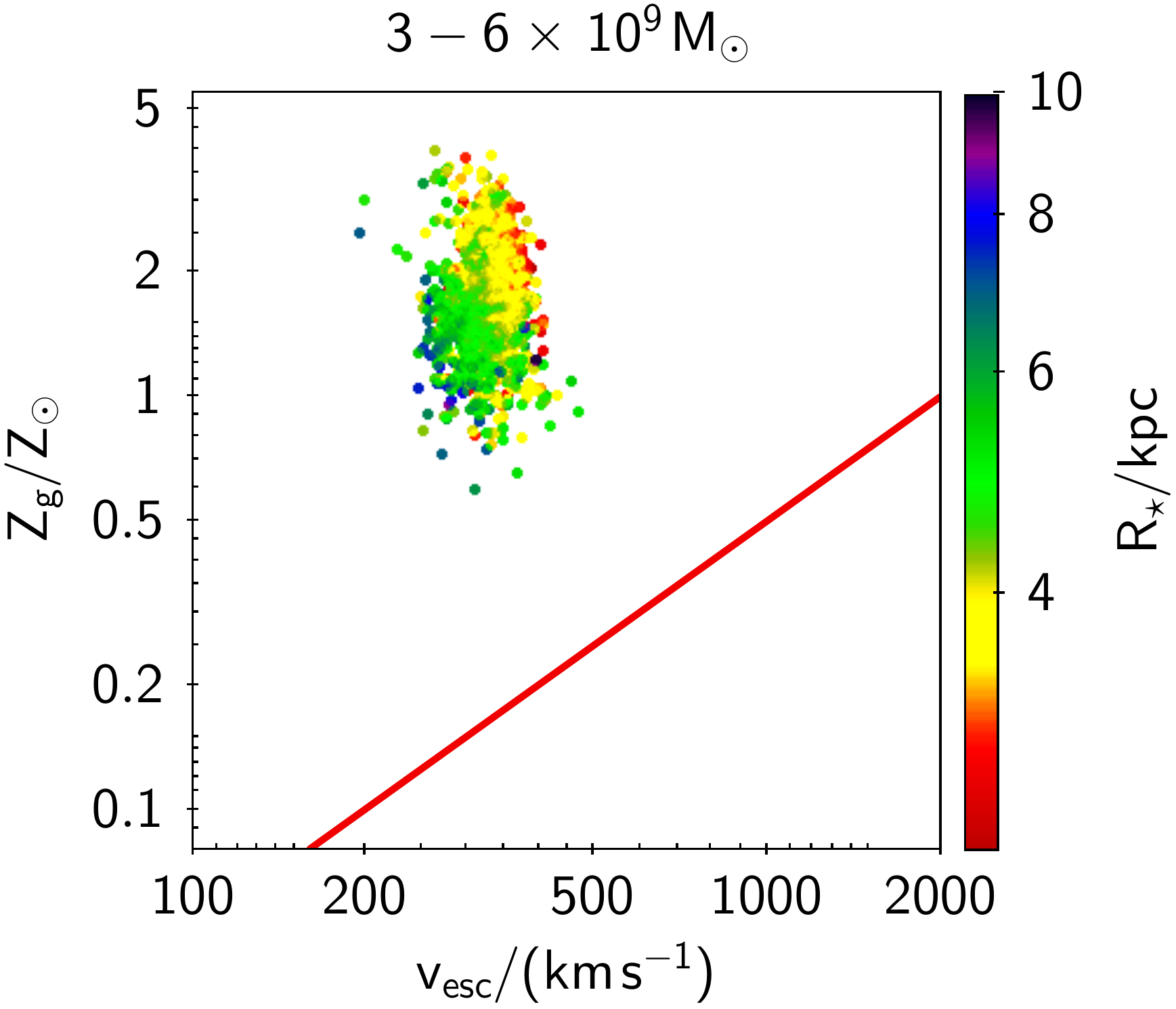}\hskip 0.2cm%
\includegraphics[width=0.33\textwidth]{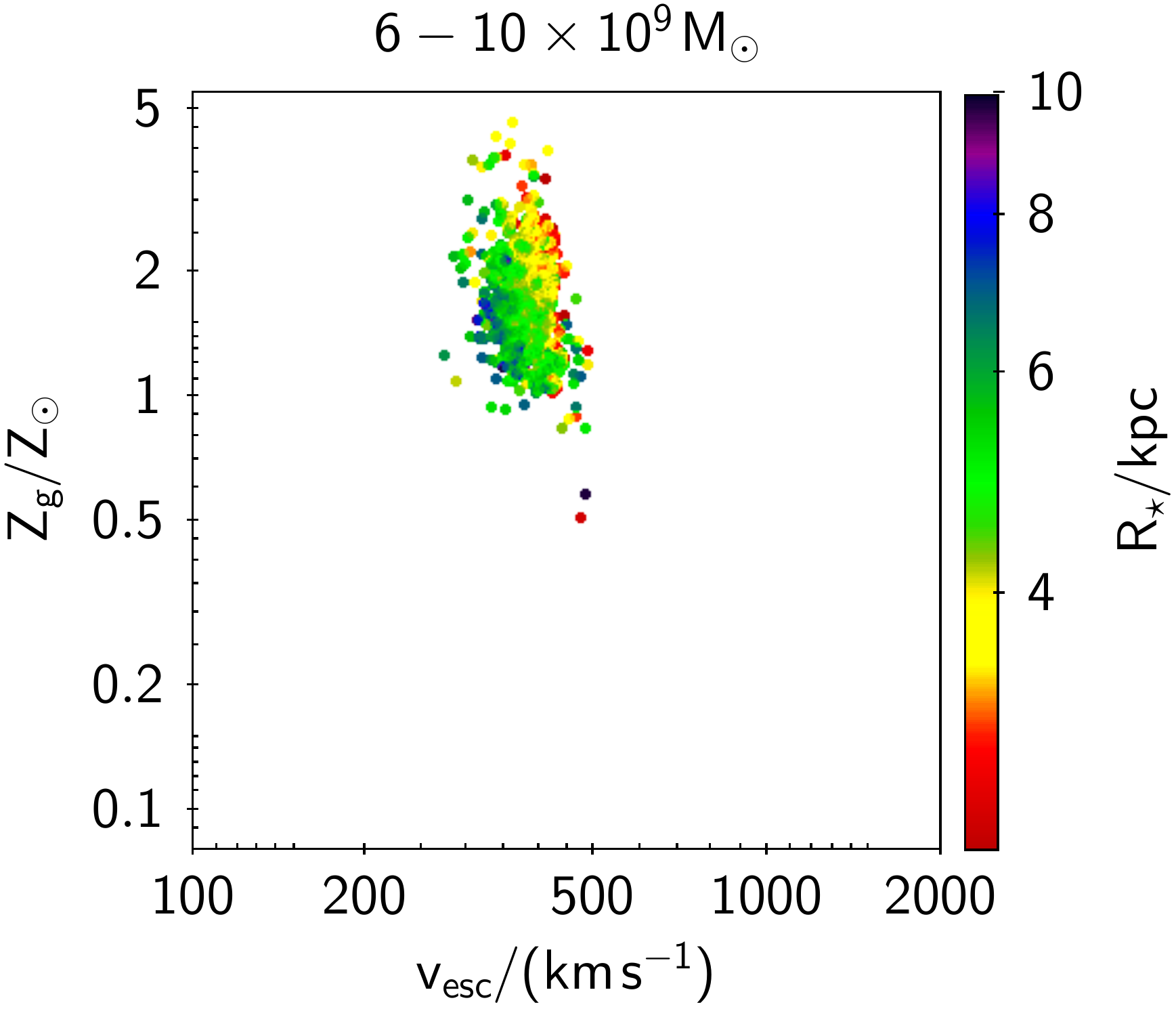}\vskip 0.2cm%
\includegraphics[width=0.33\textwidth]{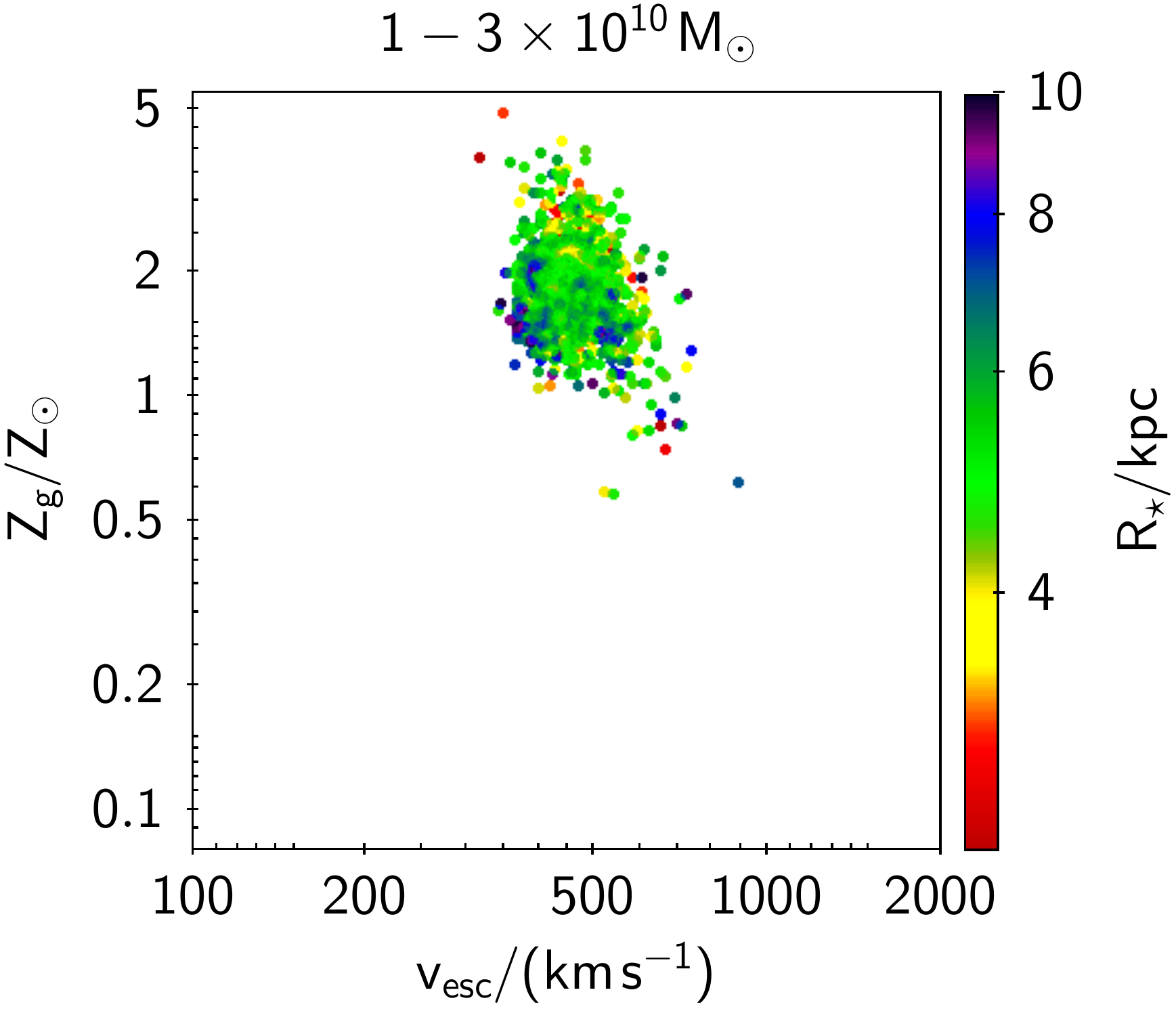}\hskip 0.2cm%
\includegraphics[width=0.33\textwidth]{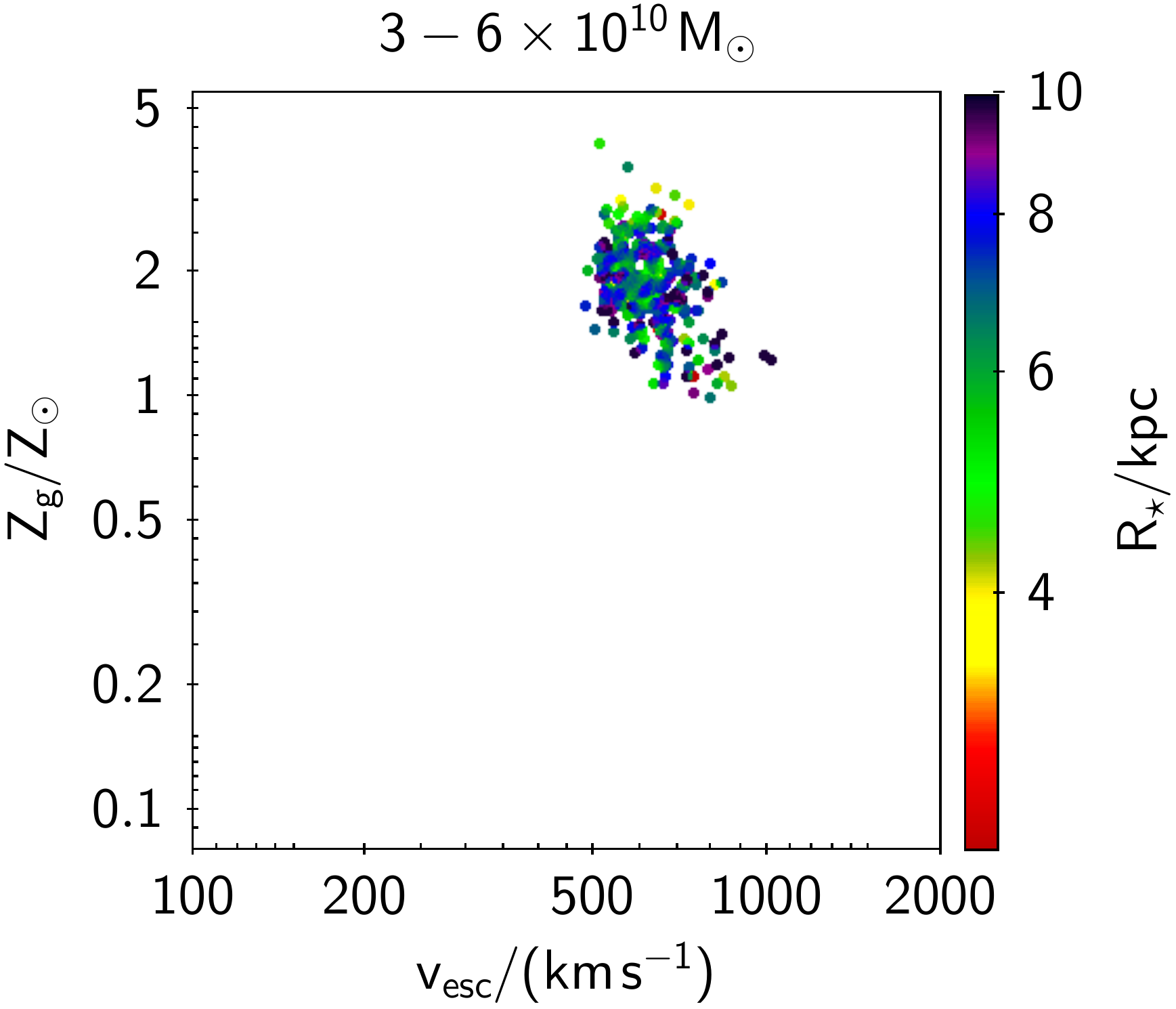}\hskip 0.2cm%
\includegraphics[width=0.33\textwidth]{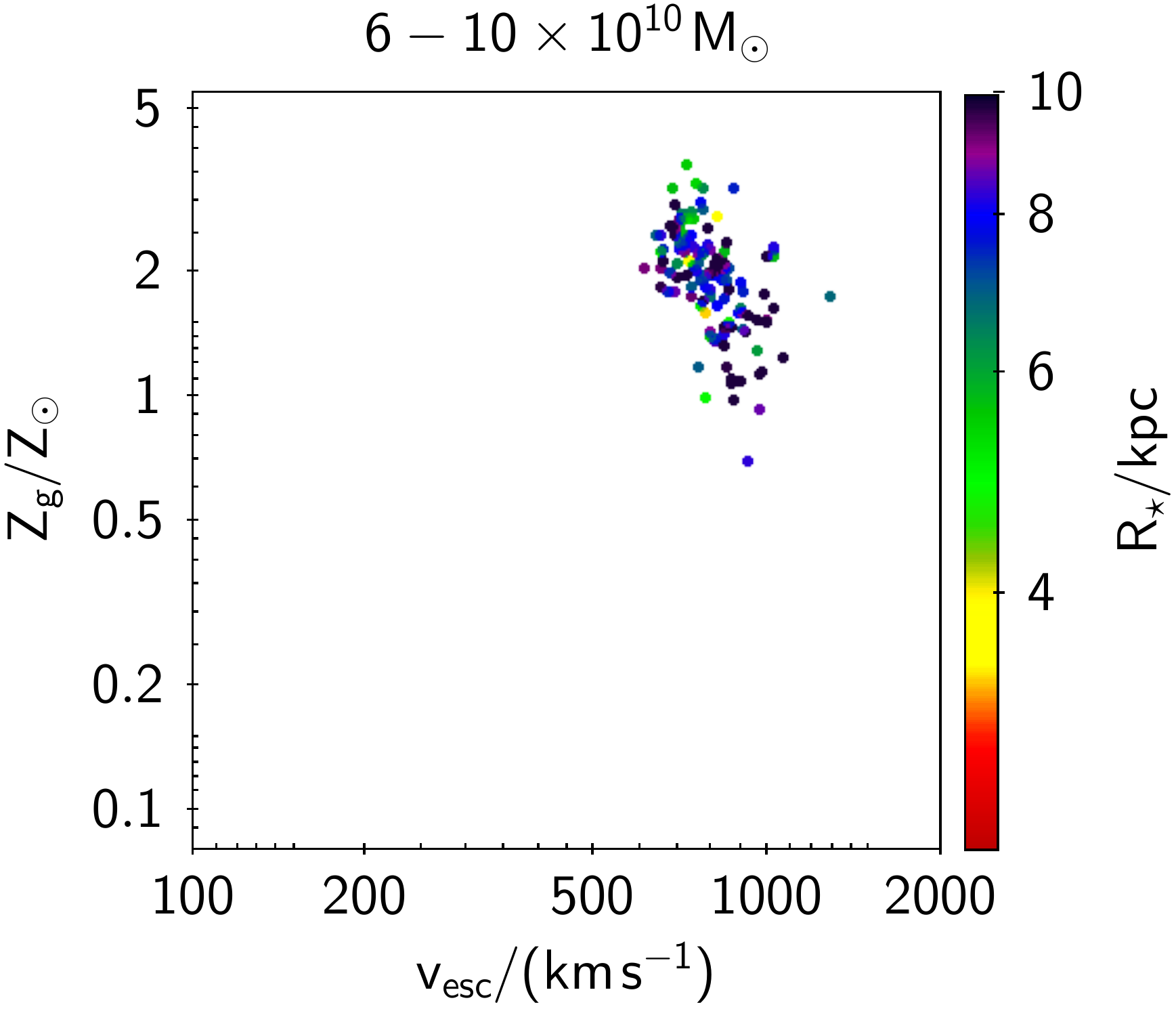}\vskip 0.2cm%
\caption{Gas-phase metallicity ($Z_g$) versus escape velocity ($v_{esc}$) color-coded with the half stellar mass radius of the galaxies ($R_\star$). The top-left panel contains the full data set. The rest of the panels show the same scatter plot selecting narrow mass bins (as labelled on top of each figure). There is no clear relation between $Z_g$ and $v_{esc}$. Maybe, in the panels corresponding to the high-mass end, there is a hint of anti-correlation. The solid line in the central panel represents the anti-correlation expected according to the toy-model worked out in the main text (upper limit set by Eq.~[\ref{eq:limit}]). The axes and the color code are identical in all panels.}
\label{fig:z_vs_vscp}
\end{figure*}

Figure~\ref{fig:ellison2_save} is similar to  Fig.~\ref{fig:ellison2o}, but this time the symbols are color-coded with the escape velocity at the half stellar mass radius.  The spread in gas-phase metallicity is uncorrelated with the escape velocity, which is mostly set by the stellar mass. For a given stellar mass the galaxies have the same escape velocity irrespectively of the metallicity of their gas  (c.f. Figs.~\ref{fig:ellison2_save} and \ref{fig:ellison2o}). This fact is even more clear in Fig.~\ref{fig:z_vs_vscp}. It shows $Z_g$ versus $v_{esc}$ for the full set of \eagle\ galaxies (top left), as well as for narrow bins in galaxy mass ($\Delta\log M_\star\sim 0.3$). Even though the full set shows a global trend for the gas-phase metallicity to increase with the escape velocity, this is just a construct (or a mirage) resulting from the mass-metallicity relation and the superposition of galaxies of all masses in the plot. The panels corresponding to single mass bins, show the scatter in metallicity to be independent of $v_{esc}$.  If the relation between size and gas-phase metallicity is due to changes in the escape velocity, the toy model in Sect.~\ref{sec:depthwell} predicts the red solid line in Fig.~\ref{fig:z_vs_vscp} (plotted in the central panel). The fact that the numerical simulations do not follow such line indicates that the metallicity is not set by $v_{esc}$. If anything, there seems to be a weak anti-correlation between $Z_g$ and $v_{esc}$ in the bins of higher mass (the three bottom panels in Fig.~\ref{fig:z_vs_vscp}). Such anti-correlation is contrary to the positive correlation needed to explain the observed anti-correlation between gas-phase metallicity and size. It may arise from the dependence of the gas depletion time-scale on the depth of the gravitational potential, as discussed in Sect.~\ref{sec:denstar}.  However, the predicted slope is much too shallow (see Eq.~[\ref{eq:odd}]).

\begin{figure}
\includegraphics[width=0.45\textwidth]{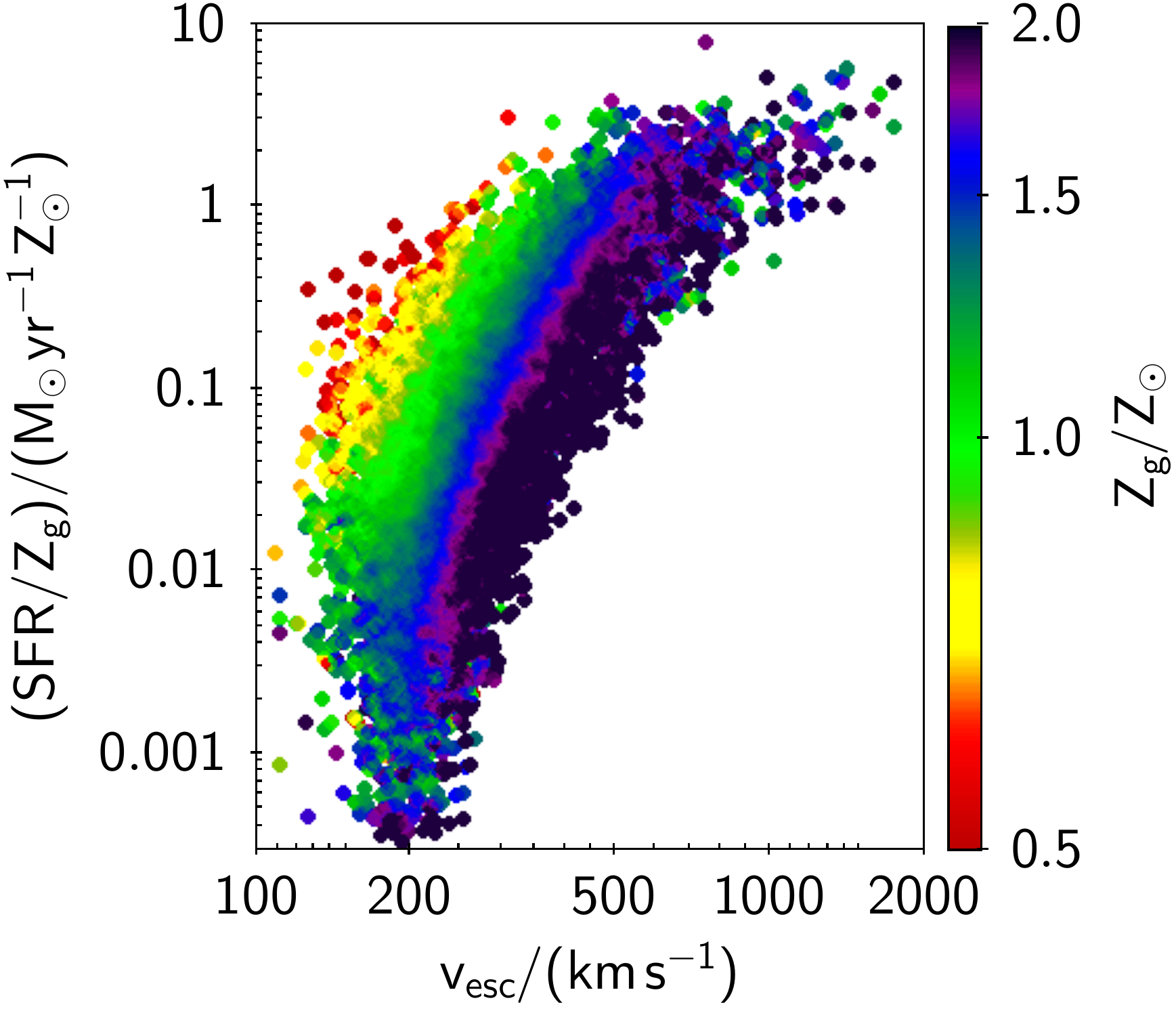}
\caption{ ${\rm SFR}/Z_g$  versus $v_{esc}$  color-coded according to $Z_g$. The quantity in ordinates should be independent of the depth of the gravitational potential and so independent of $v_{esc}$. It increases with increasing $v_{esc}$, discarding the depth of the gravitational potential as a major player in explaining the variation of gas metallicity with galaxy size.}
\label{fig:ultimate}
\end{figure}

There is one more argument against the depth of the gravitational potential setting the relation between metallicity and size. Should the variation with size be due to the depth of the gravitational potential, then one would expect the ratio ${\rm SFR}/Z_g$ to be constant at a fixed stellar mass, since in the stationary state this ratio solely depends on the gas accretion rate (Eq.~[\ref{eq:ultimate2}]). Figure~\ref{fig:ultimate} shows ${\rm SFR}/Z_g$ versus $v_{esc}$ color-coded with $Z_g$. It evidences a large variation of  ${\rm SFR}/Z_g$  with $v_{esc}$, which indicates that the depth of the gravitational potential by its own cannot be responsible for the variation of $Z_g$ with galaxy size. 


\section{Is the correlation between size and gas-phase metallicity due to differences in the mean density of the galaxies?}\label{sec:mdensity}

As it happened with the answer to the question posed in Sect.~\ref{sec:well}, the short answer is {\em no}.

\begin{figure}
\includegraphics[width=0.45\textwidth]{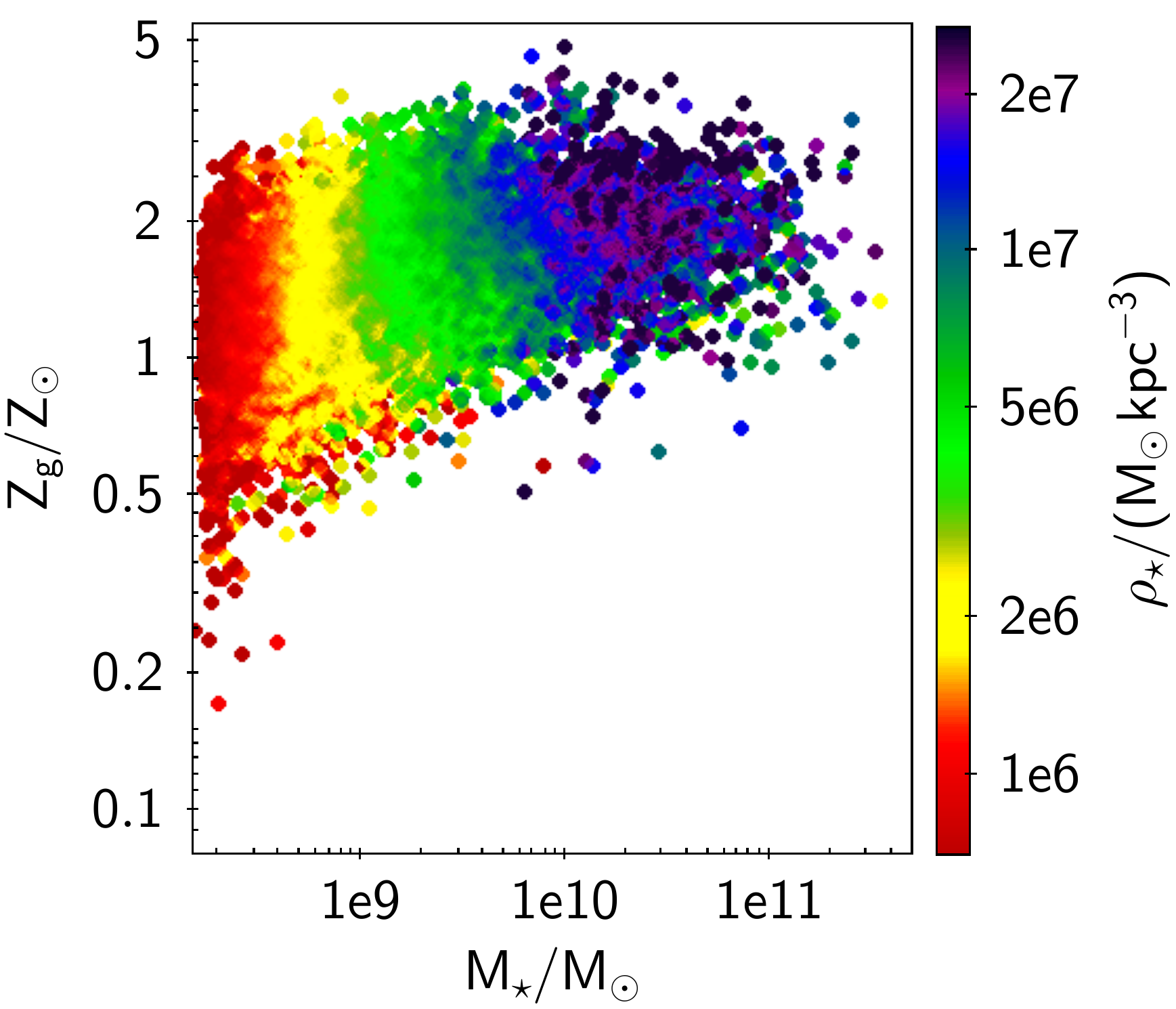} 
\caption{Gas-phase metallicity versus stellar mass color-coded with the stellar mass density. There is a strong dependence of $\rho_\star$ on $M_\star$, but not so much of a dependence of $Z_g$ on $\rho_\star$ for constant $M_\star$.}
\label{fig:density}
\end{figure}
Figure~\ref{fig:density} shows $Z_g$ versus $M_\star$ color-coded with the stellar volume density\footnote{$\rho_\star=M_\star/(8\,\pi\,R_\star^3/3)$, keeping in mind that $R_\star$ represents the half-mass radius whereas $M_\star$ is the total stellar mass. } of the galaxies. There is a variation of $\rho_\star$ with $Z_g$ at a fixed $M_\star$ which is significantly larger than the variation with escape velocity (cf. Figs.~\ref{fig:ellison2_save} and \ref{fig:density}). Given a stellar mass, denser galaxies tend to be metal richer. However, such trend seems to be a mirage resulting from the superposition of galaxies with very different densities but the same $Z_g$ and $M_\star$. Figure~\ref{fig:z_vs_vscp} shows $Z_g$ versus $\rho_\star$ for all the galaxies (top left), and for galaxies within narrow mass bins that cover the whole range of masses from $10^{8}\,M_\odot$ to $10^{11}\,M_\odot$. The solid line in the central panel represents the expected correlation according to Eq.~(\ref{eq:for_my_guts}). It roughly agrees with trend followed by the EAGLE galaxies in this mass bin. It indicates that the variation is behaving as expected theoretically. However, it is very mild compared with the range of gas metallicities exhibited by the galaxies.
Obviously, the density scales with the size at a given mass, and since there is a correlation between metallicity and size  (Fig.~\ref{fig:ellison2o}), there should be a correlation between metallicity and density at a fixed mass. However, galaxies with the whole range of metallicities exist for every $\rho_\star$ and so density does not seem to be the primary driver of any correlation with metallicity.  
\begin{figure*}
\includegraphics[width=0.33\textwidth]{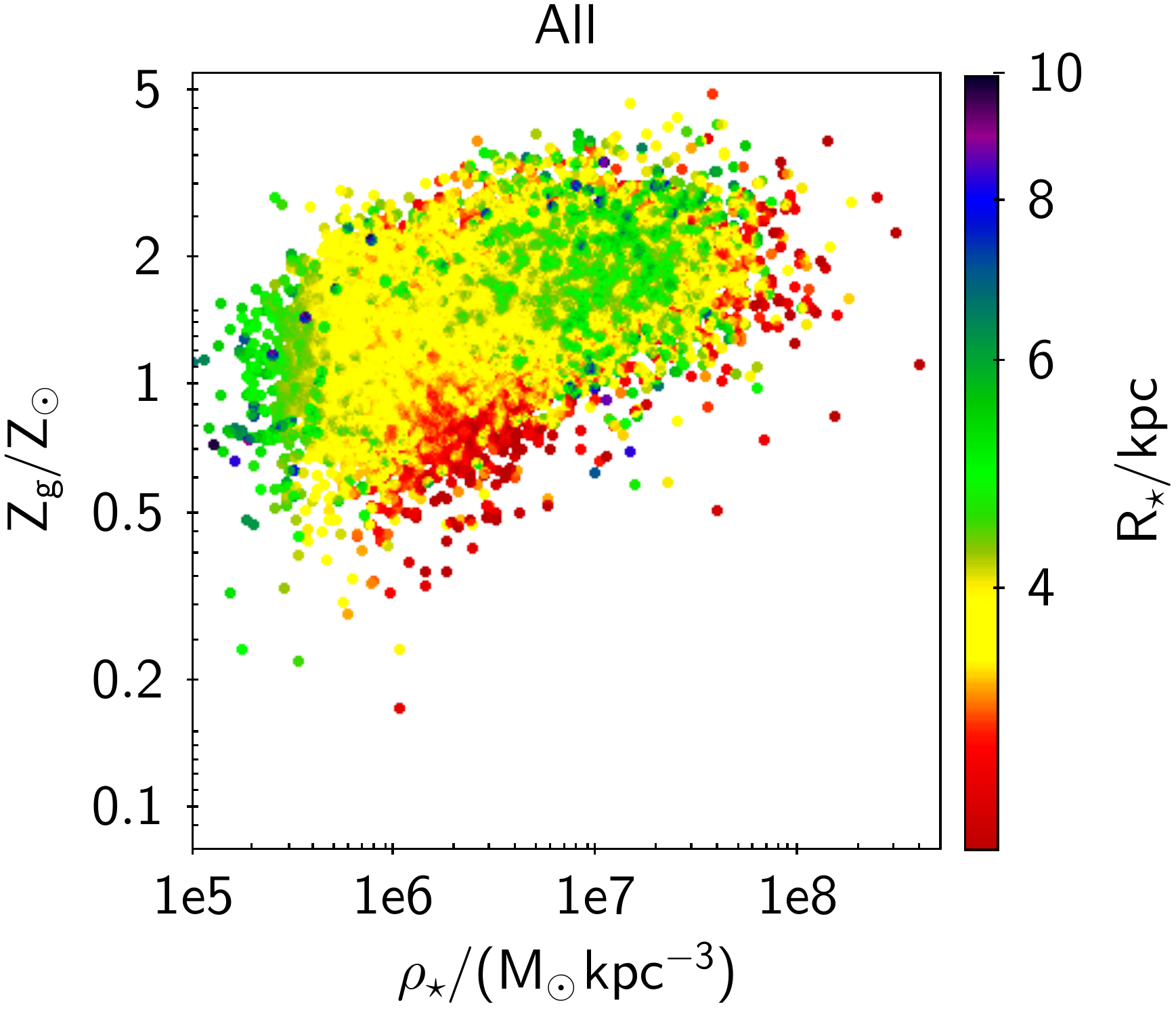}\hskip 0.2cm%
\includegraphics[width=0.33\textwidth]{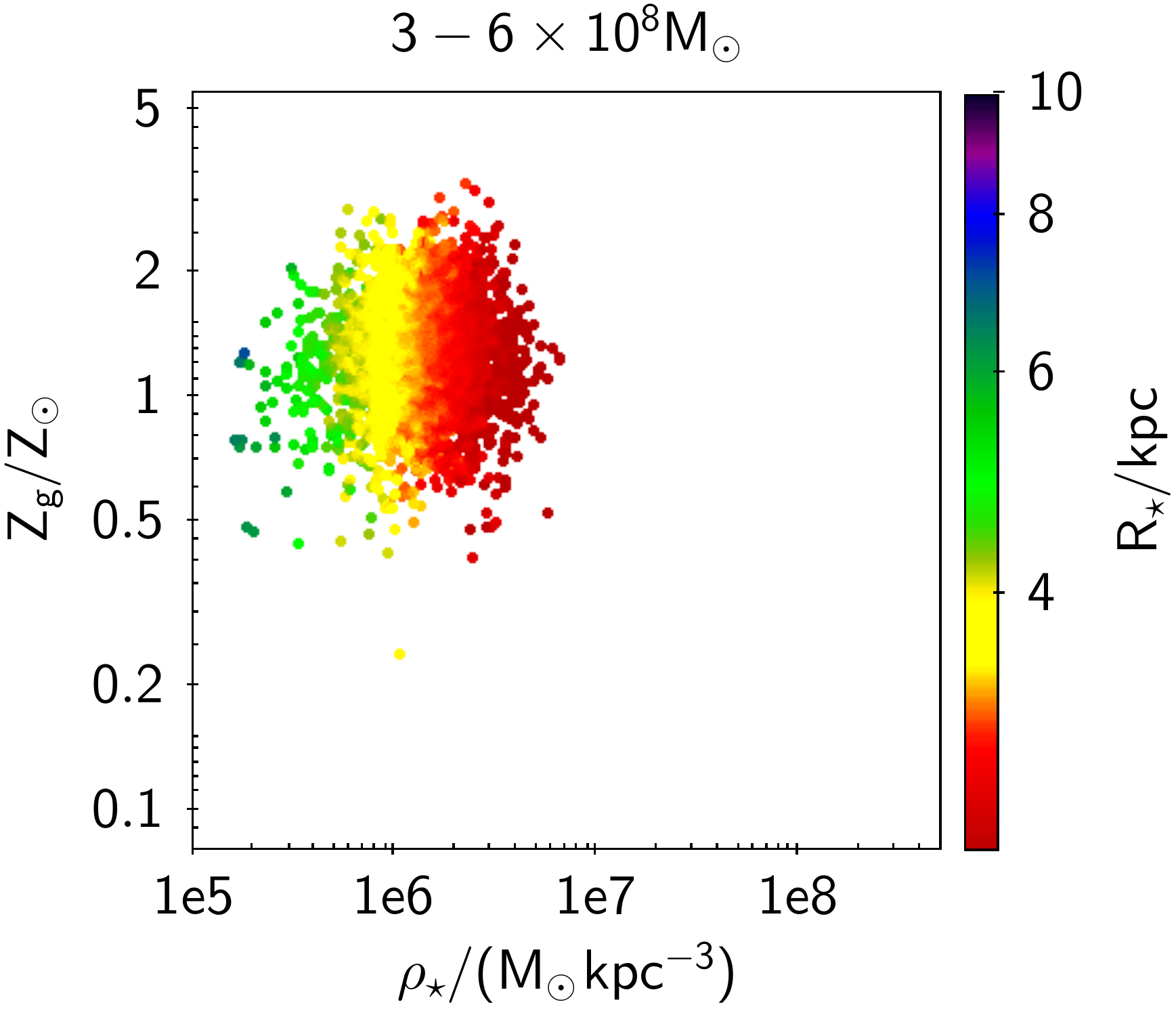}\hskip 0.2cm%
\includegraphics[width=0.33\textwidth]{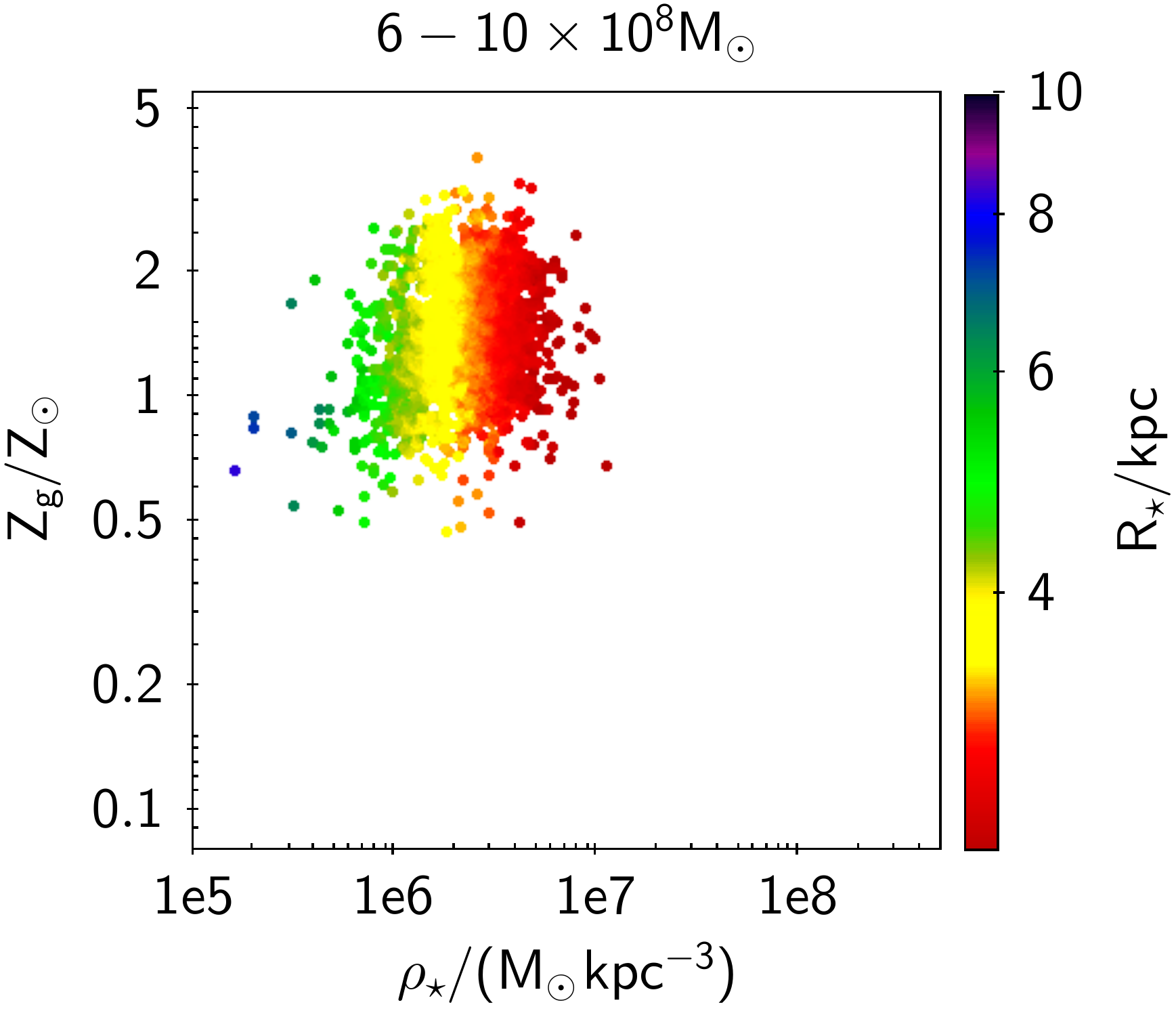}\vskip 0.2cm%
\includegraphics[width=0.33\textwidth]{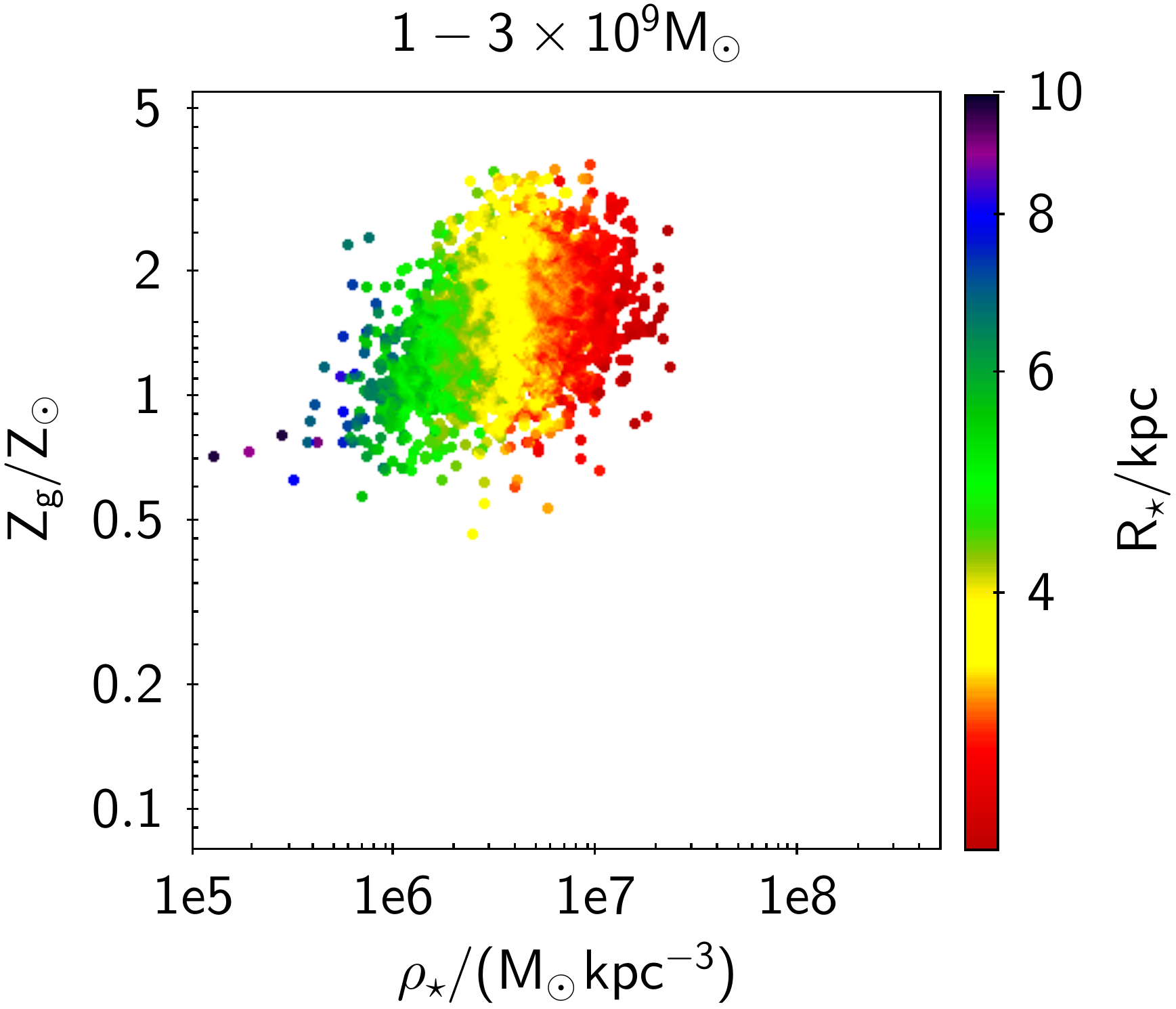}\hskip 0.2cm%
\includegraphics[width=0.33\textwidth]{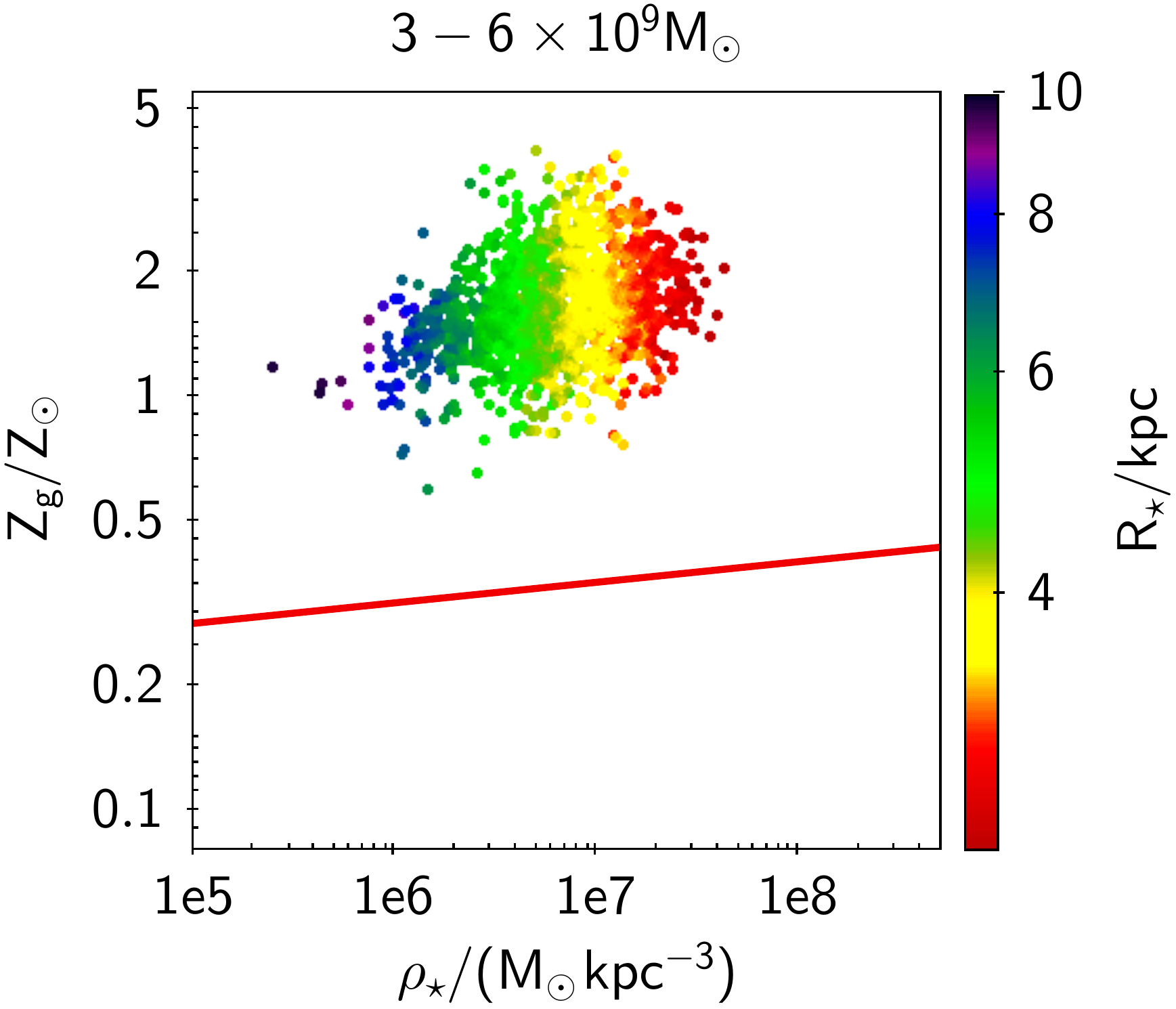}\hskip 0.2cm%
\includegraphics[width=0.33\textwidth]{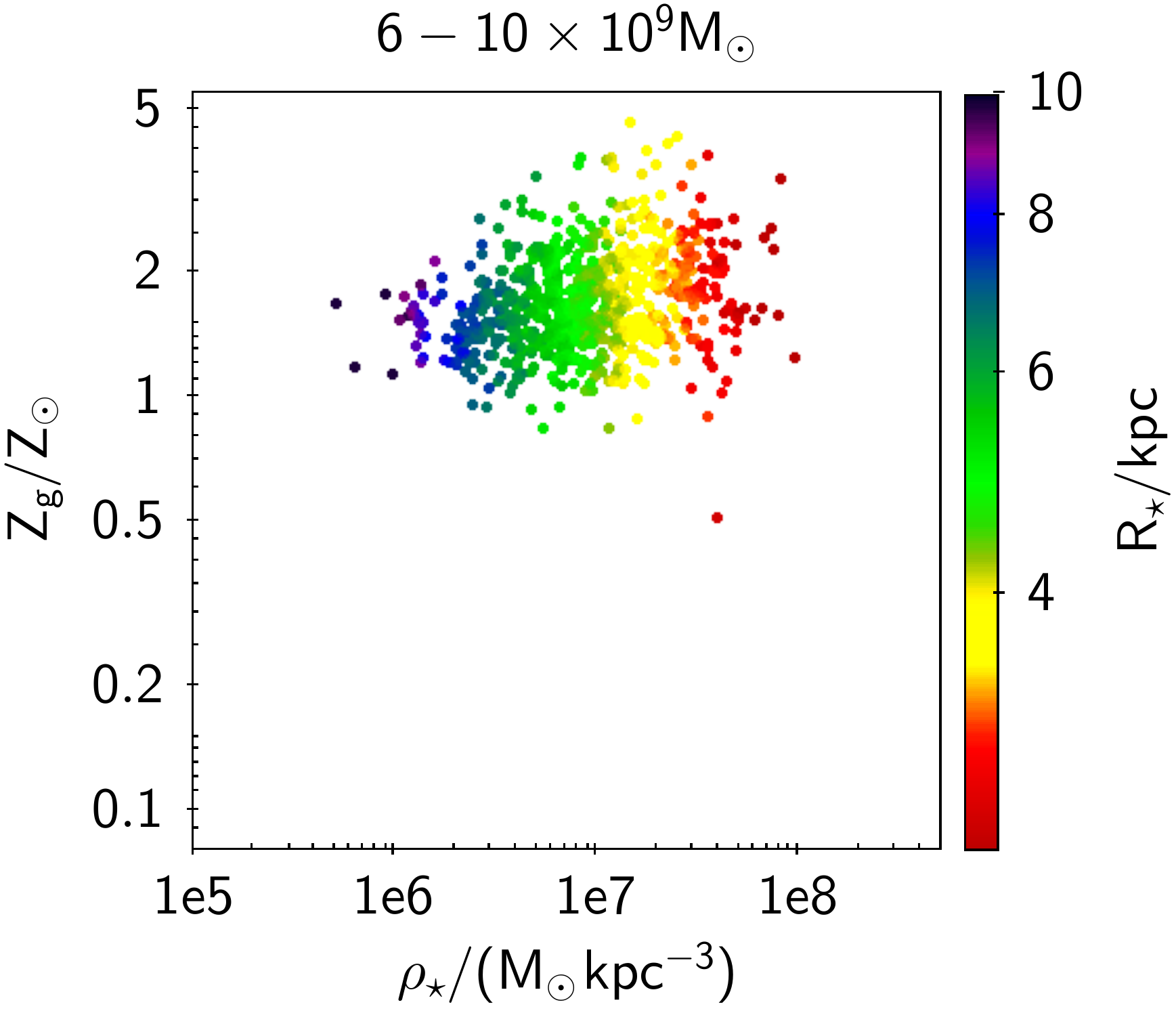}\vskip 0.2cm%
\includegraphics[width=0.33\textwidth]{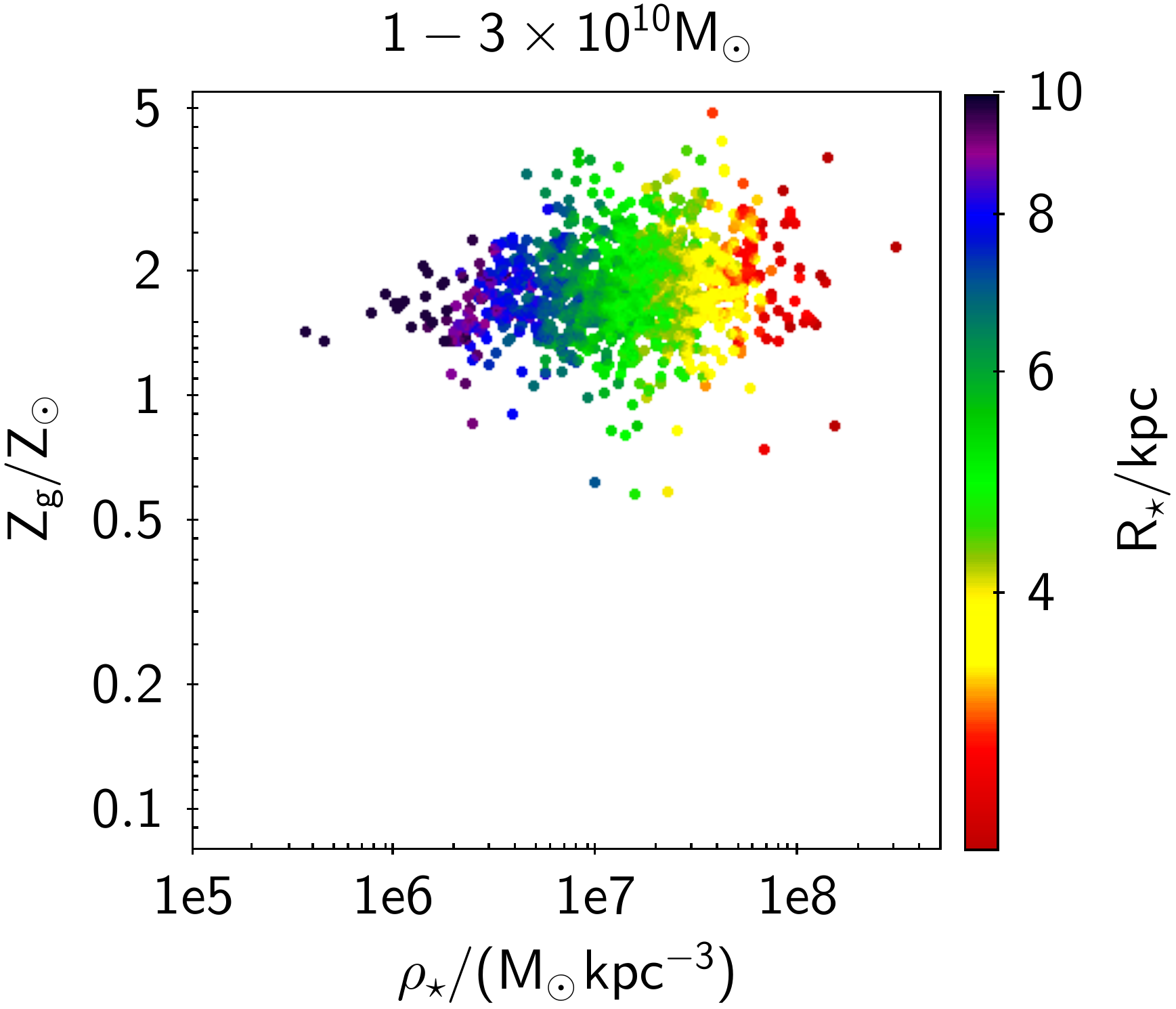}\hskip 0.2cm%
\includegraphics[width=0.33\textwidth]{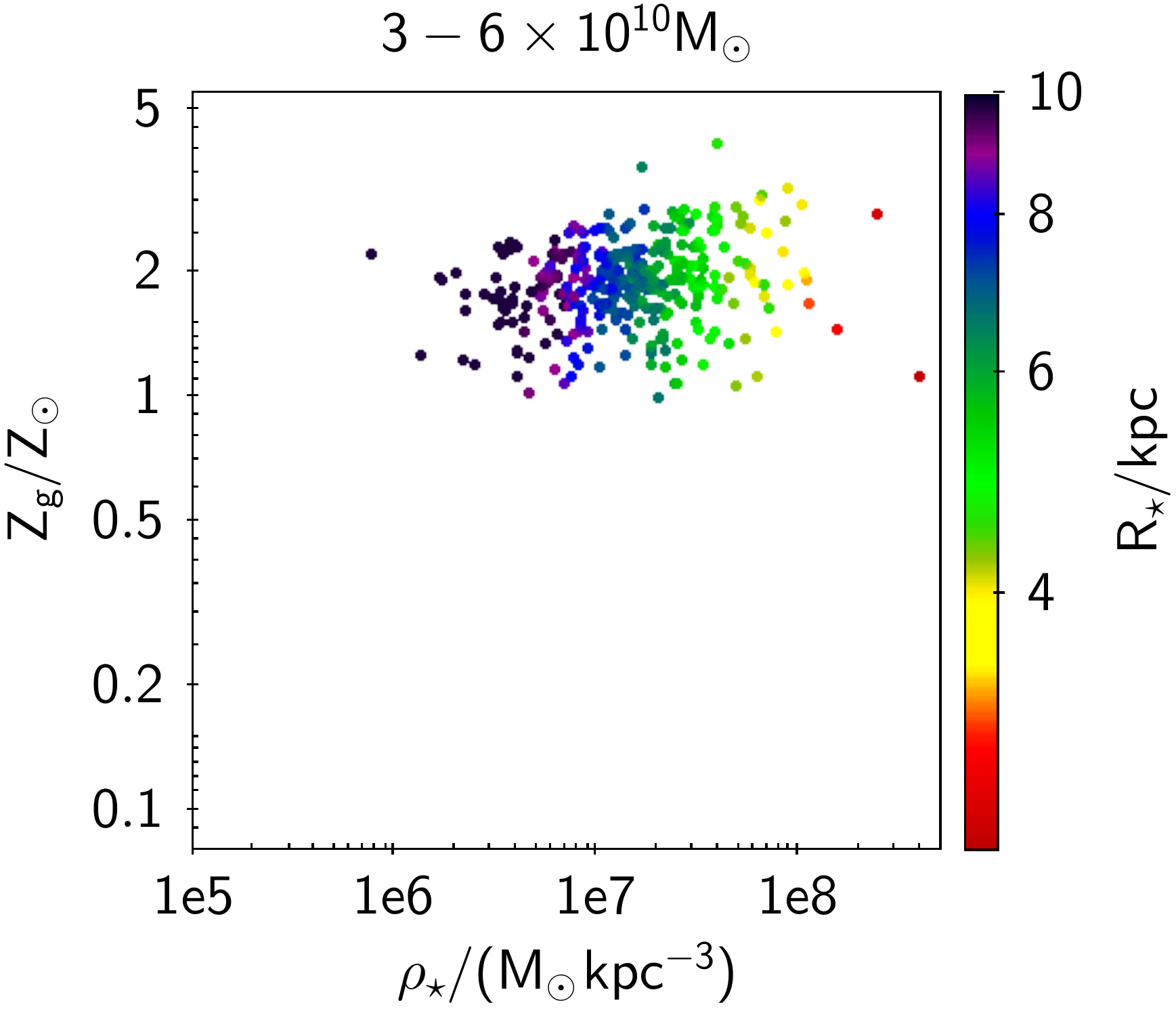}\hskip 0.2cm%
\includegraphics[width=0.33\textwidth]{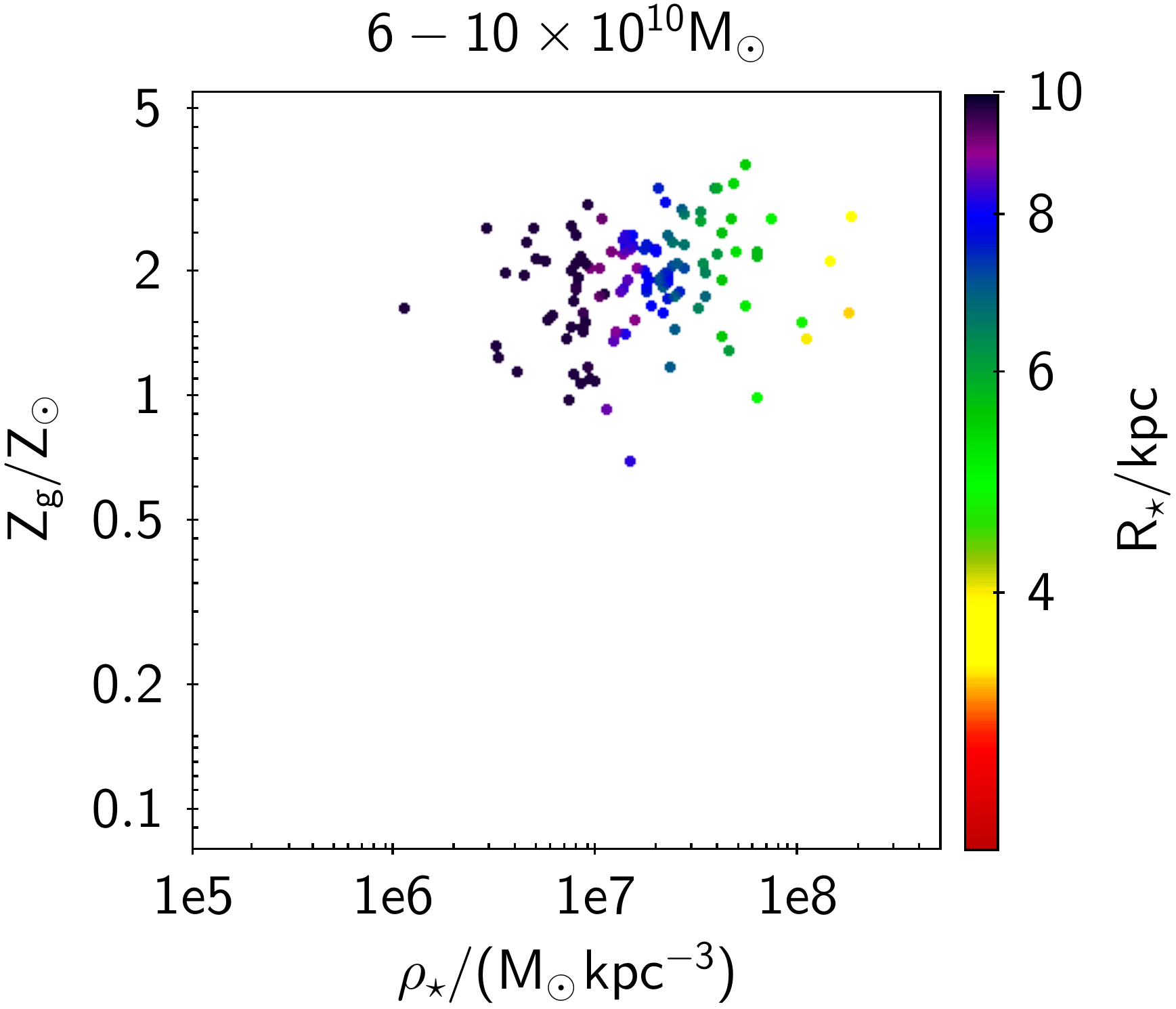}\vskip 0.2cm%
\caption{$Z_g$ versus $\rho_\star$ for all the galaxies (top left), and for galaxies within narrow mass bins that cover the whole range of masses from $10^{8}\,M_\odot$ to $10^{11}\,M_\odot$. The points are color-coded according to the mean radius of all the galaxies at each position on the plane. 
There is no strong correlation between $Z_g$ and $\rho_\star$ when considering individual mass bins. The solid line in the central panel shows the correlation expected according to the toy-model described in Sect.~\ref{sec:denstar}.}
\label{fig:z_vs_rho}
\end{figure*}

\section{Is the correlation between size and gas metallicity due to differences in the gas accretion history of the galaxies?}\label{sec:thisisit}

Disk galaxies grow in size with time by accreting gas in their outskirts. Those whose last major gas accretion episode happened earlier are now smaller and more metallic. We think that this physical process is responsible for the anti-correlation between size and gas-phase metallicity in the EAGLE galaxies.

\begin{figure*}
\includegraphics[width=0.31\textwidth]{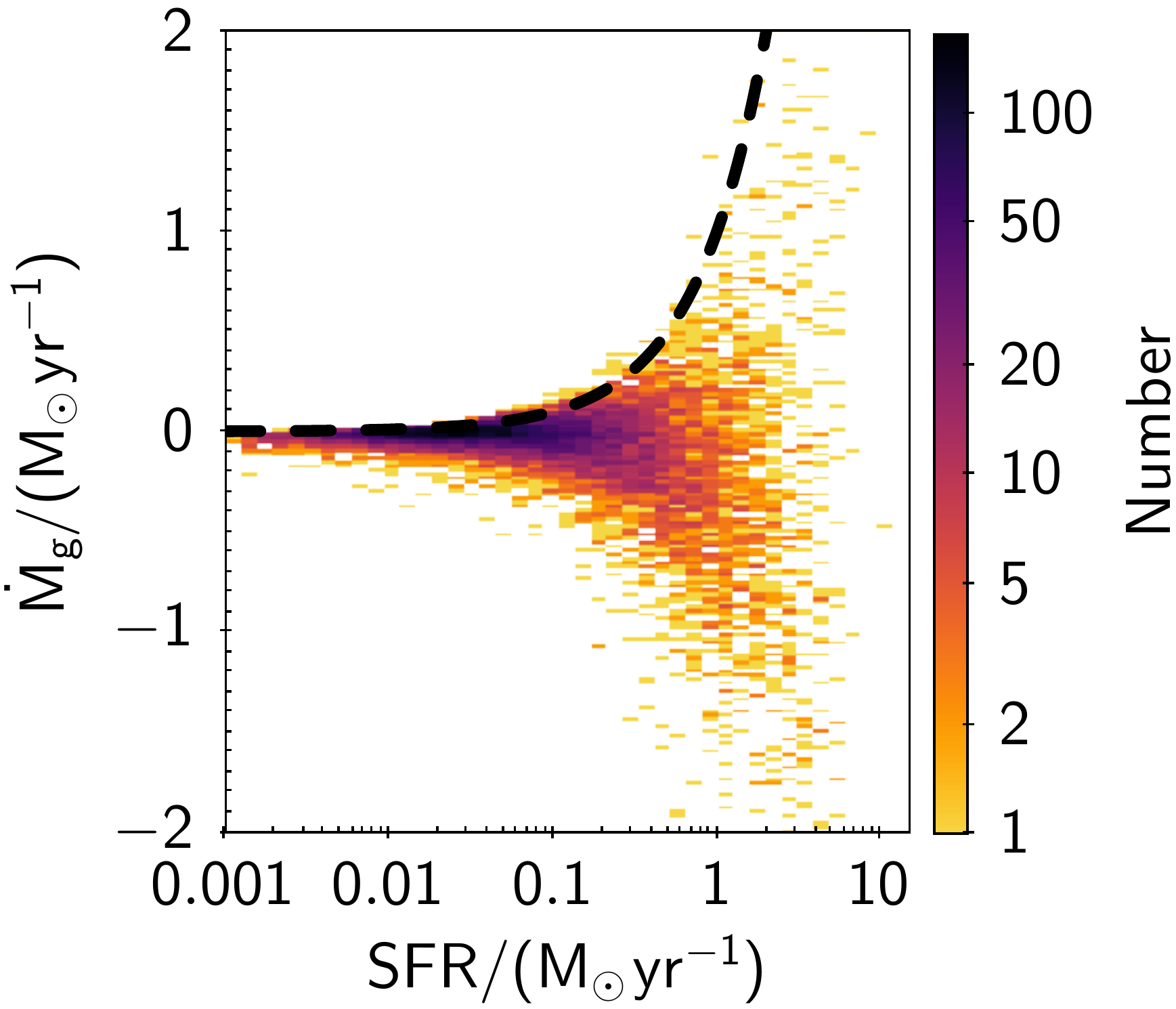}\hskip 0.5 cm%
\includegraphics[width=0.31\textwidth]{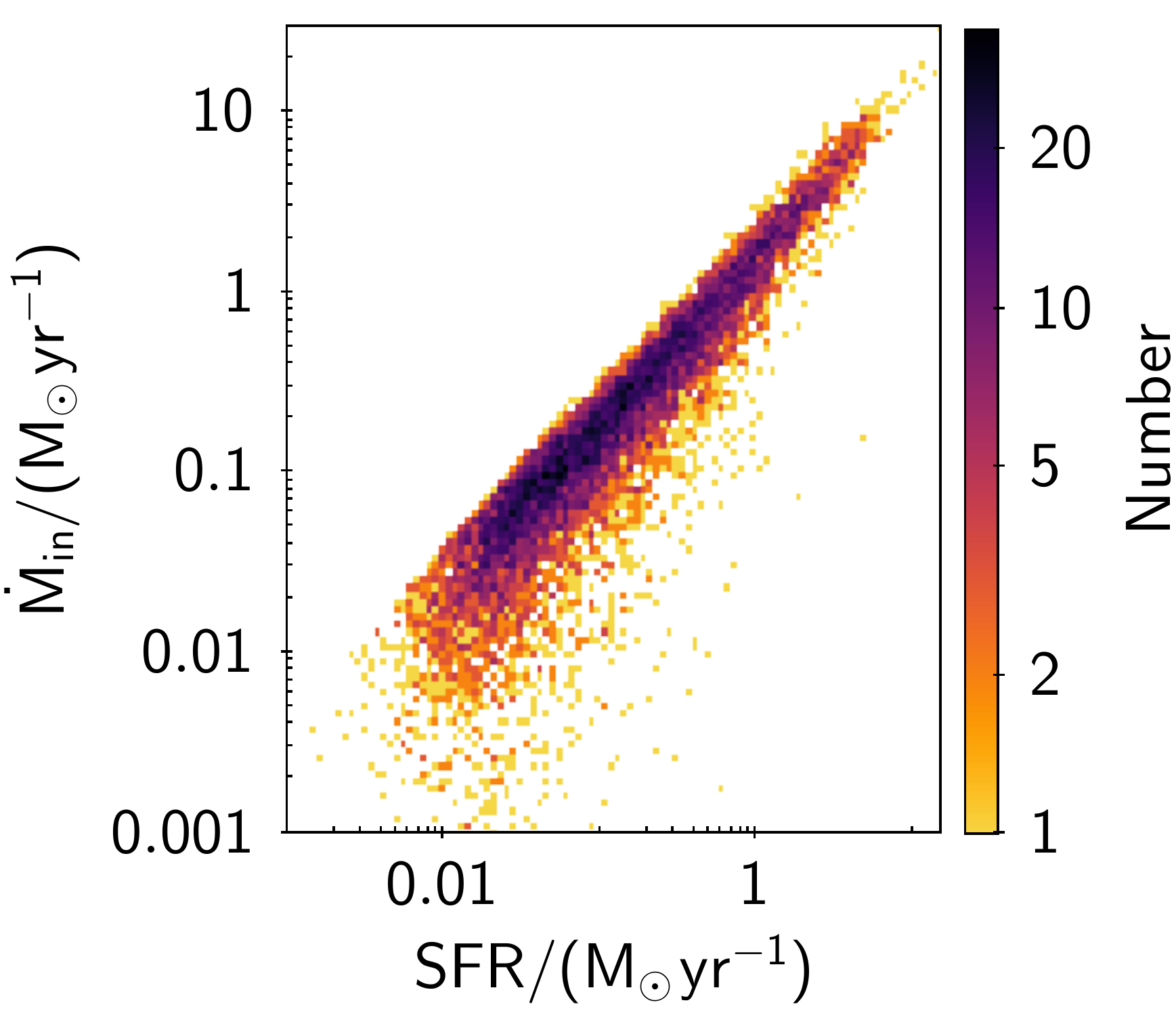}\hskip 0.5 cm%
\includegraphics[width=0.31\textwidth]{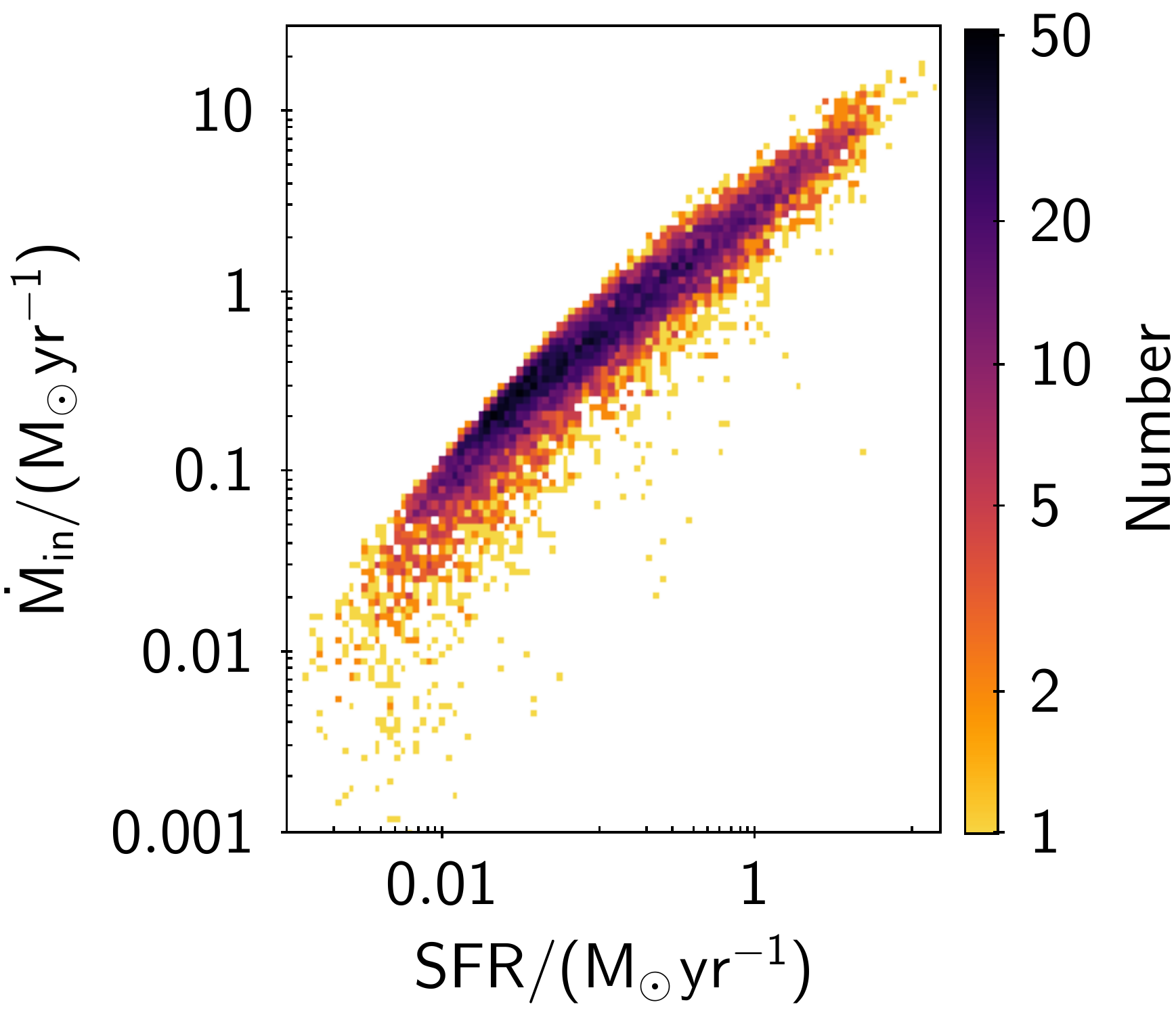}%
\caption{Left panel: time derivative of the gas mass versus SFR. Only the star-forming gas is considered. The dashed line corresponds to the one-to-one relation, so that most galaxies have $\dot{M}_g < {\rm SFR}$. Center panel: $\dot{M}_{in}$ versus SFR, estimated with $w=1$.  Right panel: $\dot{M}_{in}$ versus SFR for $w$ varying with stellar mass as parameterized by \citeauthor{2013MNRAS.430.2891D}~(\citeyear{2013MNRAS.430.2891D}; $w$ goes from 10 to 0.7  for $M_\star$ from $1.5\times 10^{8}\,M_\odot$ to $5\times 10^{11}\,M_\odot$). $\dot{M}_{in}$ is roughly proportional to SFR, independently of the actual $w$ in use. All panels are color-coded according to the number of galaxies in each point of the plane.}
\label{fig:mass_accretion_rate}
\end{figure*}

Ideally, one would like to study the dependence of the gas metallicity on the present gas accretion rate $\dot{M}_{\rm in}$.  The EAGLE database does not provide the gas accretion rate of the galaxies directly. We have estimated it from the gas mass and the SFR. The database provides the gas mass and thus its variation in time $\dot{M}_g$. We compute the difference of gas mass in the two last snapshots (redshifts 0 and 0.1), and then divide it by the difference in look-back time (1.35\,Gyr). Mass conservation coupled with Eq.~(\ref{eq:massload}) guarantees  
\begin{equation}
\dot{M}_{in}=\dot{M}_g+(1-R+w)\,{\rm SFR},
\label{eq:justification}
\end{equation}
so that $\dot{M}_{in}$ can be inferred from $\dot{M}_g$,  SFR, and $w$. However, $w$ is unknown. Fortunately, this fact is not critical since  SFR is usually larger than $\dot{M}_g$, and its contribution completely dominates $\dot{M}_{in}$. This is shown in Fig.~\ref{fig:mass_accretion_rate}. The left panel in the figure represents $\dot{M}_g$ versus SFR, and the galaxies tend to be below the one-to-one line (the dashed line). One reaches a similar conclusion by computing $\dot{M}_{in}$ for various $w$ and then noting that, independently of its actual value,  $\dot{M}_{in}\propto {\rm SFR}$.
The central panel in Fig.~\ref{fig:mass_accretion_rate} includes the scatter plots  $\dot{M}_{in}$ versus ${\rm SFR}$ for $w=1$. It shows a clear scaling between  $\dot{M}_{in}$ and SFR despite the fact that this case is particularly unfavorable. The smaller the value of $w$ is the larger the possible differences between $\dot{M}_{in}$ and SFR  (Eq.~[\ref{eq:justification}] yields $\dot{M}_{in}\propto {\rm SFR}$ when $w \gg 1$), and $w$ is typically larger than one \citep[e.g.,][]{2011MNRAS.416.1354D,2014A&ARv..22...71S}. If a more realistic $w=w(M_\star)$ is included, the relation tightens even further since most of the scatter at small SFRs in the central panel of Fig.~\ref{fig:mass_accretion_rate} comes from low-mass galaxies, where $w \gg 1$. The improvement is shown in the right panel of Fig.~\ref{fig:mass_accretion_rate}. It has been computed with the semi-empirical   $w=w(M_\star)$ worked out by \citet{2013MNRAS.430.2891D} to reproduce the observed mass-metallicity relation. It was chosen  because this parametrization of  $w=w(M_\star)$ yields the whole range of expected values, from low values at the high-mass end of the EAGLE mass distribution ($w\simeq 0.7$ at $M_\star=5\times 10^{11} M_\odot$) to large values at the low-mas end   ($w\simeq 10$ at $M_\star=1.5\times 10^{8} M_\odot$).
The gas mass used in Fig.~\ref{fig:mass_accretion_rate} corresponds to the star-forming gas, i.e., gas dense enough to contain a molecular phase cradling stars\footnote{The molecular phase is not resolved in the simulation, so that the process of transforming gas into stars is taken care of by  sub-grid physics. For details, see  Sect.~4.3 in \citet[][]{2015MNRAS.446..521S}.}. However, one reaches the same conclusion even if the mass of all the gas is used in this calculation. 

\begin{figure}
\includegraphics[width=0.45\textwidth]{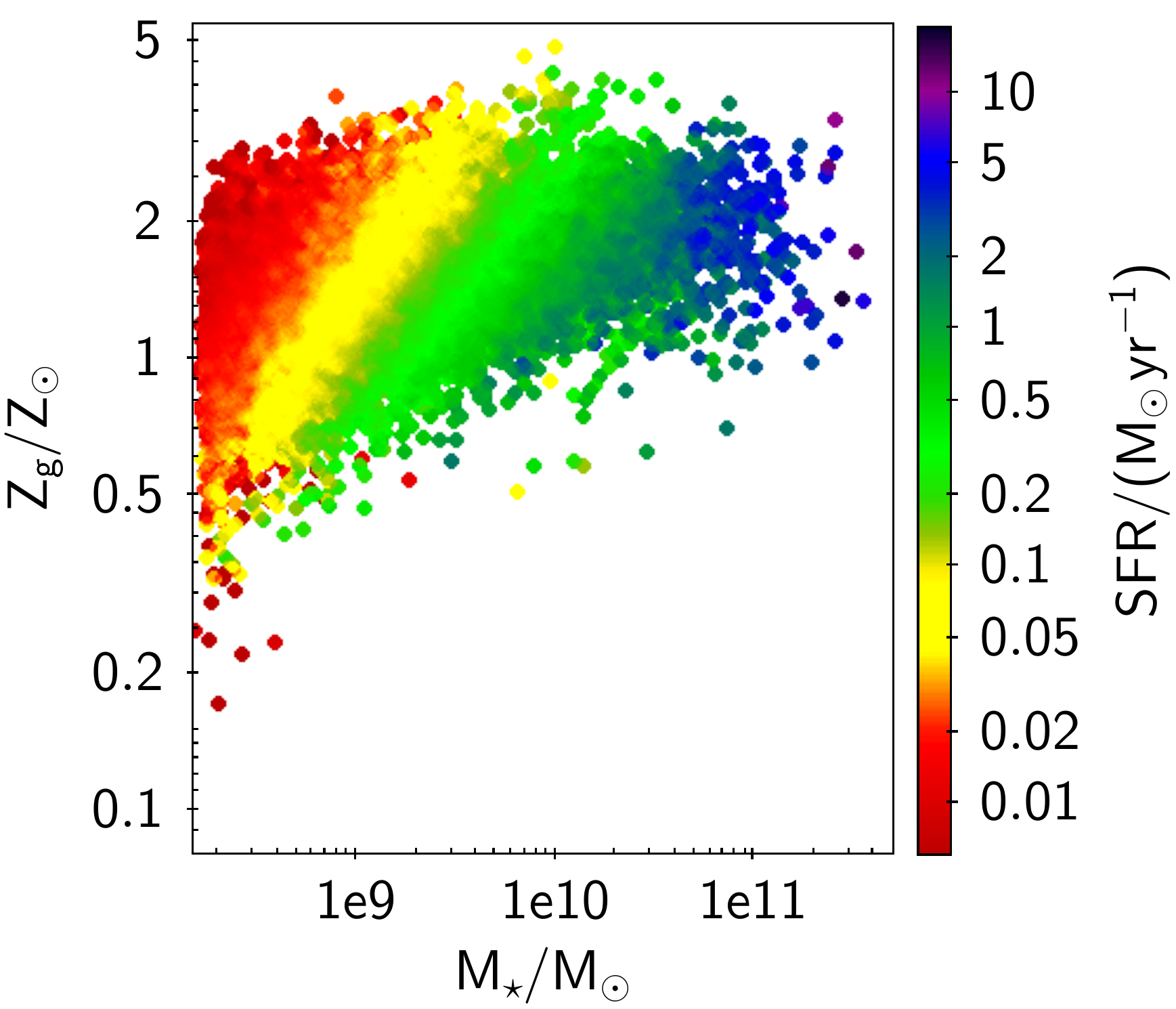}
\caption{Gas-phase metallicity versus stellar mass for the galaxies of the EAGLE simulation. The symbols are color-coded according to the SFR, which is used as a proxy for gas accretion rate. There is a clear anti-correlation between $Z_g$ and SFR at a given stellar mass.  Masses are given in $M_\odot$, metallicities in $Z_\odot$, and timescales in yr.}
\label{fig:ellison2_save4_sfr}
\end{figure}

\begin{figure*}
\includegraphics[width=0.33\textwidth]{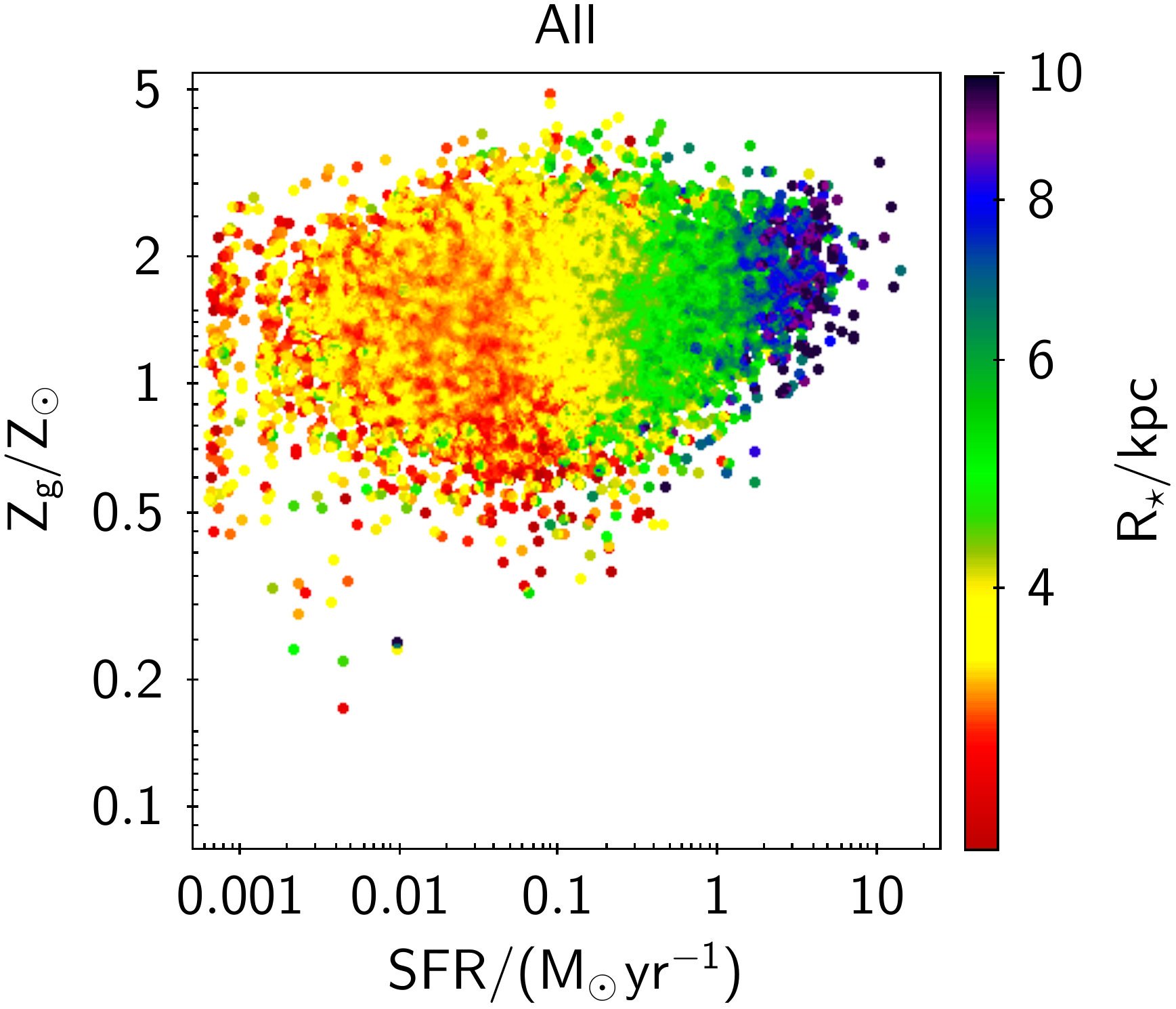}\hskip 0.2cm%
\includegraphics[width=0.33\textwidth]{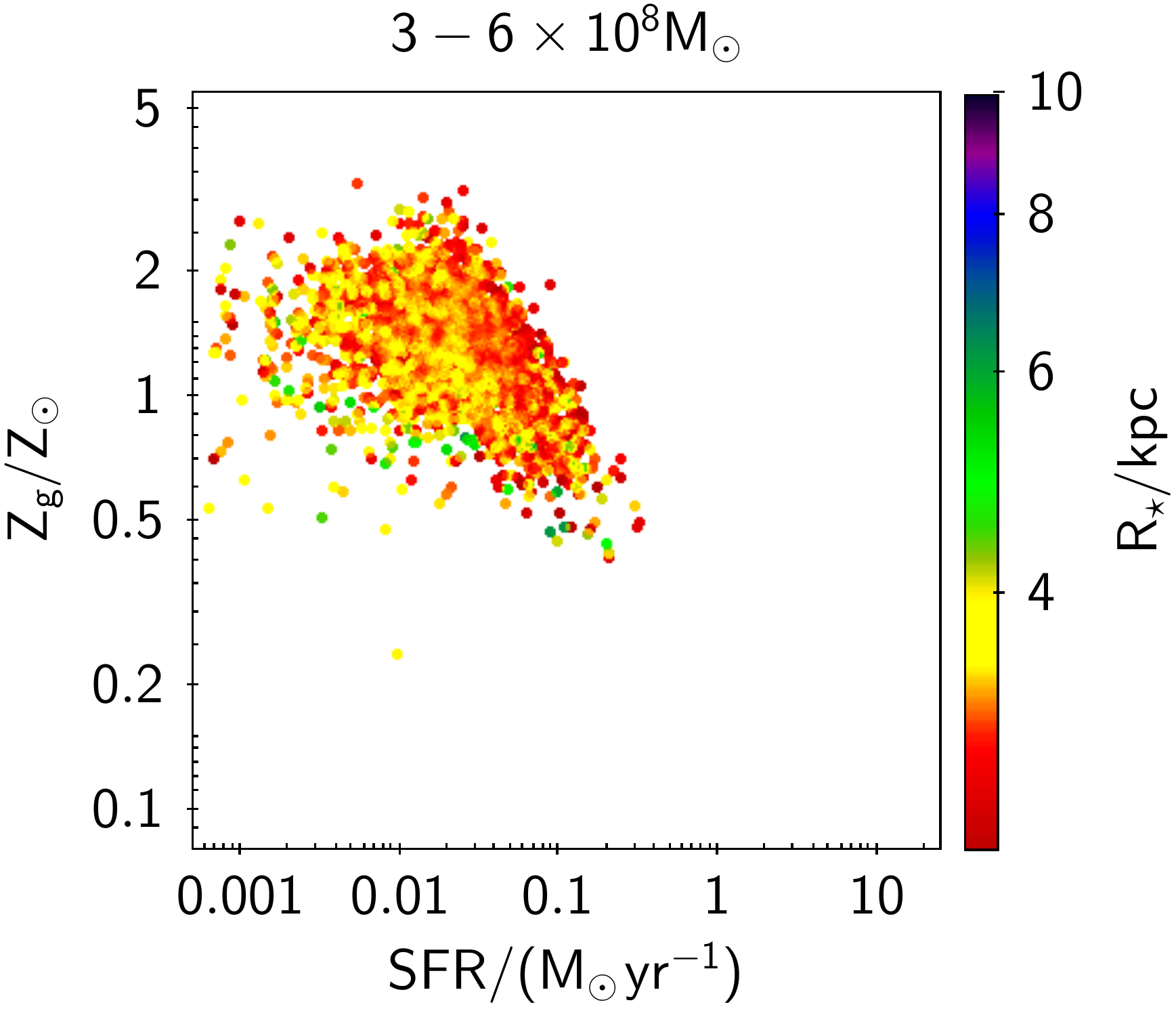}\hskip 0.2cm%
\includegraphics[width=0.33\textwidth]{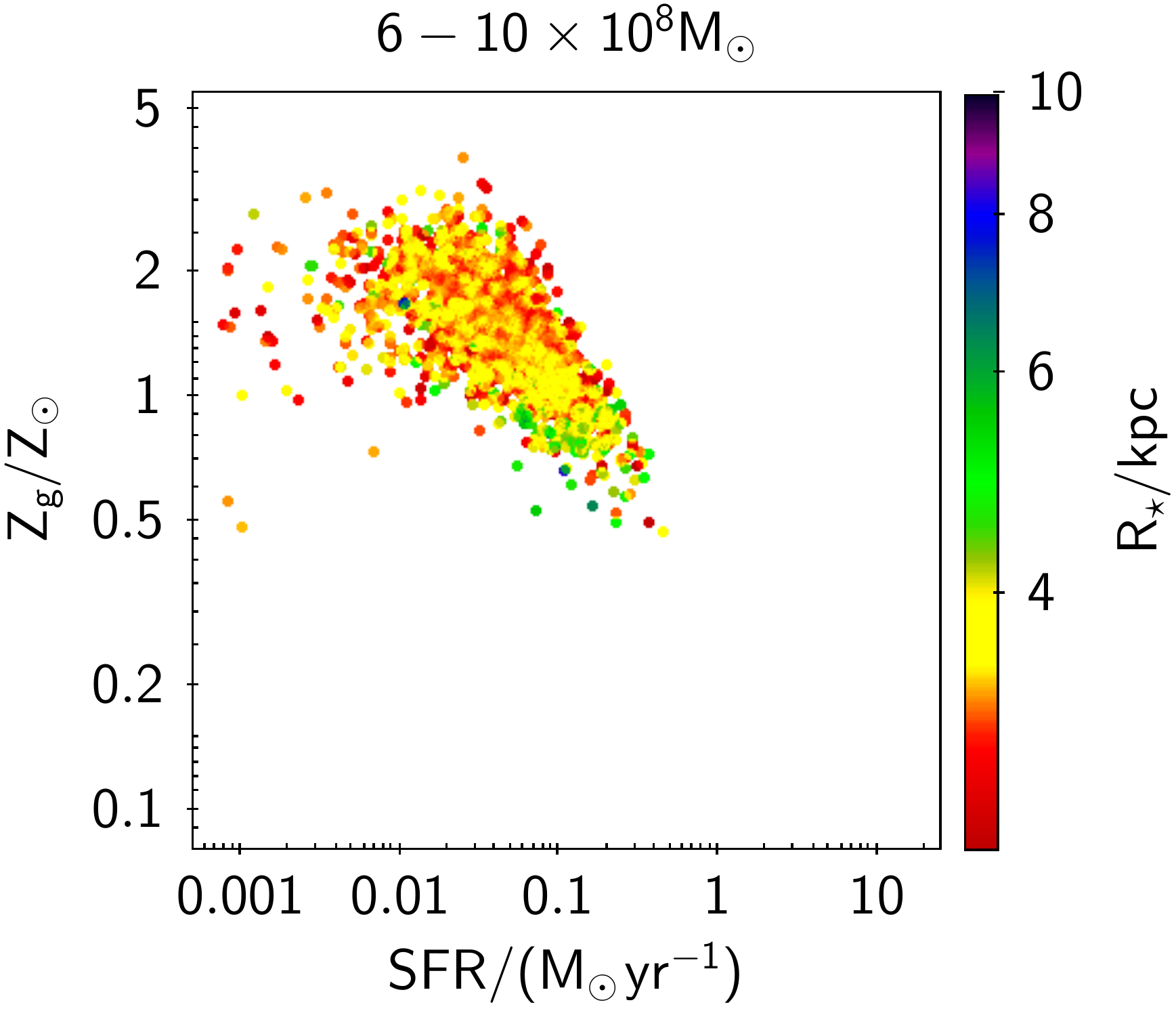}\vskip 0.2cm%
\includegraphics[width=0.33\textwidth]{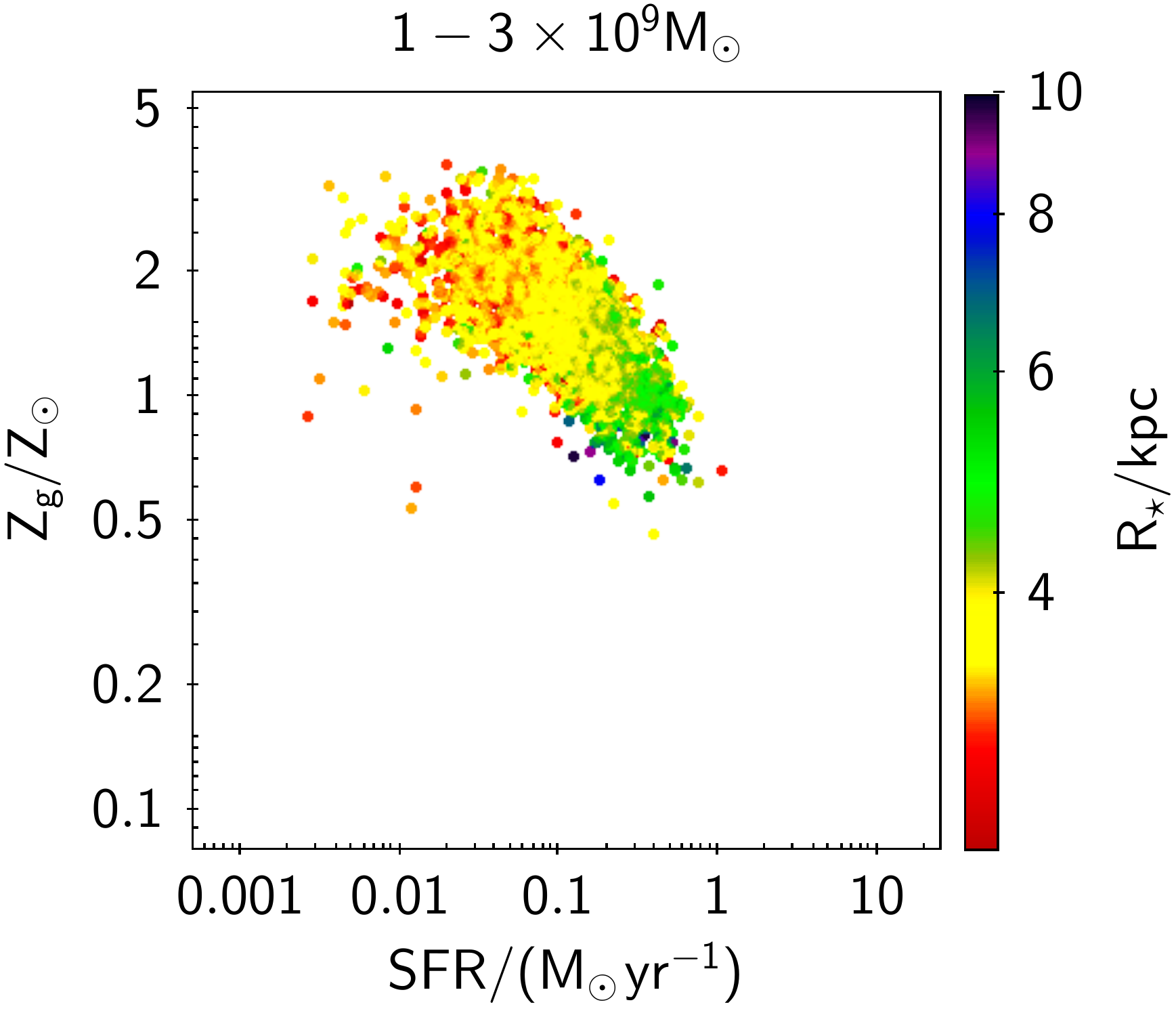}\hskip 0.2cm%
\includegraphics[width=0.33\textwidth]{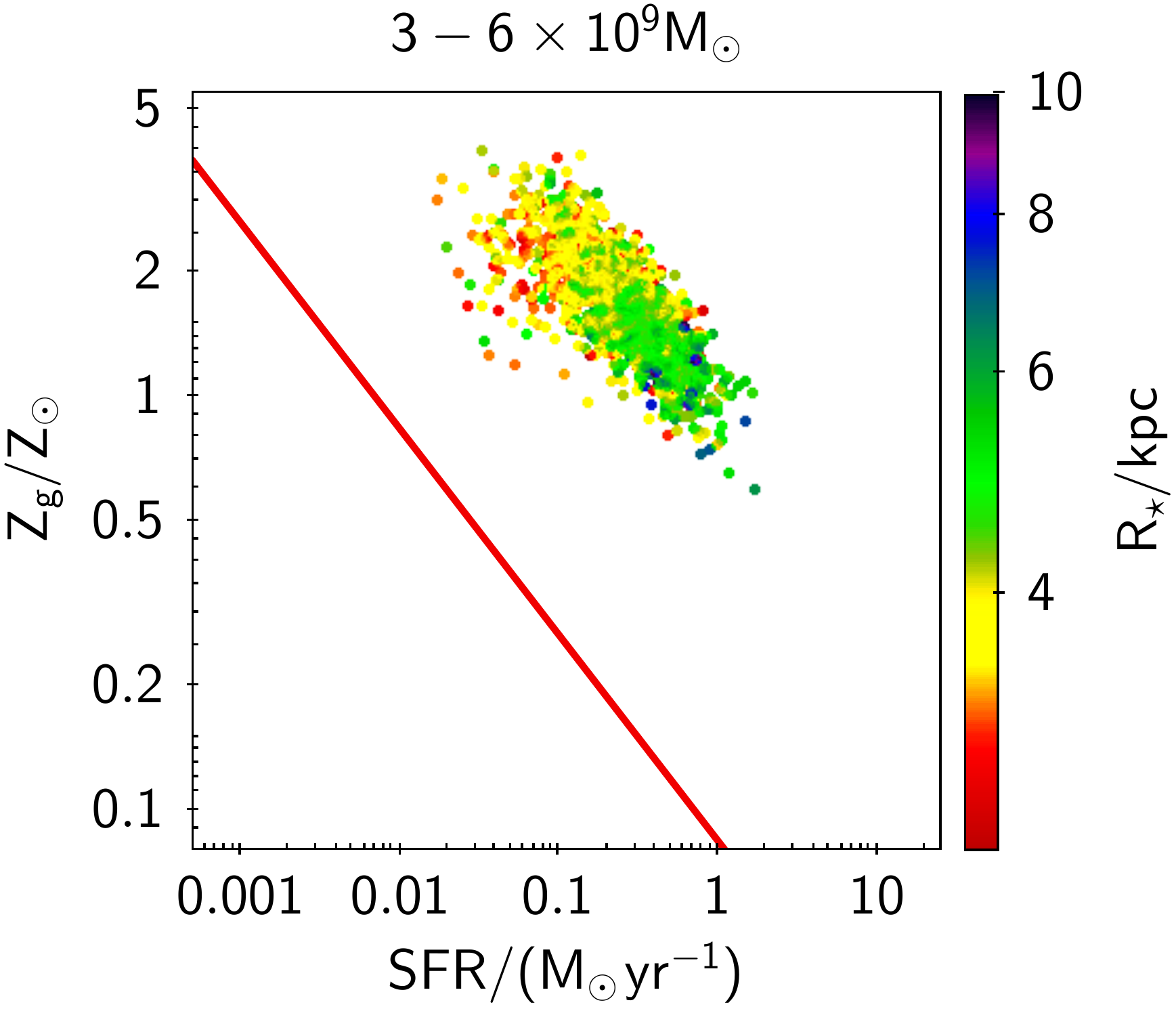}\hskip 0.2cm%
\includegraphics[width=0.33\textwidth]{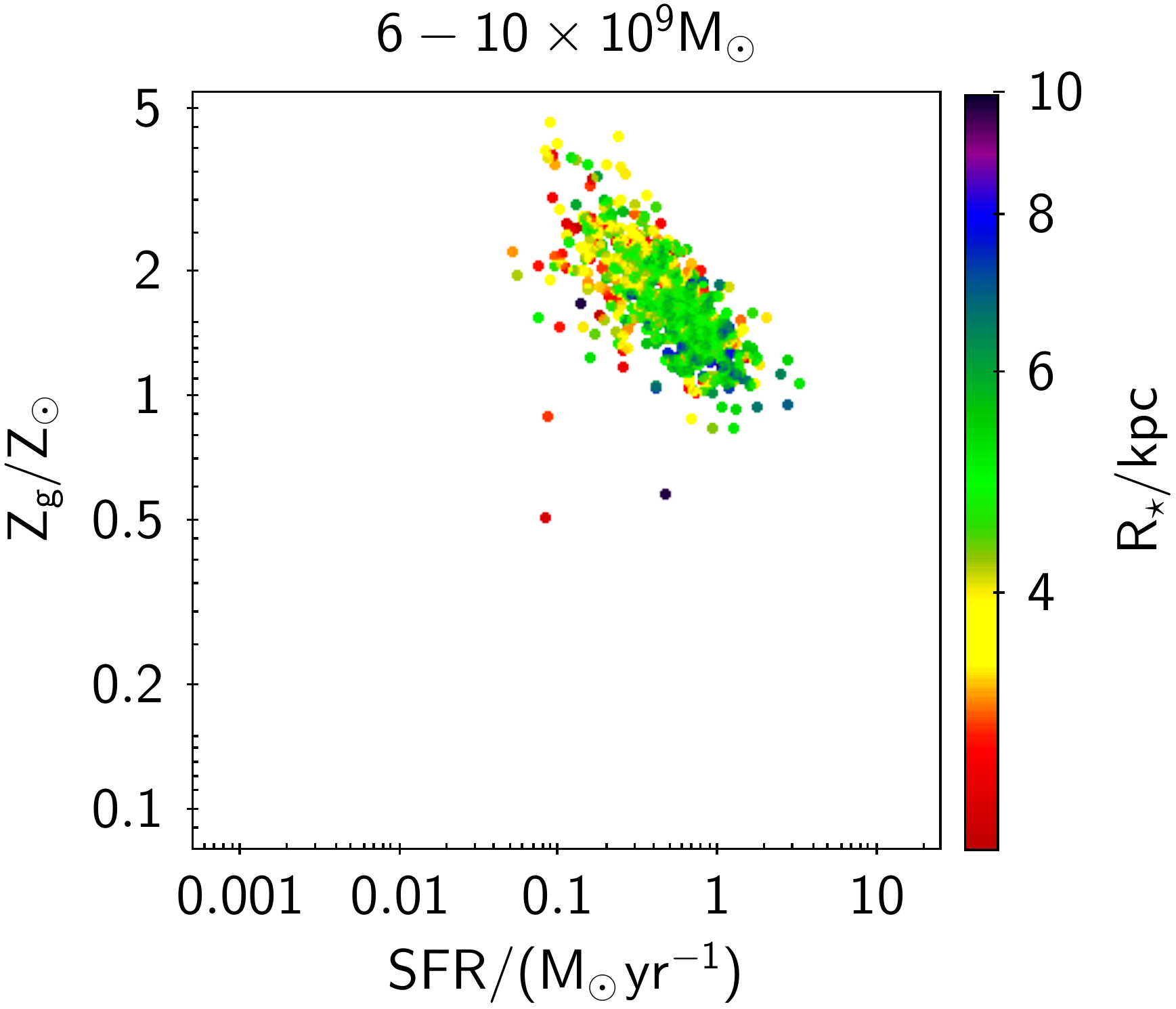}\vskip 0.2cm%
\includegraphics[width=0.33\textwidth]{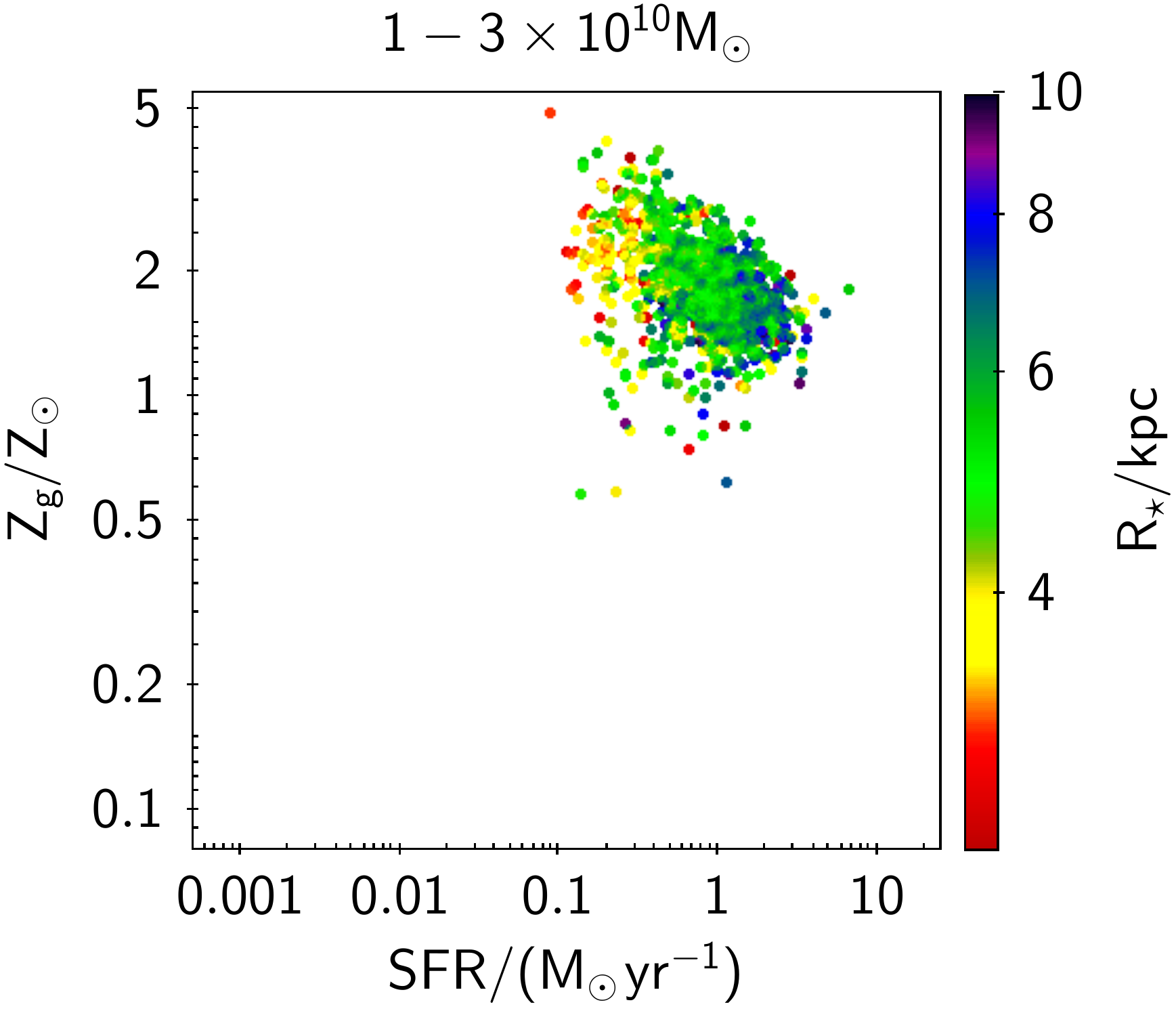}\hskip 0.2cm%
\includegraphics[width=0.33\textwidth]{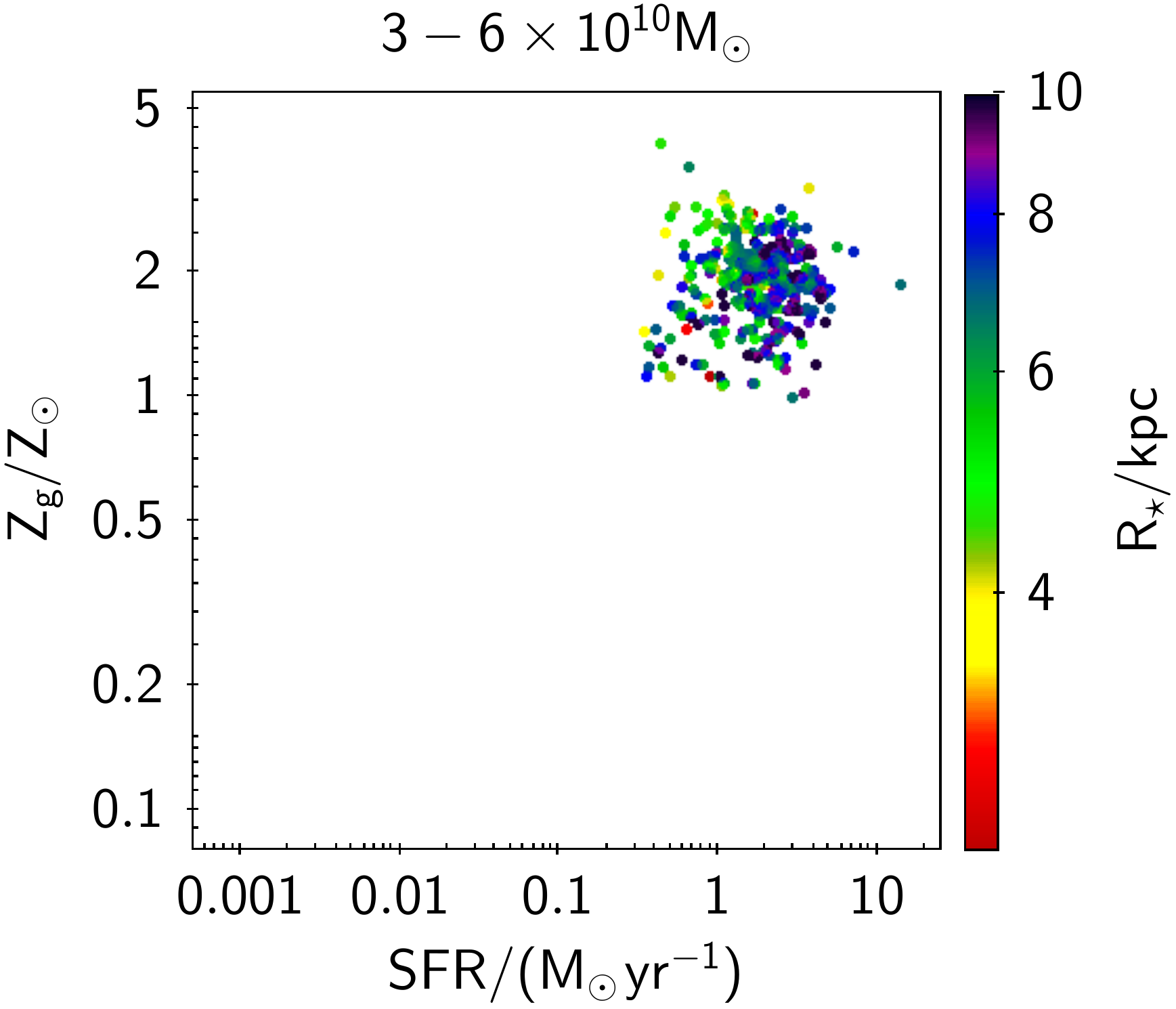}\hskip 0.2cm%
\includegraphics[width=0.33\textwidth]{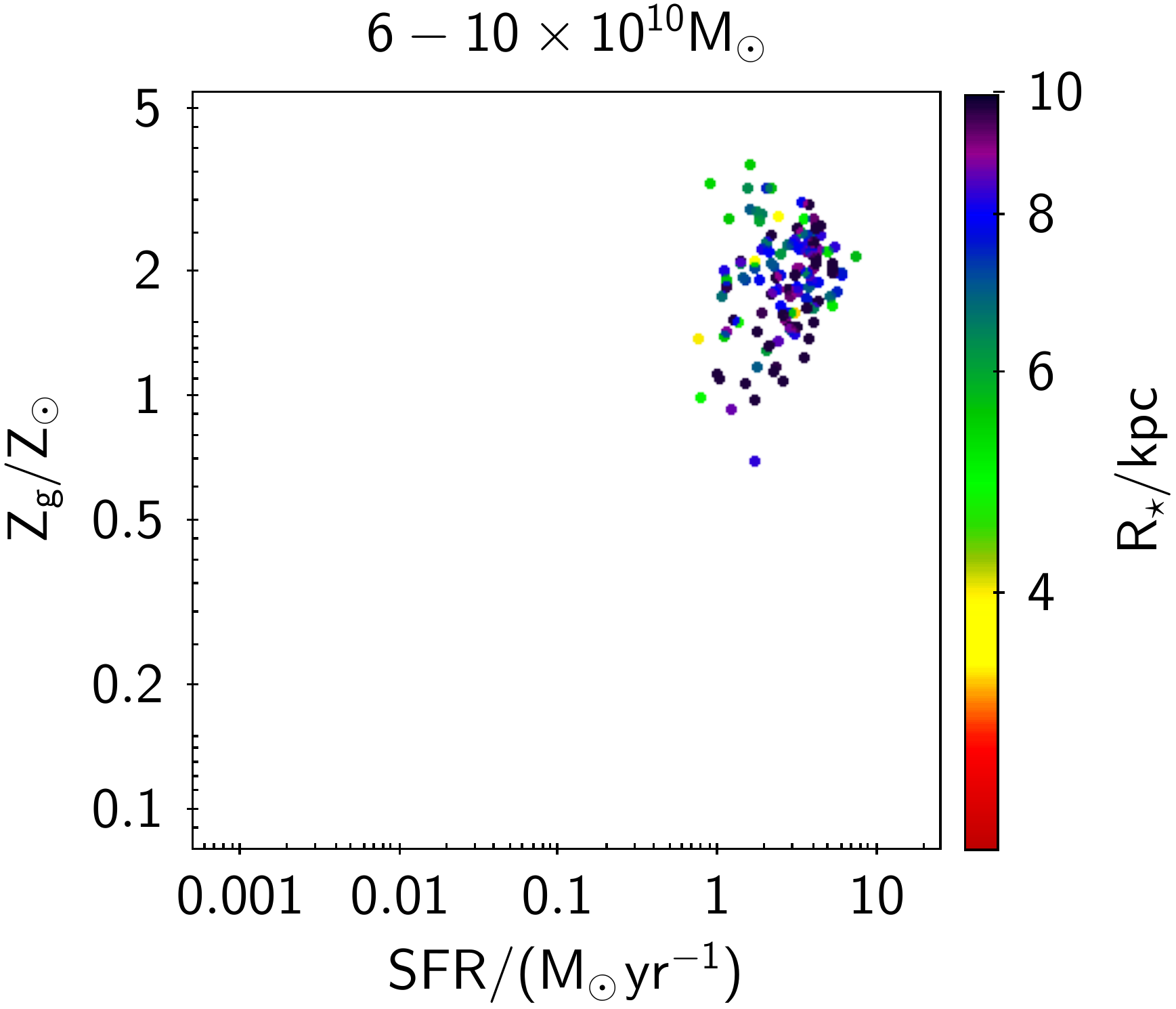}\vskip 0.2cm%
\caption{Gas-phase metallicity ($Z_g$) versus SFR color-coded with the half stellar mass radius of the galaxies ($R_\star$). The top-left panel displays the full data set. The rest of the panels show the same scatter plot selecting narrow mass bins (as labelled on top of each one). There is a clear anti-correlation  between ordinates and abscissae. The slope of the line shown in the $3 - 6\times 10^9\,M_\odot$ panel indicates the correlation expected according to the toy-model worked out in the main text. The axes and the color code are identical in all panels. SFR is used here as a proxy for gas accretion rate.}
\label{fig:z_vs_sfr}
\end{figure*}

In view of the uncertainties in $w$, and due to the good scaling between the two quantities, from now on we use SFR as a proxy for  $\dot{M}_{in}$. Figure~\ref{fig:ellison2_save4_sfr} shows the gas-phase metallicity versus stellar mass color-coded with the mean SFR. Unlike what happens when the color-code reflects $v_{esc}$ or $\rho_\star$ (Figs.~\ref{fig:ellison2_save} and \ref{fig:density}, respectively), this time there is a clear dependence of $Z_g$ on the SFR for a fixed mass. This trend is even more clear in Fig.~\ref{fig:z_vs_sfr}. It contains  the scatter plot of metallicity ($Z_g$) versus SFR color-coded with the half mass-stellar radius of the galaxies. The top-left panel represents the full data set. The rest of the panels show the same scatter plot selecting narrow mass bins (as labelled on top of each figure). There is a clear relation between ordinates and abscissae.  The solid line in the panel corresponding to masses between 3 and $6 \times 10^9\,M_\odot$ shows the anti-correlation expected according to the toy-model worked out in the main text (Eq.~[\ref{eq:logder}] with $\dot{M}_{\rm in}\propto {\rm SFR}$). Given a stellar mass, galaxies of higher SFR are also larger (Fig.~\ref{eq:ellison2_save_ssfr}, central panel, but see also the color coding of the panels in Fig.~\ref{fig:z_vs_sfr}). This relation between size, SFR and $Z_g$ is the one that, once averaged over the full population of galaxies, gives rise to the correlation between size and gas metallicity we are trying to explain.   

Note that the correlation between metallicity and SFR, so clear in the panels of the individual mass bins, washes out when considering all the galaxies together (Fig.~\ref{fig:z_vs_sfr}, top left). This is due to the fact that the {\em negative} correlation between $Z_g$ and SFR at fixed $M_\star$, turns into a {\em positive} correlation between $Z_g$ and SFR when the variation with mass is considered. Both the mean $Z_g$ and the mean SFR increase with increasing $M_\star$ (see Fig.~\ref{eq:ellison2_save_ssfr}, left panel). The two tendencies tend to cancel out when averaging over galaxies of all masses, resulting in a lack of correlation.

\begin{figure*}
\includegraphics[width=0.33\textwidth]{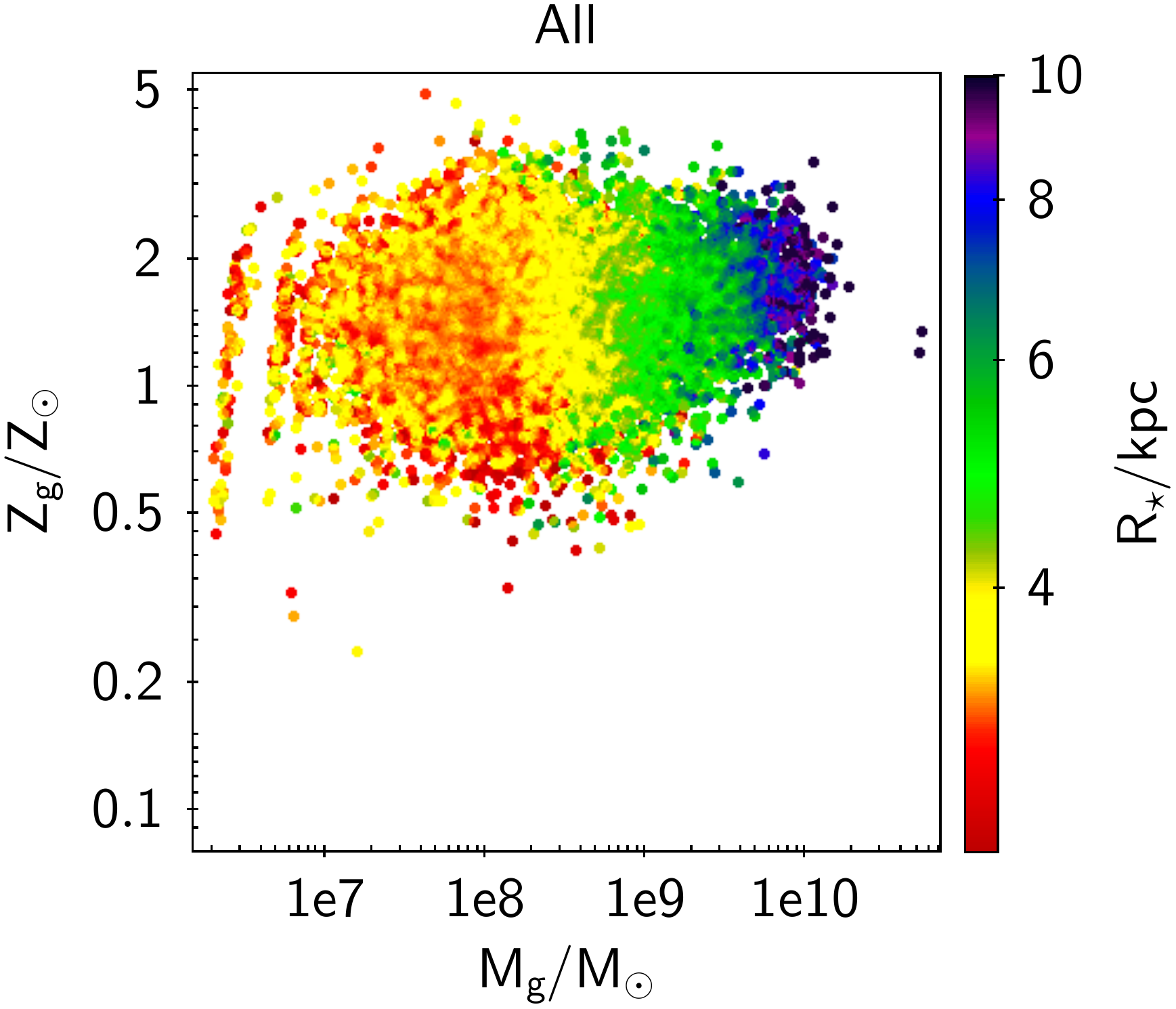}\hskip 0.2cm%
\includegraphics[width=0.33\textwidth]{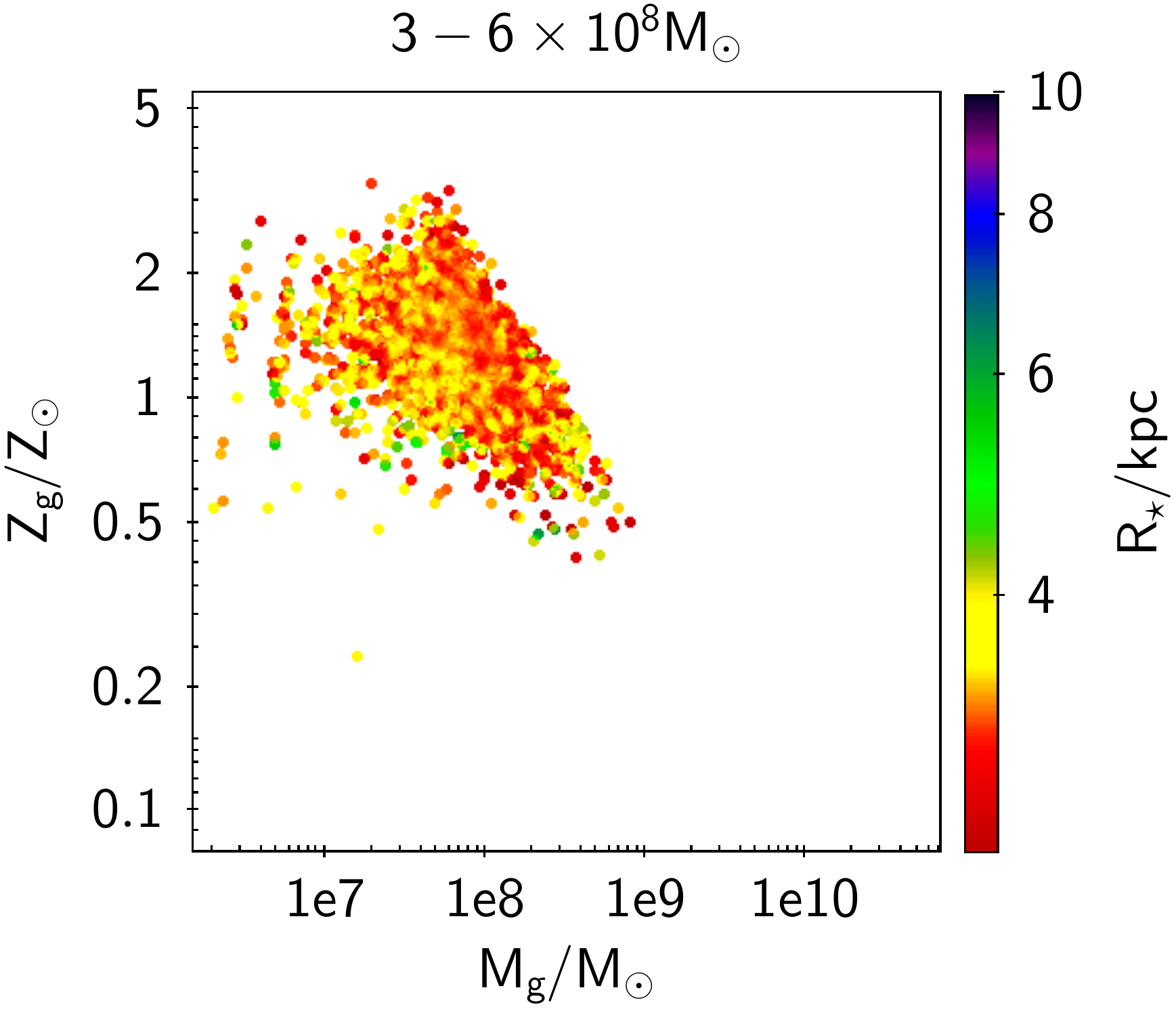}\hskip 0.2cm%
\includegraphics[width=0.33\textwidth]{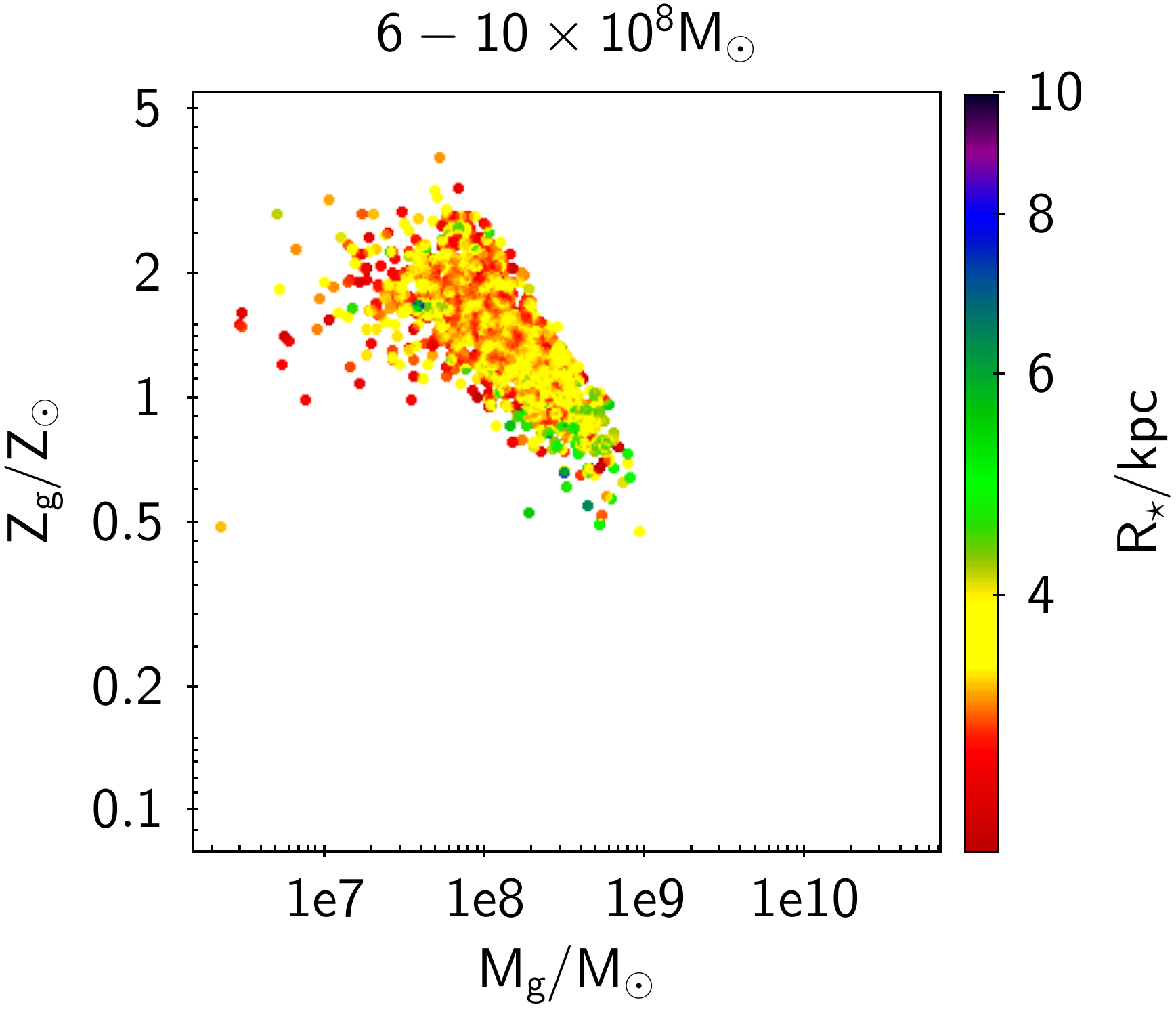}\vskip 0.2cm%
\includegraphics[width=0.33\textwidth]{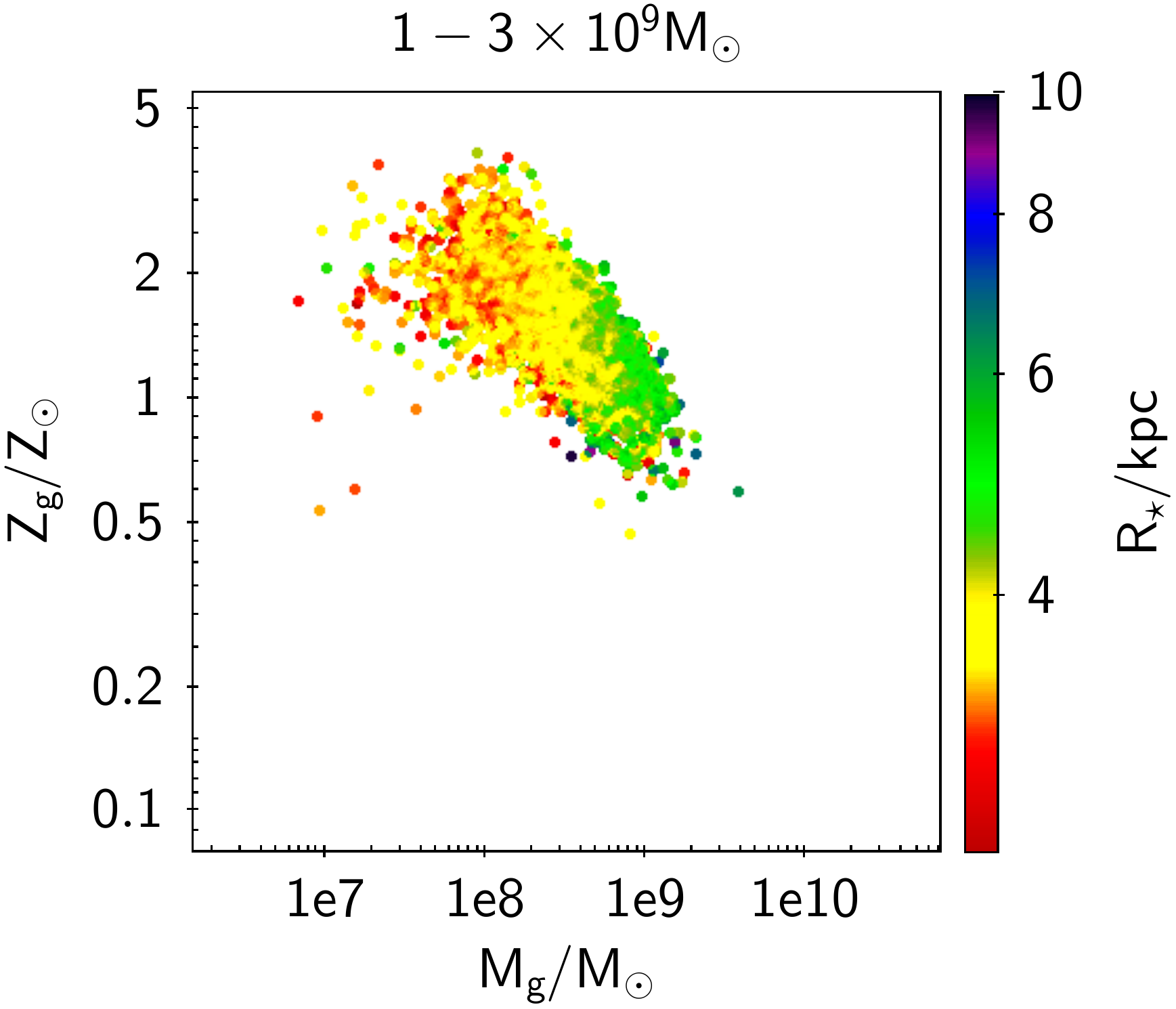}\hskip 0.2cm%
\includegraphics[width=0.33\textwidth]{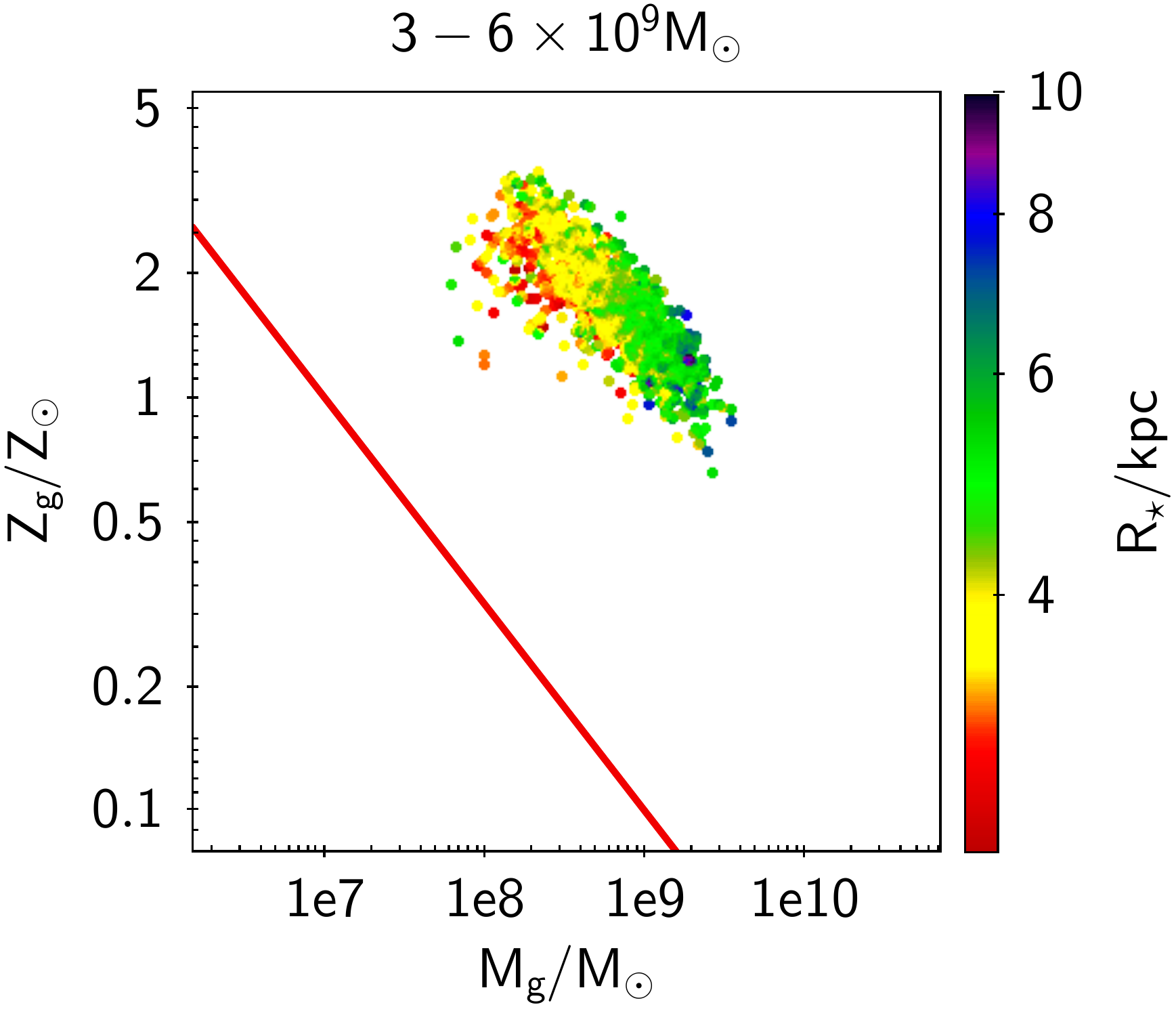}\hskip 0.2cm%
\includegraphics[width=0.33\textwidth]{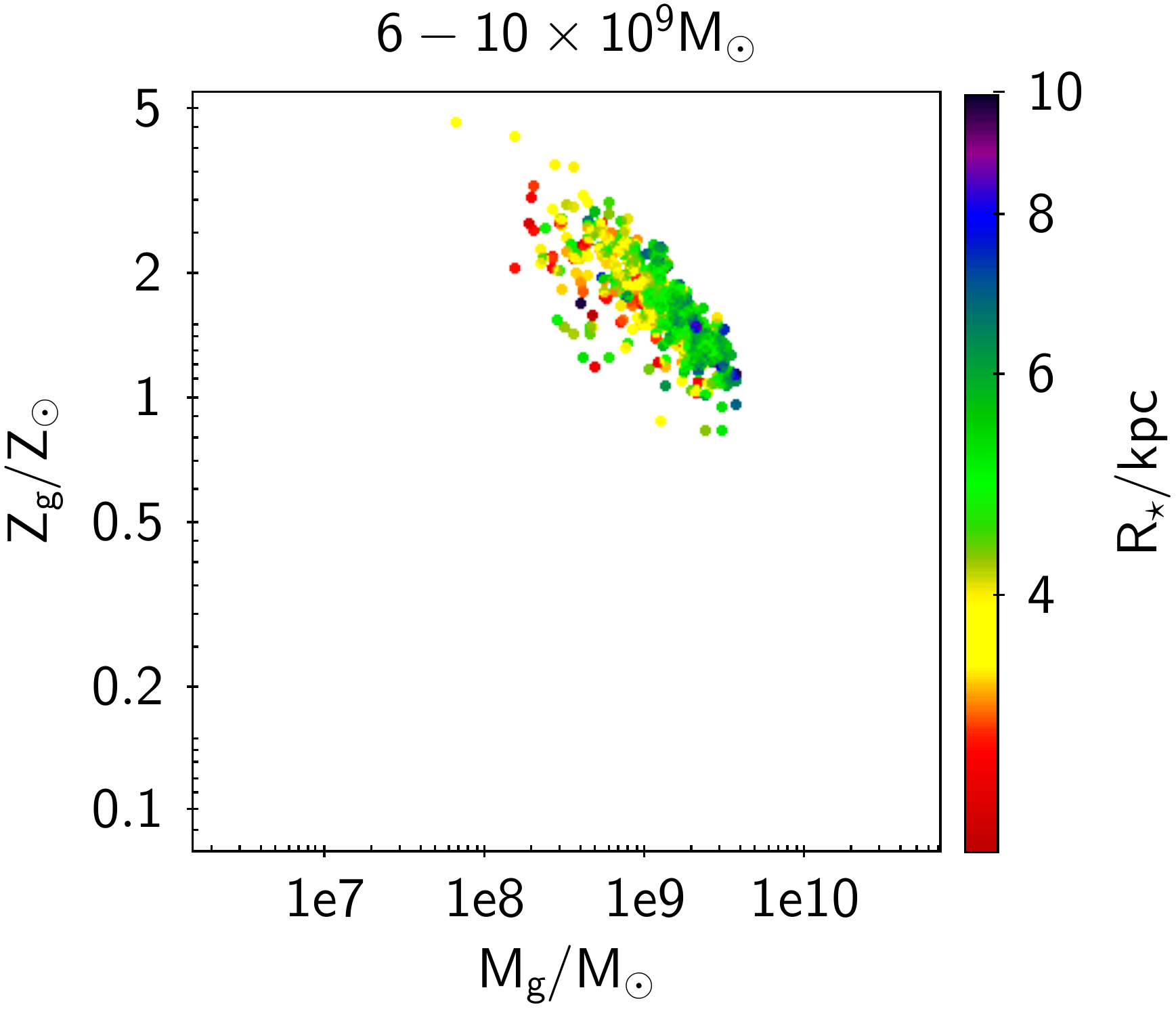}\vskip 0.2cm%
\includegraphics[width=0.33\textwidth]{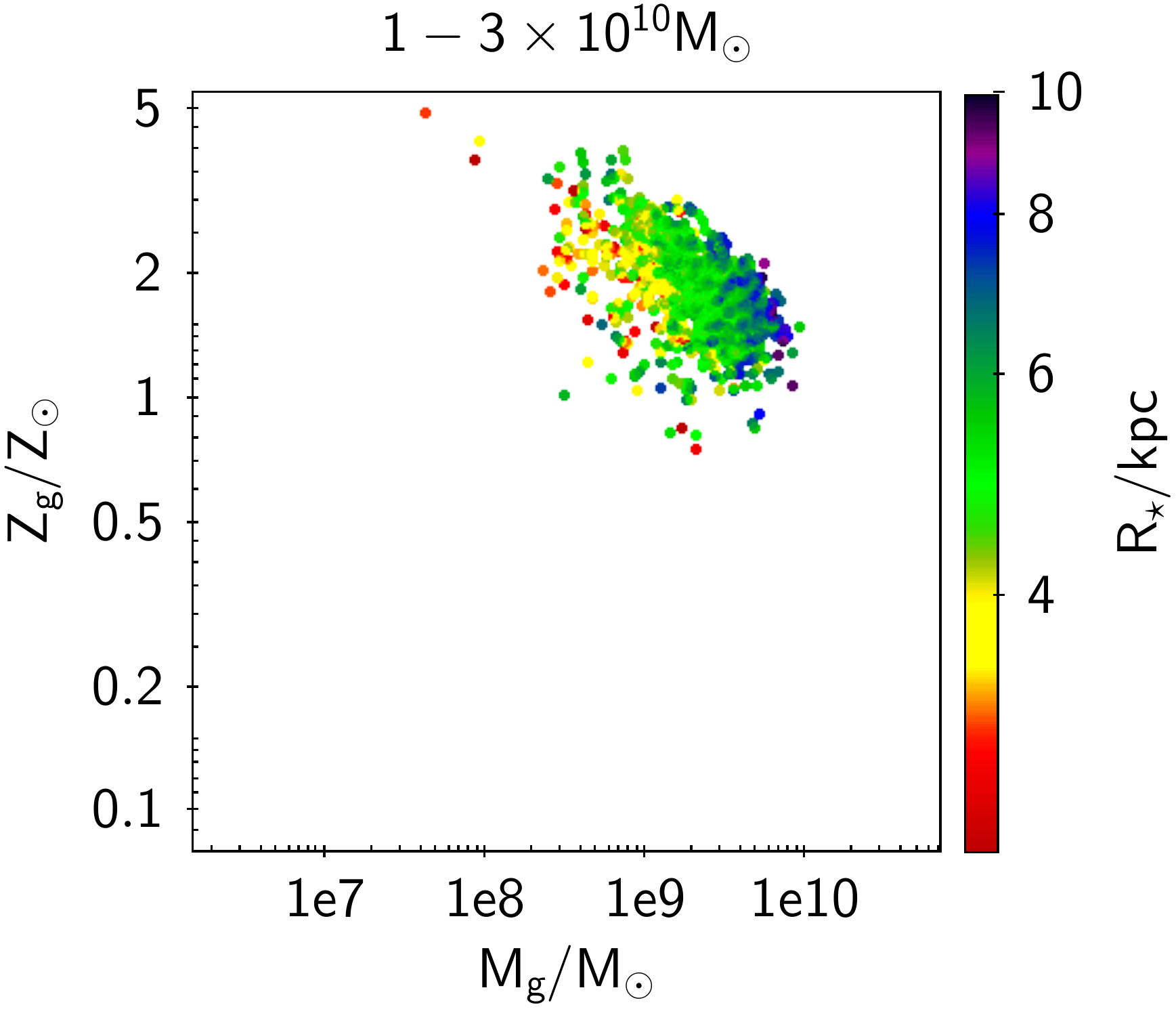}\hskip 0.2cm%
\includegraphics[width=0.33\textwidth]{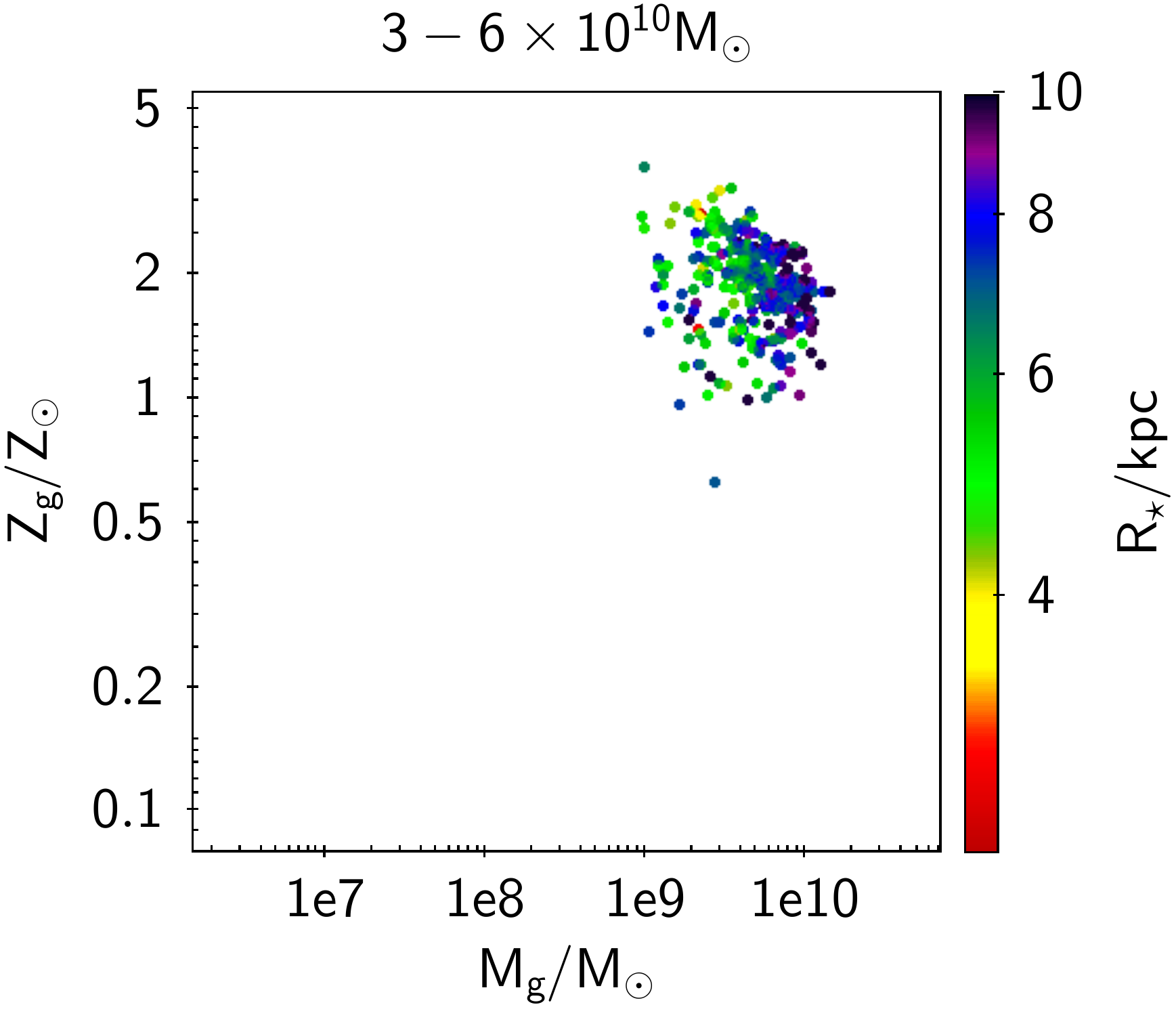}\hskip 0.2cm%
\includegraphics[width=0.33\textwidth]{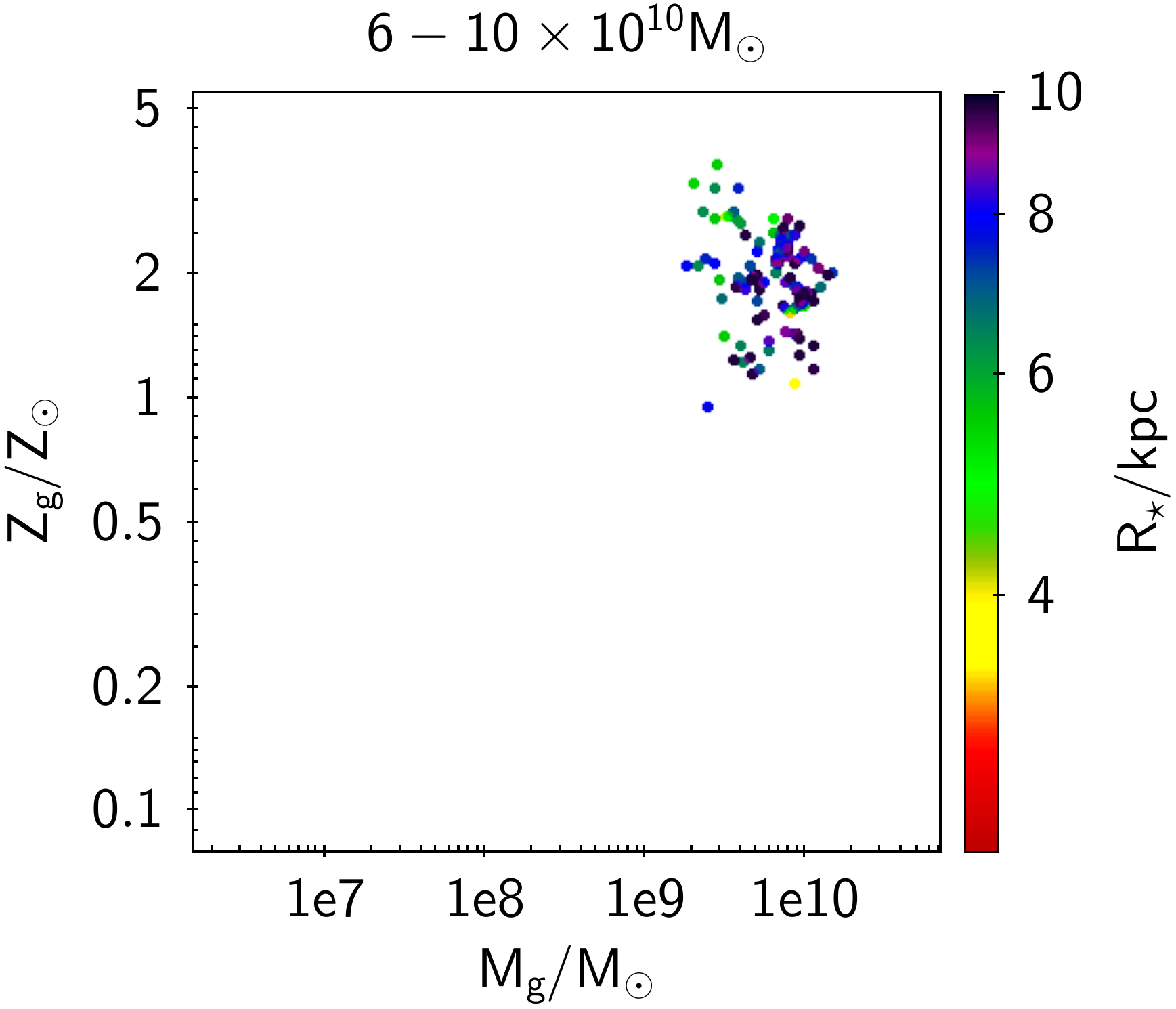}\hskip 0.2cm%
\caption{Gas-phase metallicity ($Z_g$) versus $M_g$ color-coded with the half stellar mass radius of the galaxies ($R_\star$). The top-left panel displays the full data set. The rest of the panels show the same scatter plot selecting narrow mass bins (as labelled on top of each one). There is a clear anti-correlation between ordinates and abscissae.  The axes and the color code are identical in all panels. The behavior is similar to that shown by the SFR in Fig.~\ref{fig:z_vs_sfr} -- the slope of the solid red line shown in the central panels of the two figures is the same. 
}
\label{fig:z_vs_mg}
\end{figure*}

There is a tight correlation between $M_g$  and SFR in the EAGLE galaxies. Therefore, the above discussion could have been made in terms of $M_g$ rather than SFR. However, we would have reached exactly the same conclusion because $M_g$ is also a proxy for $\dot{M}_{in}$. Both are proportional in the stationary state solution (see Eq.~[\ref{eq:mg0}]) and, even in general, $M_g$ represents a time-average of $\dot{M}_{in}$ over a time-lapse $\tauin$; see Eq.~(\ref{2ndeq}). Figure~\ref{fig:z_vs_mg} is similar to Fig.~\ref{fig:z_vs_sfr} but replacing SFR with $M_g$, and the behavior and the trends coincide. 


\subsection{Sanity check: age of the stellar populations}

If the relation between size and gas metallicity is due to the growth of galaxies with time, then there should be a tight correlation between galaxy size and age of the stellar population. This is indeed the case.

\begin{figure*}
\begin{center}
\includegraphics[width=0.45\textwidth]{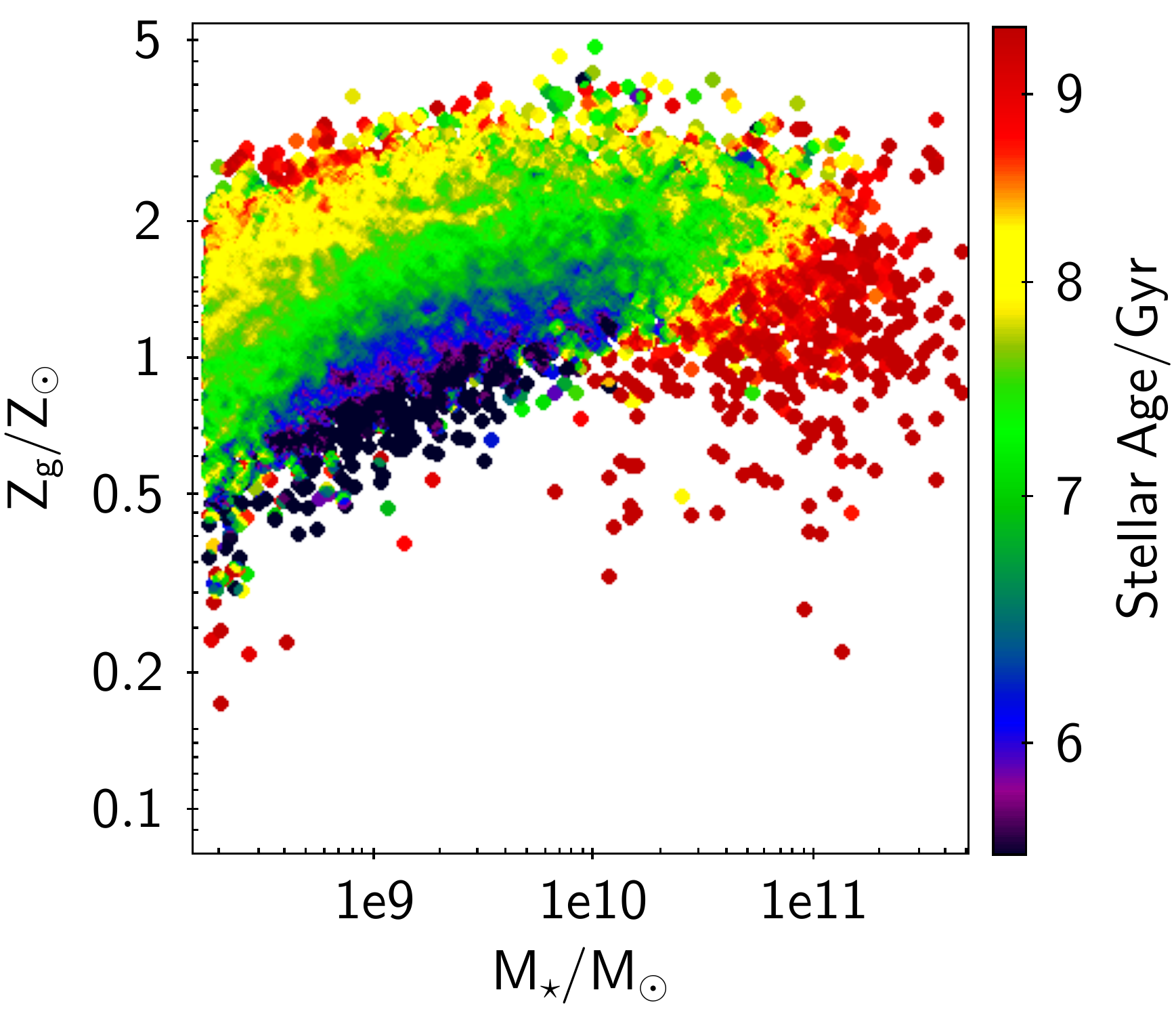}\hskip 0.5cm%
\includegraphics[width=0.45\textwidth]{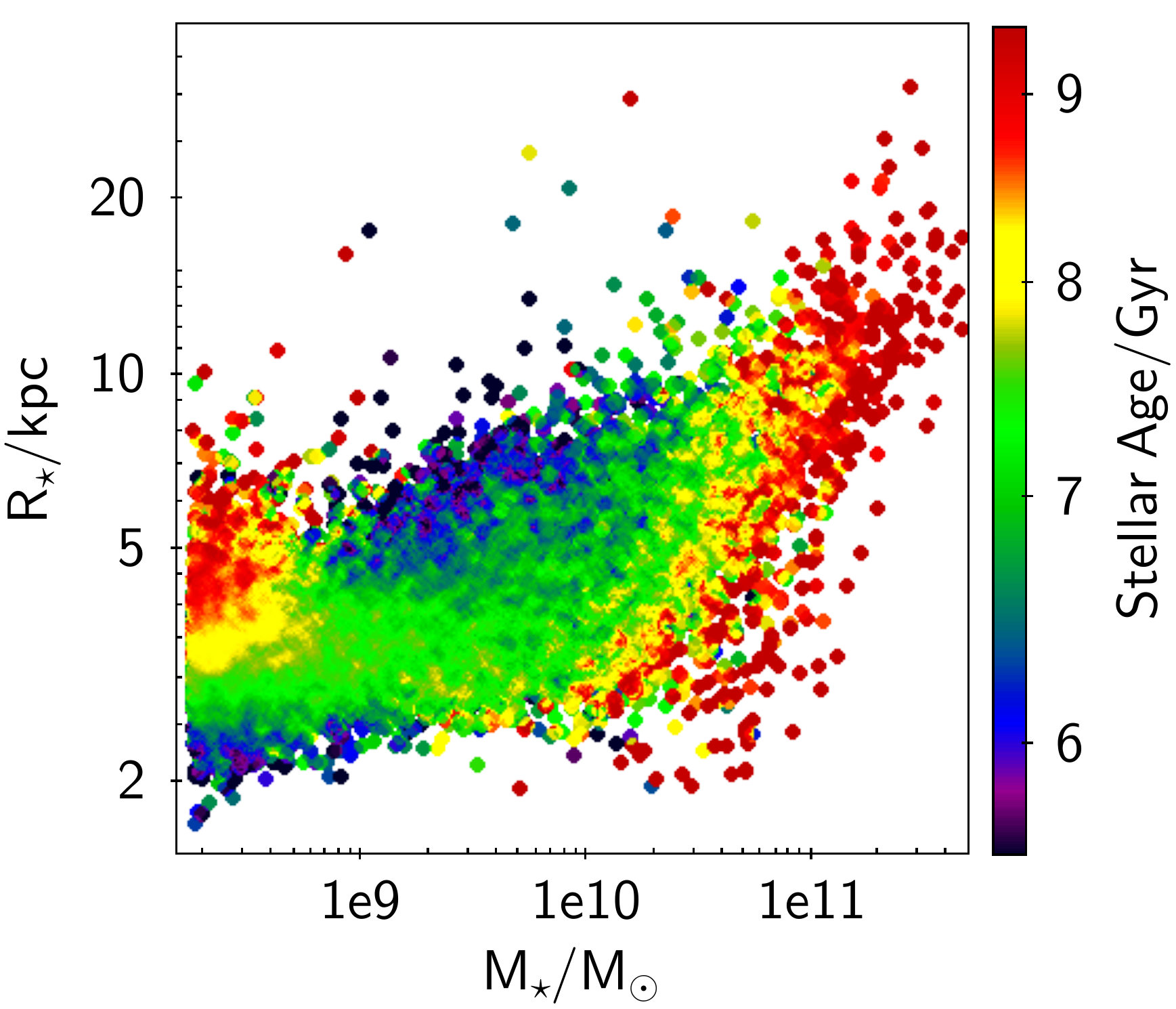}
\end{center}
\caption{Left  panel: $Z_g$ versus $M_\star$ color-coded with the mean mass-weighted age of the stars in each galaxy. Galaxies with older populations have metal richer gas. Ages are given in Gyr. 
The color code has been inverted with respect to the rest of the figures so that the oldest galaxies appear as red points.
Right panel: galaxy radius versus stellar mass color coded with stellar age. For galaxies with $M_\star > 5\times 10^8\,M_\odot$, bigger galaxies of the same $M_\star$ have younger stellar populations.}
\label{fig:sanity}
\end{figure*}

The \eagle\ database  provides the mean age of the stars in each galaxy, weighted by birth mass. This age estimate is not biased toward recent star formation, and it shows a tight correlation with $Z_g$: see Fig.~\ref{fig:sanity}, left panel. The panel on the right-hand side of Fig.~\ref{fig:sanity} shows galaxy radius versus stellar mass color-coded with mean stellar age. For galaxies with $M_\star > 5\times 10^8\,M_\odot$, bigger galaxies of the same $M_\star$ have also younger stellar populations. Again this is consistent with the idea that smaller galaxies formed earlier, at least when the stellar mass is $M_\star > 5\times 10^8\,M_\odot$, a mass limit that coincides with the onset of the relation size versus metallicity in the \eagle\ galaxies (see Figs.~\ref{fig:ellison4}a and \ref{fig:ellison4}c). 

The \eagle\ database also provides colors for the model galaxies, which show that  galaxies with metal richer gas have redder stellar populations. They are redder because they have evolved longer and, therefore, the colors are also consistent with larger galaxies having younger stellar populations.

\section{The relation between size and gas-phase metallicity at high redshift}\label{sec:redshift}

\begin{figure*}
\includegraphics[width=0.33\textwidth]{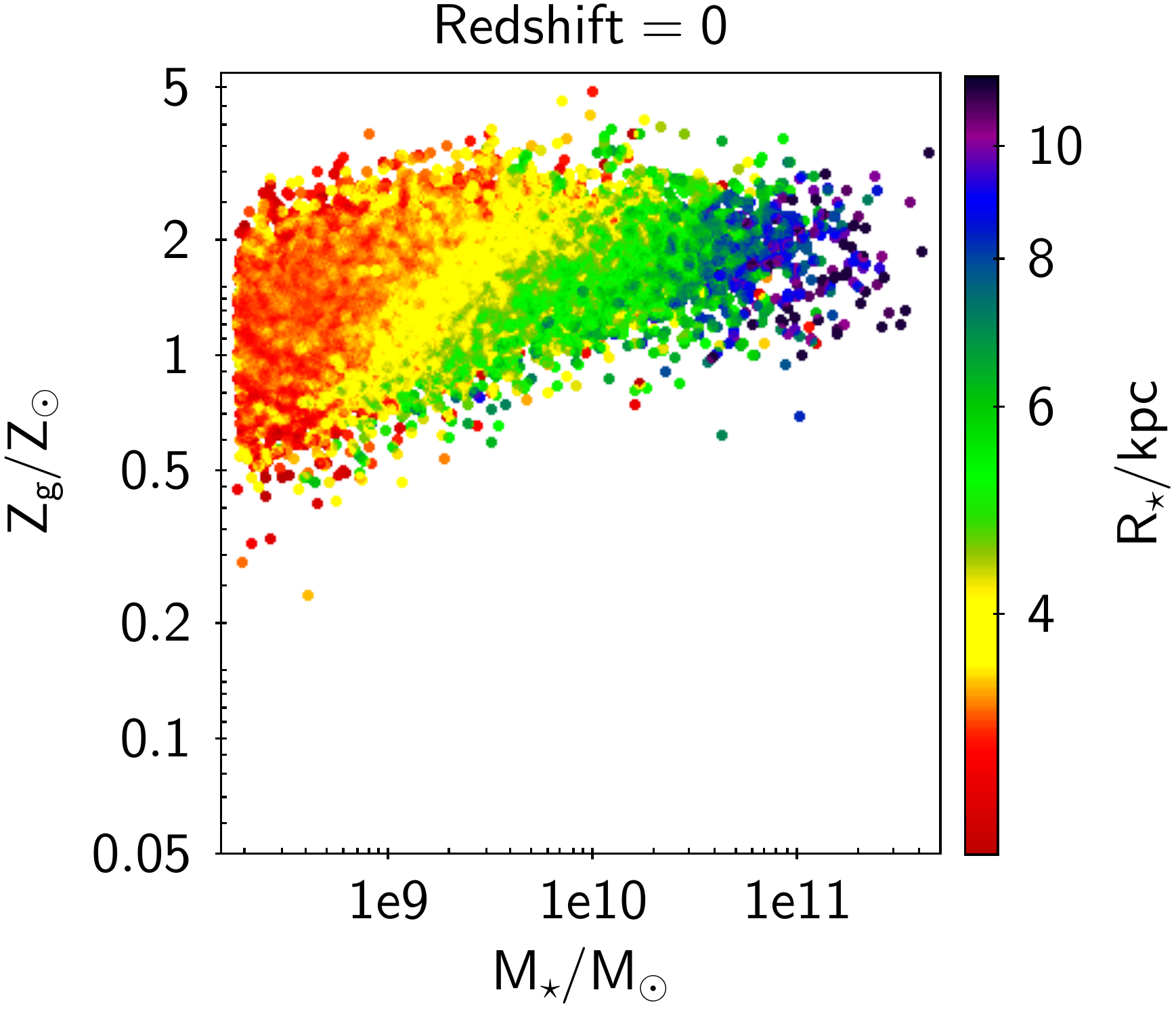}\hskip 0.2cm%
\includegraphics[width=0.33\textwidth]{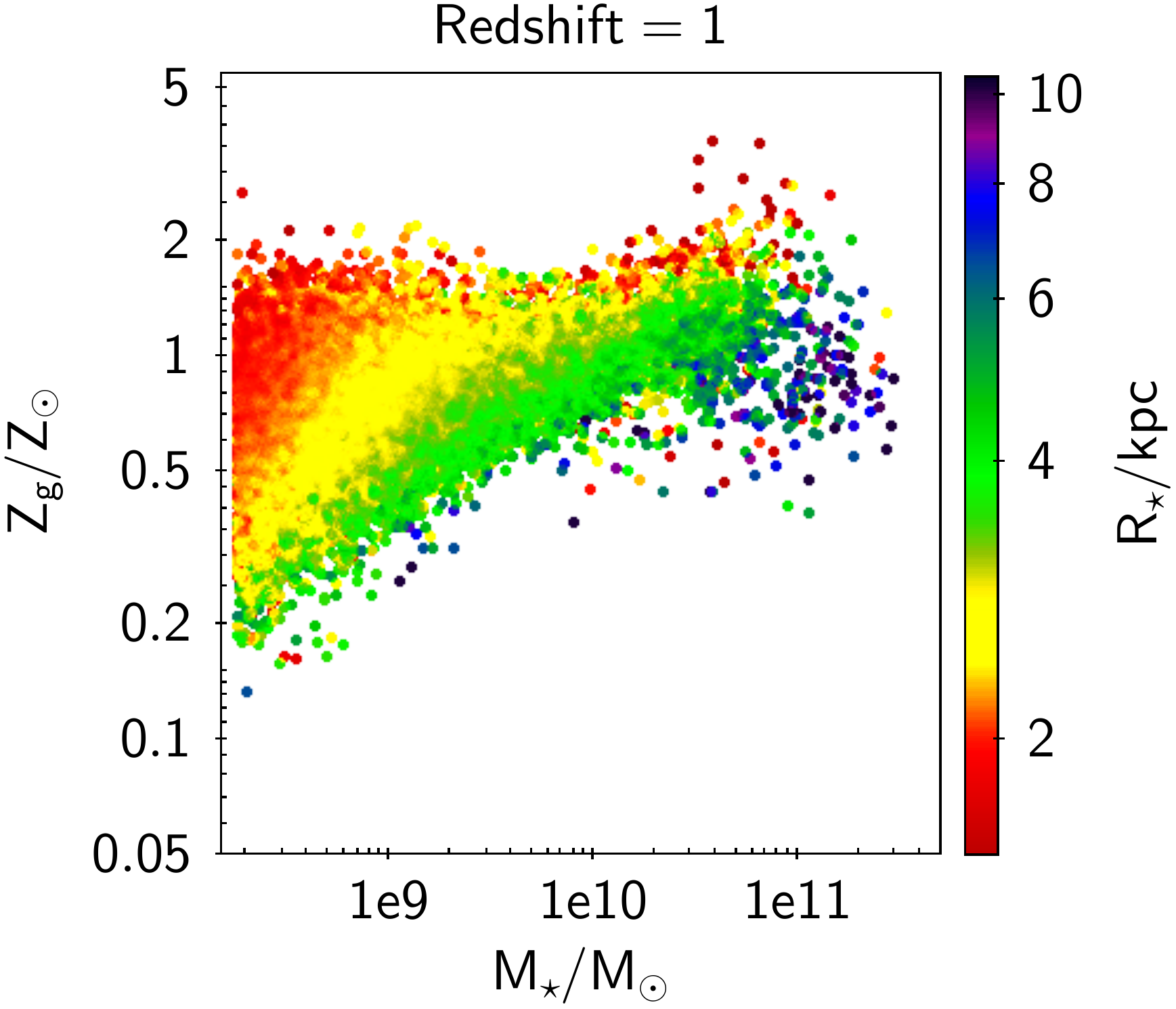}\hskip 0.2cm%
\includegraphics[width=0.33\textwidth]{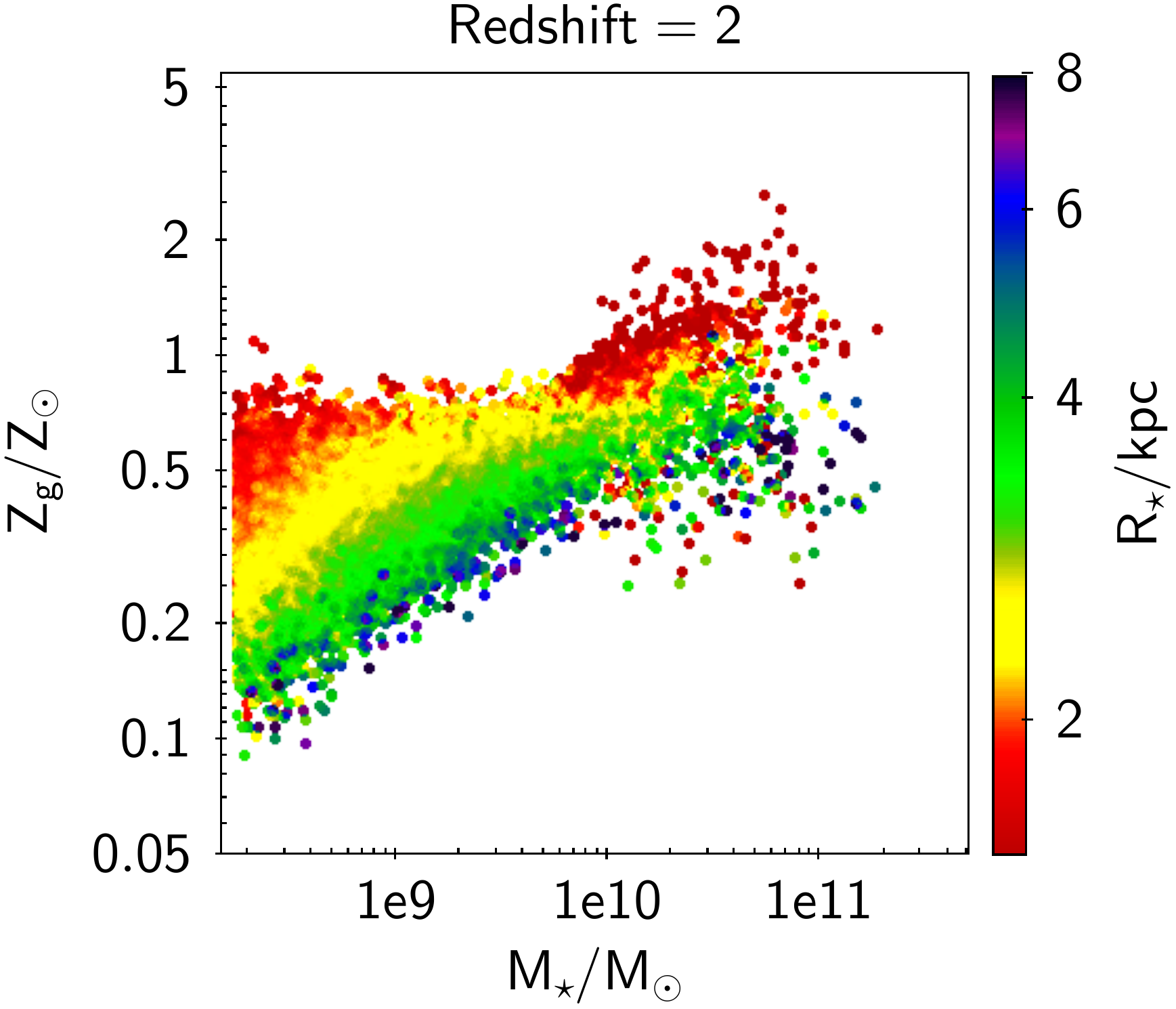}\vskip 0.2cm%
\includegraphics[width=0.33\textwidth]{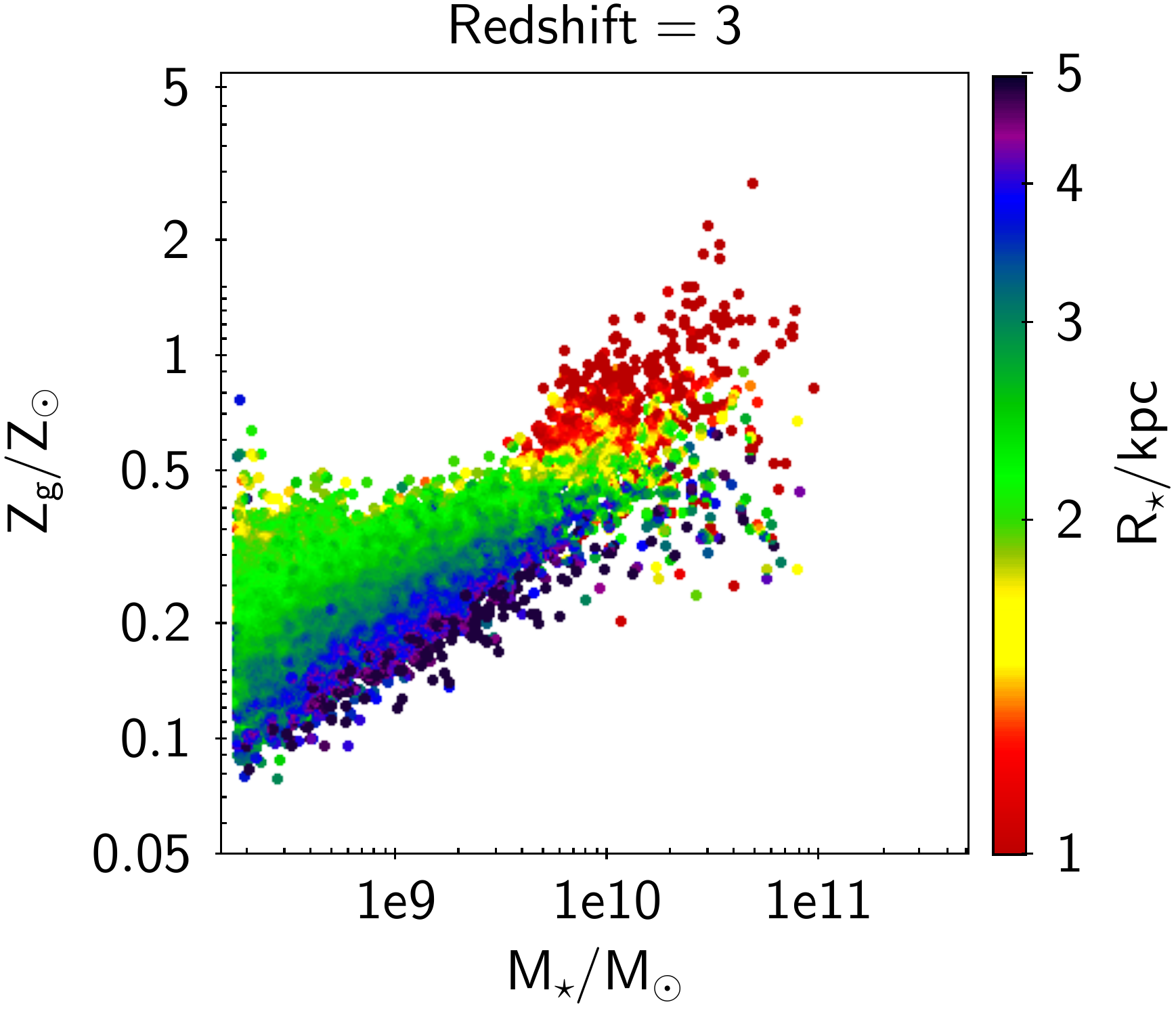}\hskip 0.2cm%
\includegraphics[width=0.33\textwidth]{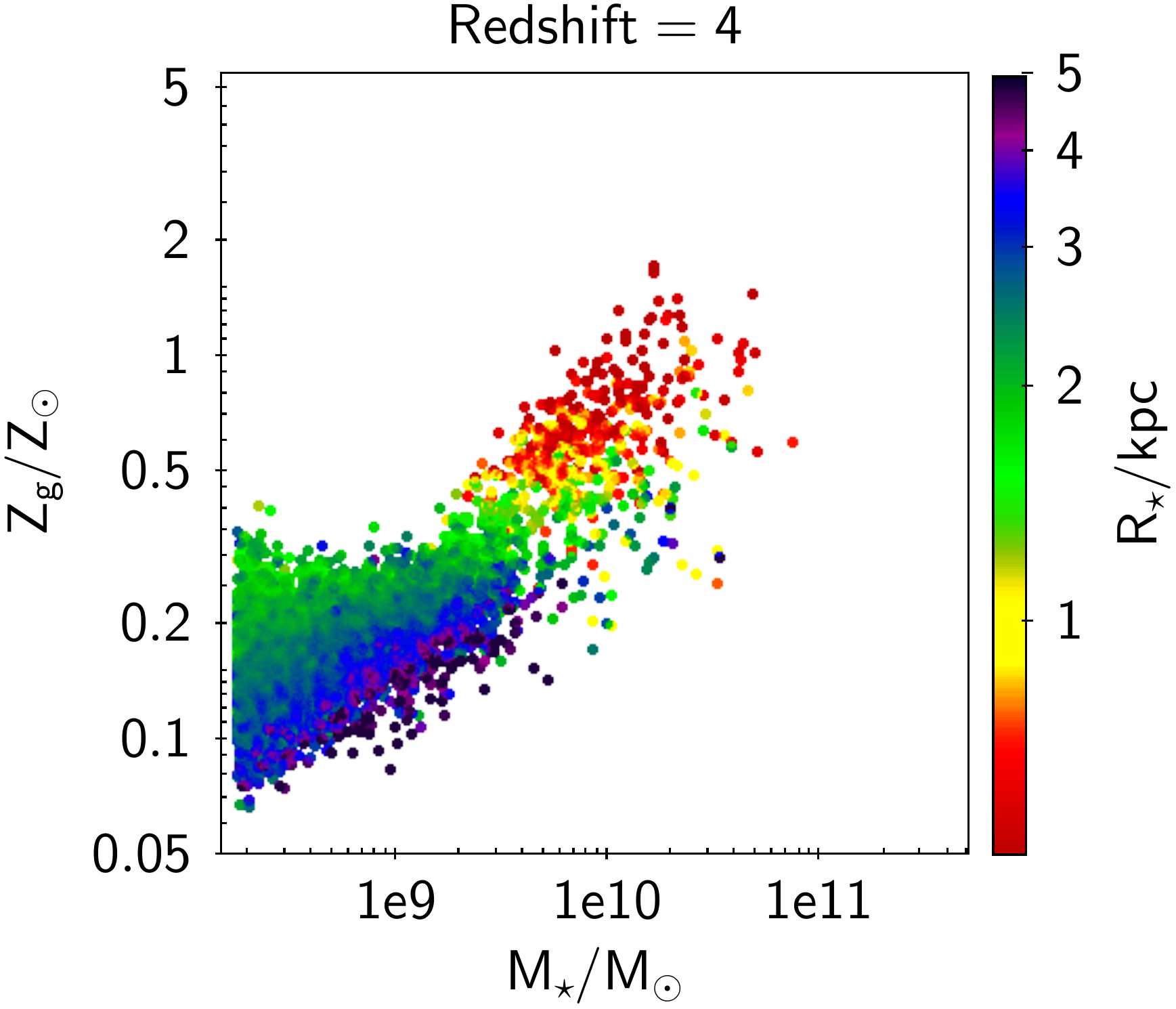}\hskip 0.2cm%
\includegraphics[width=0.33\textwidth]{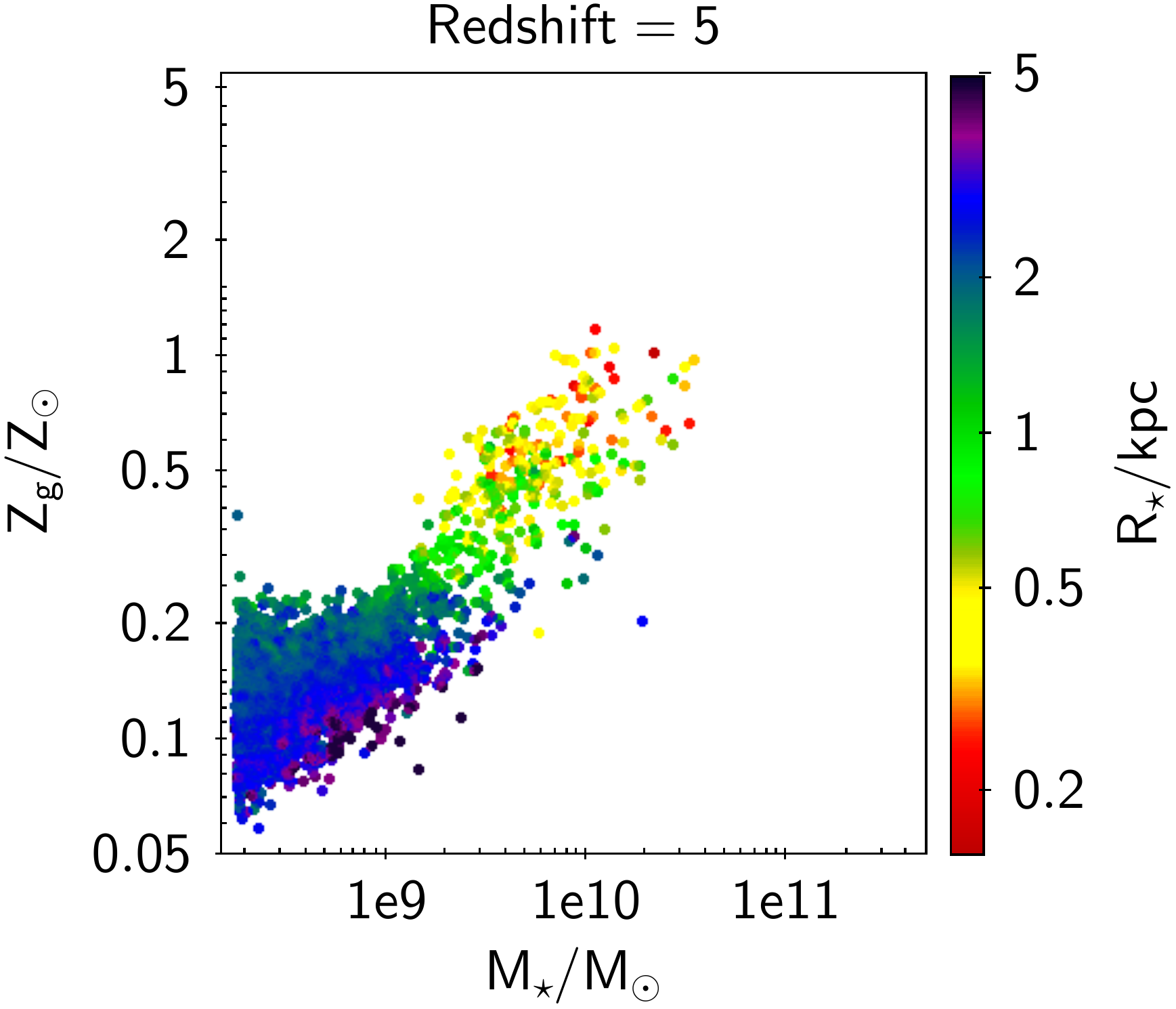}\vskip 0.2cm%
\includegraphics[width=0.33\textwidth]{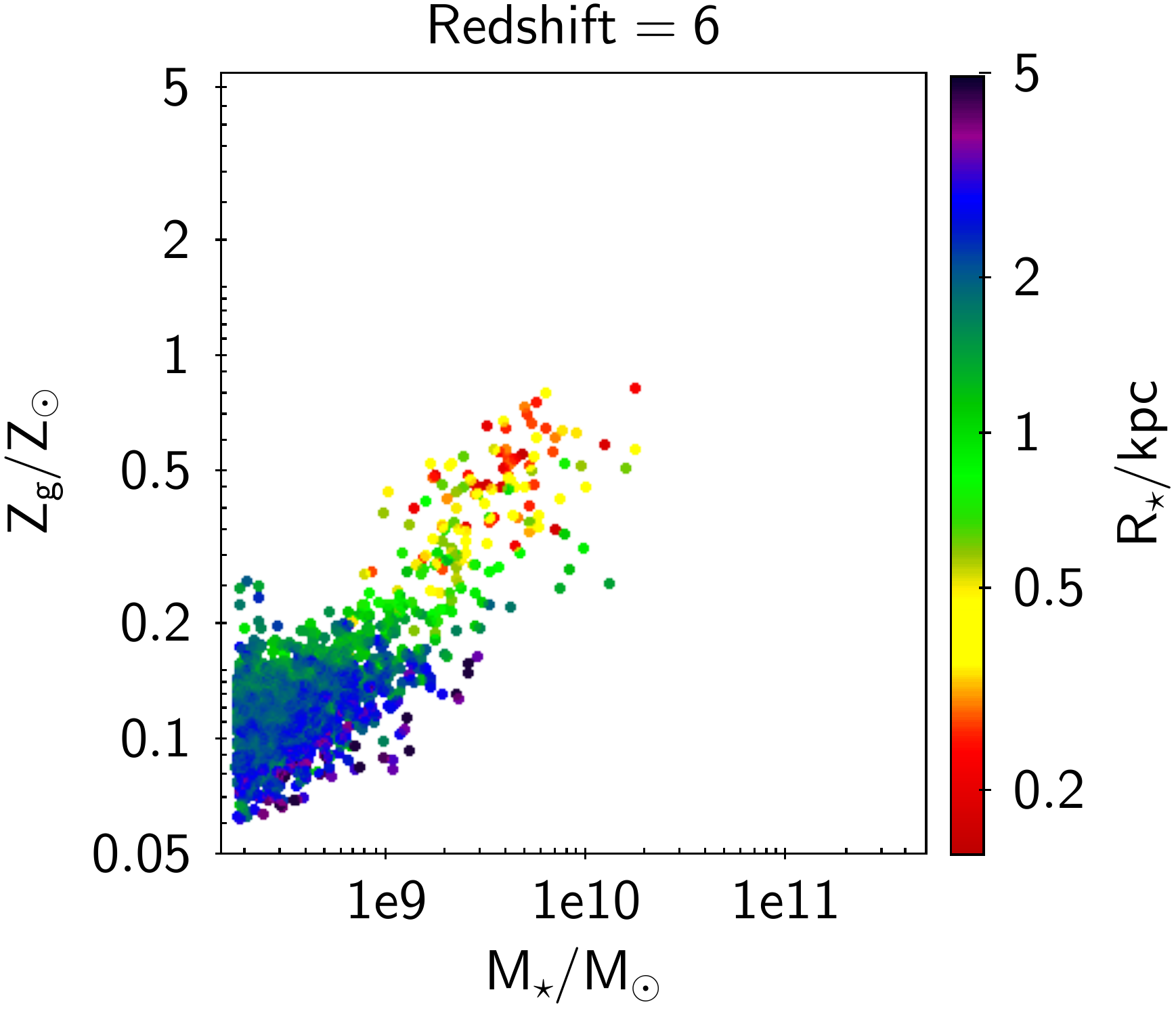}\hskip 0.2cm%
\includegraphics[width=0.33\textwidth]{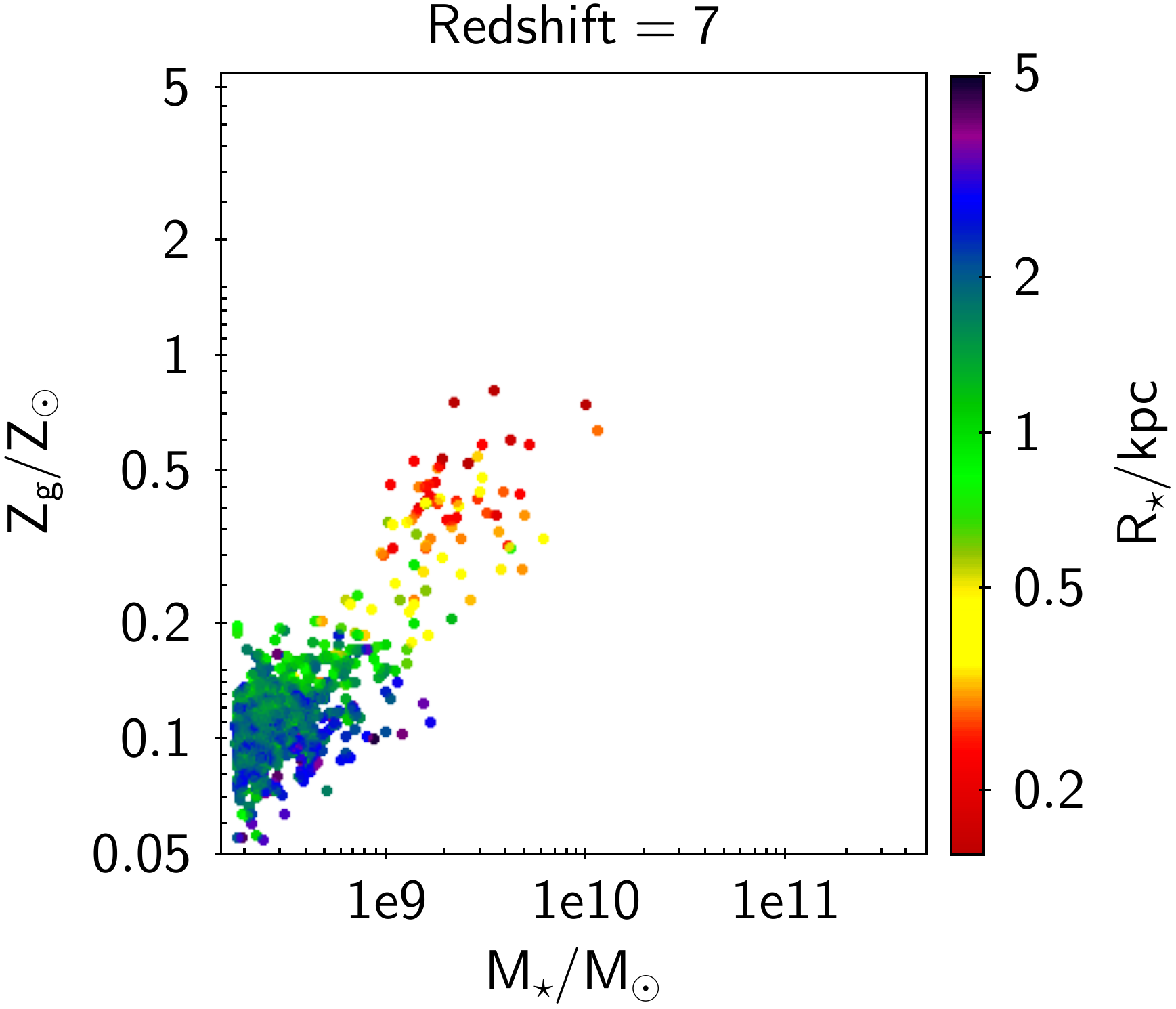}\hskip 0.2cm%
\includegraphics[width=0.33\textwidth]{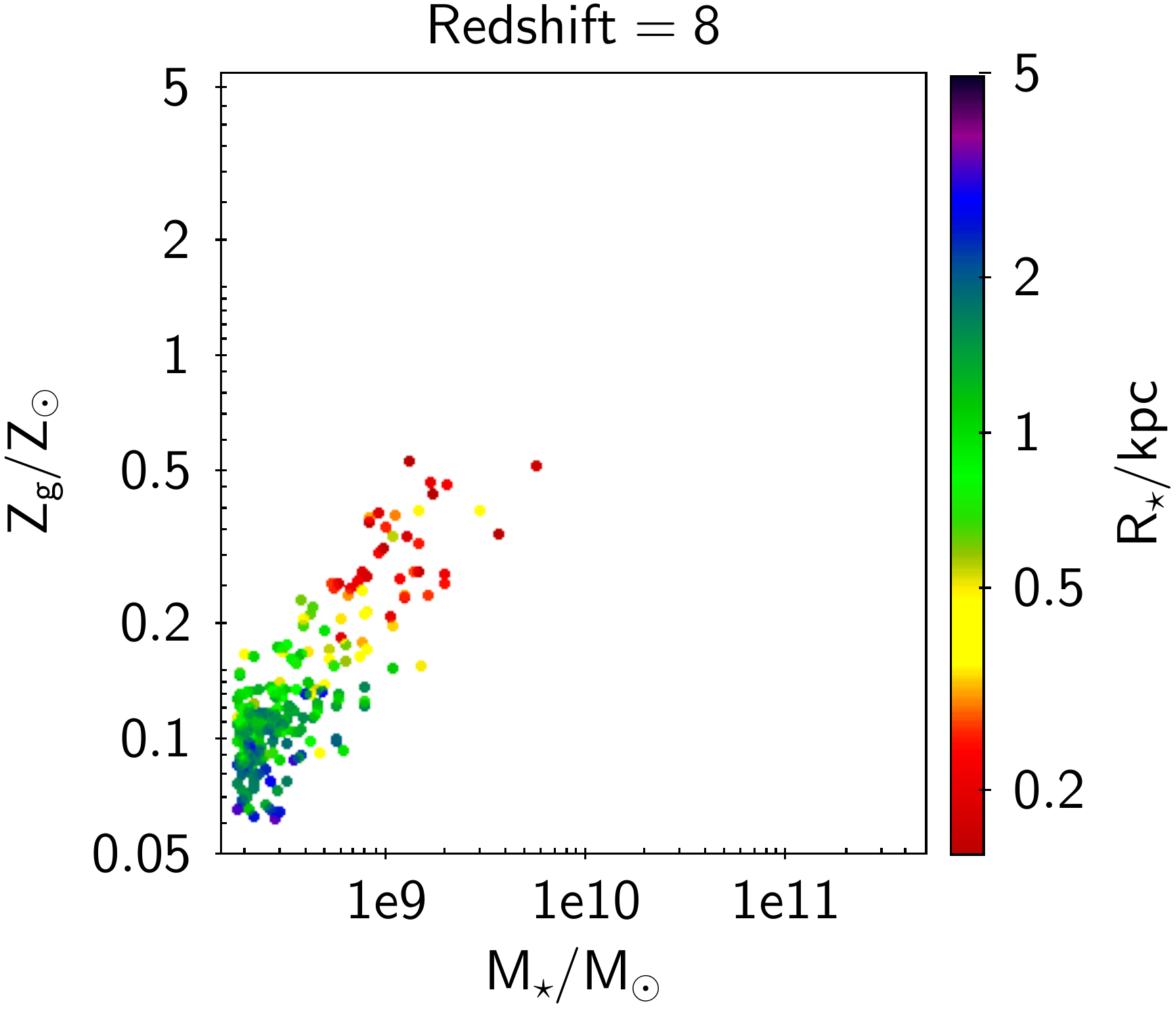}\vskip 0.2cm%
\caption{Variation with redshift of the relation between galaxy size and gas metallicity. Each panel shows, at a different redshift, the scatter plot metallicity versus stellar mass color-coded with the stellar radius of the galaxy. Top left: a copy of Fig.~\ref{fig:ellison2o} shown here for reference. Only star-forming galaxies are represented. Redshift grows from left to right and top to bottom, as indicated by the label on top of each panel. The range of masses and metallicities is the same in all panels. The range of sizes, coded by color, varies strongly, so that galaxies decrease in size with increasing redshift.}
\label{fig:ellison2zz}
\end{figure*}

The anti-correlation between size and metallicity remains at high redshift.  Figure~\ref{fig:ellison2zz} shows $Z_g$ vs $M_\star$ for the \eagle\ galaxies at redshifts from 0 to 8. (Higher redshifts are not shown because the number of objects decreases drastically, but the trends shown at redshift 8 still remain.)  Galaxies of the same $M_\star$ have different $Z_g$ according to their radii. The relation changes qualitatively at different redshifts, as does the relation between gas metallicity and mass:  the metallicities drop with increasing redshift, and the dependence of metallicity on $M_\star$ strengthens. However, all these changes conspire to accentuate the global trend between size an metallicity existing at redshift 0. High redshift galaxies tend to have their sizes and masses poorly correlated (galaxies of the same size may have a large range of masses, and vice-versa). This enhances the dependence of metallicity on mass (see, e.g., the redshift 3 plot in Fig.~\ref{fig:ellison2zz}).   

Another notable change of the relation is the emergence of a population of very compact and massive objects at redshift of one and larger
(the red points at high mass in all the panels, which are absent at redshift 0). These galaxies have to be identified with the so-called {\em blue nuggets}, which are thought to be an extreme starburst phase leading to the compaction and eventual quenching of massive galaxies at high redshift \citep[e.g.,][]{2015MNRAS.450.2327Z,2016MNRAS.457.2790T}. The end products are compact quenched {\em red nuggets}, which seem to be  precursors of  local massive ellipticals \citep{2009ApJ...697.1290B,2009MNRAS.398..898H} and/or bulges of massive spirals and S0s \citep{2013pss6.book...91G,2016MNRAS.457.1916D}.



\section{Discussions and conclusions}\label{sec:conclusions}

\citet{2008ApJ...672L.107E} found that, for the same stellar mass, physically smaller star-forming galaxies are also more metal rich. This work explores the possible physical cause of such relation. The approach is indirect. We first show that the central star-forming galaxies in the EAGLE cosmological numerical simulation reproduce the observed relation qualitatively and quantitatively (see Sect.~\ref{sec:eagle}, where we also discuss existing differences). It is a non-trivial relation, in the sense that it does not follow from other well-known relations such as the SFR--stellar mass relation (i.e., the star-formation main sequence), the mass--metallicity relation, or the fundamental metallicity relation that connects the three variables mass, metallicity, and SFR. The EAGLE simulation was not tuned in any way to comply with Ellison et al. observations, therefore, the fact that simulated galaxies follow the observed trend is taken as a solid argument supporting that the correlation, both in simulations and observations, results from a common underlying physical cause. Thus, we study the simulation to pinpoint the origin of the correlation, taking as an ansatz that the models already include all the relevant physics. However, one has to keep in mind that the conclusions of our work rely on the validity of this hypothesis.

We consider the three obvious possibilities that may change the metallicity of the star-forming gas in a galaxy in relation to its size (Sect.~\ref{sec:framework}). (1) SF driven outflows carry away gas and metals, and the  effectiveness of this process is related to the  depth of the gravitational well the SN driven winds escape from. This fact potentially links metallicity with galaxy size which, given the mass, sets the depth of the potential well. (2) The time-scale to deplete the gas depends on the gas density. Denser systems are more efficient transforming gas into  stars and, therefore, their gas become metal richer sooner. For a given mass, the smaller the galaxy the denser it is, which provides yet another potential connection between size and gas metallicity. (3) Finally, galaxies systematically grow in size with time, so  galaxies with late SF are systematically bigger than those formed earlier. Delayed SF means still having metal-poor gas, which provides a connection between size and gas-phase metallicity. 
These predictions are analyzed using a simple toy-model in Sects.~\ref{sec:depthwell}, \ref{sec:denstar}, and \ref{sect:lateacc}, which provides physical insight and allows us to estimate the magnitude of the expected effects resulting from each one of these mechanisms. 

Aided with this interpretative framework, we analyze the actual \eagle\ galaxies in Sects.~\ref{sec:well}, \ref{sec:mdensity}, and \ref{sec:thisisit}. Outflows are discarded as the cause of the correlation because, even though there is a trend for the more metal rich galaxies to have deeper gravitational potential (Fig.~\ref{fig:z_vs_vscp}, top left panel), this trend washes out when galaxies of the same mass are considered (the rest of the panels in Fig.~\ref{fig:z_vs_vscp}). In a sense, the global trend is a mirage resulting from the global increase of both the depth of the 
gravitational well and the metallicity with increasing galaxy mass.  Varying SF efficiency with mean density is also discarded as the underlying mechanism causing the anti-correlation between metallicity and size (Sect.~\ref{sec:mdensity}). We use as proxy for gas density the stellar density. Even though the global trend is similar to that expected from the toy-model (Fig.~\ref{fig:z_vs_vscp}, central panel), its amplitude is insufficient to explain the range of variability in gas-phase metallicity of the \eagle\ galaxies (Fig.~\ref{fig:z_vs_vscp}). Finally, the growth of galaxy size with time, coupled with the recent accretion of metal-poor gas, seems to be the cause (Sect.~\ref{sec:thisisit}). If so,  galaxies of the same mass should present an anti-correlation between the recent gas infall rate and the gas metallicity. We use as proxies for the gas infall rate the SFR and the gas mass, and both indicators show a tight anti-correlation with metallicity (see Figs.~\ref{fig:z_vs_sfr} and \ref{fig:z_vs_mg}). Moreover, as expected if the correlation between size and gas metallicity is produce by recent gas accretion, the age of the stellar population of a galaxy is tightly connected with the stellar size and the gas metallicity, so that older stellar populations are characteristic of smaller metal-richer galaxies (Fig.~\ref{fig:sanity}).  

The \eagle\ galaxies need of the growth in size with time to explain the observation by \citeauthor{2008ApJ...672L.107E} This growth in size is extensively discussed in the literature. It already appears in the classical theoretical paper by \citet{1998MNRAS.295..319M}, although based on hypotheses that may not be realistic 
\citep[e.g.,][]{2012MNRAS.423.1544S,
2017arXiv171203966G}. 
In the case of the \eagle\ simulation, the mechanism of growth depends on the galaxy mass. It is due to in-situ star formation fueled by gas accretion and minor gas-rich mergers when $\log(M_\star/M_\odot)< 10.5$, whereas dry mergers play a significant role only at the high-mass end, with $\log(M_\star/M_\odot) > 11$ \citep[see][]{2017MNRAS.464.1659Q}. Stellar migration is important in the size evolution of the \eagle\ massive red compact galaxies, common at redshift 2 and depleted below redshift 1 \citep[][]{2017MNRAS.465..722F}, but these objects and their descendants contribute little to the population of star-forming galaxies at redshift 0. In principle, stellar-migration and other secular  processes \citep[e.g.,][]{2016ApJ...820..131E} redistribute old stars within the galaxy, thus distorting an initial correlation between galaxy size and  age of the stellar population. However, these mechanisms do not seem to be effective enough to  blur the underlying trend for the star-forming galaxies to increase in size with time.
Both observations and simulations show that high redshift galaxies are smaller.  This includes passively 
evolving galaxies \citep[e.g.,][]{2005ApJ...626..680D,2006MNRAS.373L..36T,2008ApJ...687L..61B}  as well as star-forming galaxies \citep[][]{2016A&A...593A..22R,2017ApJ...834L..11A}. Numerical simulations reproduce the observed trends satisfactorily \citep[e.g.,][]{2015MNRAS.449..361W,2017arXiv170705327G,2017MNRAS.465..722F}. 
Observations also show a clear relation between size and age, so that smaller galaxies generally have older stellar populations.
This result holds for passively evolving galaxies (redshift 0.2 to 0.8 and $M_\star < 10^{11}\,M_\odot$, \citeauthor{2016ApJ...831..173F}~\citeyear{2016ApJ...831..173F}; redshift $\sim 1.2$, \citeauthor{2017ApJ...838...94W}~\citeyear{2017ApJ...838...94W}; redshift 0 and $M_\star < 3\times 10^{10}\,M_\odot$, \citeauthor{2010MNRAS.403..117S}~\citeyear{2010MNRAS.403..117S}), as well as for late-type galaxies (e.g., \citeauthor{2010MNRAS.404.2087B}~\citeyear{2010MNRAS.404.2087B}). The same kind of relation between size and  age probably  explains the difference in size between star-forming and passively evolving galaxies. Given $M_\star$, the later are systematically smaller than the former \citep[e.g.,][Figs.~5 and 6]{2014ApJ...788...28V}, and the stellar populations in early-type galaxies are systematically older than  in late-types. Star formation histories  derived from spatially resolved spectra clearly show the inside-out growth of the galaxies, with younger stellar populations in the outskirts \citep[e.g.,][]{2013ApJ...764L...1P,2017MNRAS.466.4731G,2017A&A...608A..27G}.

It is important to realize that the above explanation is posible only if the galaxies have not reached equilibrium with the average mass accretion rate. Otherwise, the gas-phase metallicity is set only by  stellar physics and wind strength, and it is independent of the mass infall rate, the gas mass, or the SFR  (Eq.~[\ref{eq:ellison2_save_ssfr}]). Most galaxies have to be in a transient phase where the gas obtained during the last major gas accretion episode is still in use. A natural way for the galaxies to be systematically out of equilibrium is if the accretion turns out to be very bursty, with discrete accretion events followed by long gas-starved periods in between.          

We explore the variation with redshift of the relation between metallicity and size in the \eagle\ simulation. It is maintained up to at least redshift 8 (Fig.~\ref{fig:ellison2zz}). This fact remains to be tested observationally, but it results encouraging that the relation remains in place at redshift $\sim 1.4$ according to \citet{2012PASJ...64...60Y,2014MNRAS.437.3647Y}.


%
%
%
%
%
%


\acknowledgments
Thanks are due to the EAGLE team for making the simulations publicly available. 
This work was carried out while JSA was visiting the University of California at Santa Cruz (UCSC). He thanks David Koo for his hospitality during this period, and Guillermo Barro, Francesco Belfiore, Nicolas Bouch\'e, and Daniel Ceverino for comments on particular aspects of the analysis. 
Thanks are also due to an anonymous referee for helping us refining some of the arguments put forward in the paper.
The work has been partly funded by the Spanish Ministry of Economy and  Competitiveness (MINECO), projects {\em Estallidos} AYA2013--47742--C04--02--P and AYA2016-79724-C4-2-P, as well as by the {\em Severo Ochoa Excellence Program} granted to the Instituto de Astrof\'\i sica de Canarias by  MINECO (SEV-2015-0548).  
CDV acknowledges financial support from MIMECO through grants AYA2014-58308-P and RYC-2015-18078. 
%
%

\vspace{5mm}


\software{TOPCAT, EAGLE interface \citep{2016A&C....15...72M}
          }



\appendix


\section{Time dependence of the gas-phase metallicity in the galaxy toy-model}\label{sec:appb}

Here we assume that the galaxies have only two components: gas and stars. Using the equation of mass conservation from  chemical evolution models \citep[e.g.,][]{1980FCPh....5..287T,1990MNRAS.246..678E}, the variation with time $t$ of the mass of gas available to form  stars, $M_g(t)$, is given by,
\begin{equation}
\dot{M}_g=-(1-R)\,{\rm SFR}+\dot{M}_{in}-\dot{M}_{out},
\label{1steq}
\end{equation}
which considers the formation of stars  (1st term on the right-hand side of the equation), a gas inflow rate ${\dot{M}_{in}}(t)$, and a gas outflow rate ${\dot{M}_{out}}(t)$. As usual, dotted quantities represent time derivatives. The SFR is assumed to be proportional to the gas mass, 
\begin{equation}
{\rm SFR}=\frac{M_g}{\taug}.
\label{kslaw}
\end{equation}
with the scaling parameterized in terms of a gas consumption timescale $\taug$ (see Eq.~[\ref{eq:taug}]). (Note that Eq.~(\ref{kslaw}) is quite general since $\taug$ may depend on the physical properties of the galaxy, including the surface gas density.)  We will also assume the outflow rate to scale with the SFR,
\begin{equation}
\dot{M}_{out}=w\,{\rm SFR},
\label{mass_load}
\end{equation}
with $w$ the so-called mass-loading factor  (see Eq.~[\ref{eq:massload}]). Equation~(\ref{mass_load}) results natural  if outflows are driven by 
stellar winds or SN explosions, but may also include AGN feedback if the AGN activity is correlated with  SF.
Under the same approximations leading to Eq.~(\ref{1steq}), the metallicity of the gas that is forming stars, $Z_g$, follows the differential equation,
\begin{equation}
\Delta \dot{Z_g}=(1-R)\,y/\taug - \Delta Z_g\, \dot{M}_{in}/{M_g},
\label{deltametal}
\end{equation}
where $\Delta Z_g$ represents the difference between $Z_g$ and the metallicity of the accreted gas $Z_{in}$. The symbol $y$ stands for the stellar yield (the mass of new metals eventually ejected per unit mass locked  into stars and stellar remnants), and $R$ represents the mass return fraction  (the fraction of mass in stars that returns to the interstellar medium).

The gas mass and its metallicity follow from  integrating Eqs.~(\ref{1steq}) and (\ref{deltametal}), given Eqs.~(\ref{kslaw}) and (\ref{mass_load}), once $\dot{M}_{in}(t)$ is set. We are interested in bursty accretion, where the accretion rate can be approximated as, 
\begin{equation}
\dot{M}_{in}(t)=\dot{M}_{in0}+\delta(t-t_a)\,\Delta{M}_a,
\label{bursts}
\end{equation} 
with $\dot{M}_{in0}$ representing a background accretion rate on top of which the galaxy receives a gas clump of mass $\Delta M_a$ at $t=t_a$. The symbol  $\delta$ stands for the Dirac delta function. Provided that all scaling factors $R, w$ and $\taug$ are constant in time, and $t >> \tauin$, the general solution of Eq.~(\ref{1steq}) is 
\begin{equation}
M_g(t)=\int_0^t\,\dot{M}_{in}(t^\prime)\,{\rm \Large e}^{-(t-t^\prime)/\tau_{in}}\,dt^\prime,
\label{2ndeq}
\end{equation}
with
\begin{displaymath}
\tau_{in}=\taug\,/(1-R+w).
\end{displaymath}
Using the  accretion rate in Eq.~(\ref{bursts}), at $t >> \tauin$, the mass of gas turns out to be 
\begin{equation}
M_g(t)=M_{g0}+ \Delta {M}_a\, {\rm H}(t-t_a) \exp[-(t-t_a)/\tauin],
\label{mymgas}
\end{equation} 
with 
\begin{displaymath}
{M_{g0}}={\dot{M}_{in0}}\,\tauin ,
\end{displaymath}
and with the symbol ${\rm H}(t)$ standing for the Heaviside step function,
\begin{equation}
{\rm H}(t)= \cases{0 &if $t < 0$, \cr 1& if $t \geq 1$.}
\end{equation}
Equation~(\ref{deltametal}) is a first order linear differential equation that admits a formal
solution similar to Eq.~(\ref{2ndeq}). It can be integrated using the mass of gas in Eq.~(\ref{mymgas}) 
and the accretion rate in Eq.~(\ref{bursts}) and, after some algebra, the gas metallicity turns out to be 
\begin{equation}
\Delta Z_g(t)/\Delta Z_{g0}=\frac{M_{g0}+\Delta{M}_a\, {\rm H}(t-t_a)\,\exp[-(t-t_a)/\tauin]\,(t-t_a)/\tauin }
{M_{g0}+\Delta{M}_a\,{\rm H}(t-t_a)\,\exp[-(t-t_a)/\tauin]},
\label{my_metal2}
\end{equation}
%
with
\begin{displaymath}
\Delta Z_{g0}=y\,(1-R)/(1-R+w).
\end{displaymath}

\begin{figure}
\begin{center}
\includegraphics[width=0.5\linewidth]{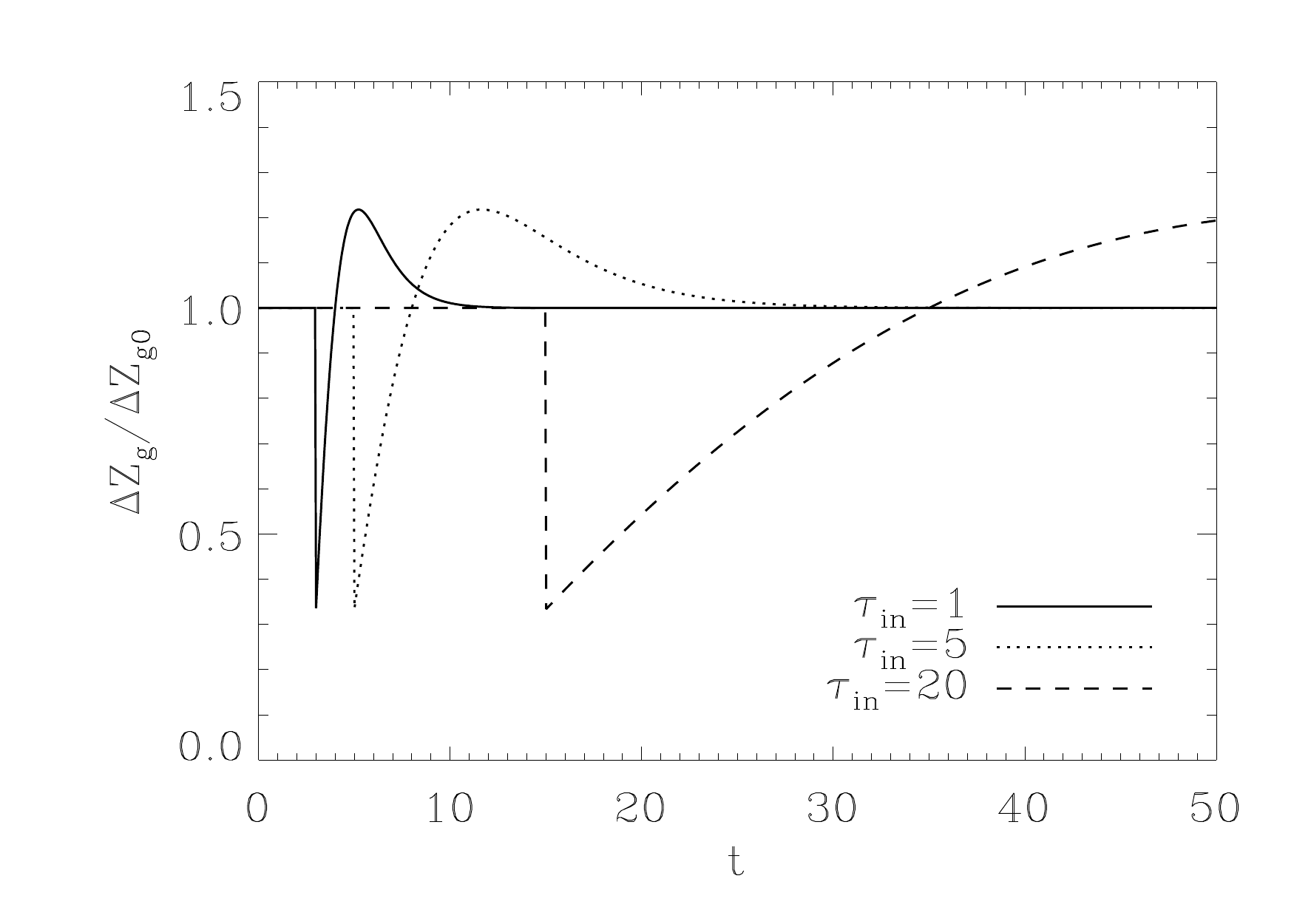}
\end{center}
\caption{
Gas-phase metallicity variation to be expected from a single gas infall event. After a sudden drop at the infall time, the metallicity increases to become the stationary state metallicity $\Delta Z_{g0}$, and keeps increasing to reach a maximum value.  The time-scale to recover the stationary state metallicity after the initial drop is just $\tauin$ whereas it takes much longer to return to the stationary state when the excess is positive. We show three events involving the same gas mass ($\Delta {\rm M_a}/{\rm M_{g0}} = 2$), occurring at different times ($t=3$, 5, and 20), and having three different gas timescales to consume the gas ($\tauin=1$, 3, and 20; see the inset). In the case where $\tauin=20$, the galaxy has not reached the stationary state metallicity yet.  The time $t$ is given in units of $\tauin$.
} 
\label{fig:single_burst}
\end{figure}

Equation (\ref{my_metal2}) describes a sudden drop in metallicity at the moment of accretion, that recovers the stationary state metallicity after $\tauin$, and then the metallicity keeps increasing, reaches a maximum, and decays again within a time-scale significantly larger than $\tauin$. The behavior is illustrated in Fig. \ref{fig:single_burst}, which represents a fairly massive burst with a mass contrast $\Delta {M_a}/{M_{g0}} = 2$. 

The stationary-state solution corresponds to $t\rightarrow\infty$, and it renders, 
\begin{equation}
{M_g}(\infty)={M_{g0}}={\dot{M}_{in0}}\,\tauin,
\label{eq:mg0}
\end{equation}
\begin{equation}
{\rm SFR}(\infty)={{\rm SFR}_0}\equiv\,{\dot{M}_{in0}}/(1-R+w),
\end{equation}
and,
\begin{equation}
\Delta Z_g(\infty)=\Delta Z_{g0}.
\label{eq:z0app}
\end{equation}  
We note that in the stationary state, the ratio between SFR and gas-phase metallicity is independent of the mass loading factor $w$, explicitly,
\begin{equation}
\frac{\rm SFR_{0}}{\Delta Z_{g0}}=\frac{\dot{M}_{in0}}{y\,(1-R)}.
\label{eq:ultimate}
\end{equation}


\section{Computing the escape velocity from the \eagle\ model galaxies}\label{sec:appa}

The escape velocity at distance $r$ from a spherically mass distribution is given by  
\begin{equation}
v_{esc}(r)=\sqrt{2\,|\Phi(r)|},
\label{eq:appa1}
\end{equation}
with $\Phi(r)$ the gravitational potential \citep[e.g.,][Eq.~{[2.31]}]{2008gady.book.....B}. In the case of a NFW profile,
\begin{equation}
\Phi(r)=-4\pi\,G\,\rho_0\,R_s^2\,\frac{\ln (1+r/R_s)}{r/R_s},
\label{eq:appa2}
\end{equation}
\citep[e.g.,][Eq.~{[2.67]}]{2008gady.book.....B}, where $R_s$ and $\rho_0$ are the two free parameters of the NFW density profile,
\begin{equation}
\rho(r)=\frac{\rho_0}{(r/R_s)\,(1+r/R_s)^2}.
\end{equation}
Once $R_s$ and $\rho_0$ are known, then the escape velocity is given by Eqs.~(\ref{eq:appa1}) and (\ref{eq:appa2}). The EAGLE database does not provide $R_s$ and $\rho_0$, but instead it provides the total mass of the halo when the mean density is 200 of the critical density $\rho_c$, $M_{200}$, as well as the corresponding half-mass radius $R_{1/2}$.  The former parameters can be obtained from the later parameters considering that the mass enclosed within a radius $r$ is \citep[e.g.,][Eq.~{[2.66]}]{2008gady.book.....B},
\begin{equation}
M(r)=4\pi\,\rho_0\,R_s^3\,\Big[\ln\Big(\frac{r+R_s}{R_s}\Big)-\frac{r}{r+R_s}\Big].
\label{eq:appa4}
\end{equation} 
By definition, $M(R_{1/2})=M_{200}/2$, therefore,
\begin{equation}
\ln\Big(\frac{R_{200}+R_s}{R_{s}}\Big)-\frac{R_{200}}{R_{200}+R_s}=\frac{1}{2}\,
\Big[\ln\Big(\frac{R_{1/2}+R_s}{R_{s}}\Big)-\frac{R_{1/2}}{R_{1/2}+R_s}\Big],
\label{eq:appa3}
\end{equation}
with
\begin{equation}
R_{200}^3=M_{200}\Big/\Big(\frac{4\pi}{3}\,200\,\rho_c\Big).
\label{eq:appa5}
\end{equation}
Since $\rho_c$, $R_{1/2}$ and $M_{200}$ are known, Eqs.~(\ref{eq:appa3}) and (\ref{eq:appa5}) allow us to infer $R_s$. We solve it iteratively.  Then Eq.~(\ref{eq:appa4}) at $r=R_{1/2}$ and $M=M_{200}/2$ provides $\rho_0$.

\end{document}